\documentclass[letterpaper]{JHEP3}
\usepackage{epsfig,multicol,amsmath}

\newcommand{\roughly}[1]{\mathrel{\raise.3ex\hbox{$#1$\kern-0.85em
\lower1ex\hbox{$\sim$}}}}

\newcommand{\lsim}{\roughly<}

\def\exd{{\hbox{d}}}

\def\endignore{}
\def\ignore #1\endignore{} % use to "comment out" text

\def\ba{\begin{eqnarray}}
\def\ea{\end{eqnarray}}
\def\be{\begin{equation}}
\def\ee{\end{equation}}

\def\ssA{{\scriptscriptstyle A}}
\def\ssB{{\scriptscriptstyle B}}

\def\AB{{\scriptscriptstyle AB}}

\def\A{\mathcal{A}}

\def\F{\mathcal{F}}

\def\M{\mathcal{M}}
\def\O{\mathcal{O}}
\def\R{\mathcal{R}}

\def\nn{\nonumber}

\def\({\left(}
\def\){\right)}

\def\pref#1{(\ref{#1})}

\def\Press{p}
\def\Dens{\varrho}
\def\PressC{p_0}
\def\PressCmax{p_0^{\rm max}}

\title{
Semi-Analytic Stellar Structure in Scalar-Tensor Gravity
}

\author{M.W.~Horbatsch${}^1$ and C.P.~Burgess${}^{1,2}$\\

$^1$ Dept. of Physics \& Astronomy, McMaster University \\
 \qquad 1280 Main St. W, Hamilton, Ontario, Canada, L8S 4L8.\\
\\
$^2$ Perimeter Institute for Theoretical Physics \\
 \qquad 31 Caroline St. N, Waterloo, Ontario, Canada  N2L 2Y5.\\
}

\abstract{ Precision tests of gravity can be used to constrain the
properties of hypothetical very light scalar fields, but these
tests depend crucially on how macroscopic astrophysical objects
couple to the new scalar field. We study the equations of stellar
structure using scalar-tensor gravity, with the goal of seeing how
stellar properties depend on assumptions made about the scalar
coupling at a microscopic level. In order to make the study
relatively easy for different assumptions about microscopic
couplings, we develop quasi-analytic approximate methods for
solving the stellar-structure equations rather than simply
integrating them numerically. (The approximation involved assumes
the dimensionless scalar coupling at the stellar center is weak,
and we compare our results with numerical integration in order to
establish its domain of validity.) We illustrate these methods by
applying them to Brans-Dicke scalars, and their generalization in
which the scalar-matter coupling slowly runs -- or `walks' -- as a
function of the scalar field: $a(\phi) \simeq a_{s} + b_{s}\phi$.
(Such couplings can arise in extra-dimensional applications, for
instance.) The four observable parameters that characterize the
fields external to a spherically symmetric star are the
stellar radius, $R$, mass,
$M$, scalar `charge', $Q$, and the scalar's asymptotic value,
$\phi_{\infty}$. These are subject to two relations because of the
matching to the interior solution, generalizing the usual
mass-radius, $M(R)$, relation of General Relativity. Since
$\phi_\infty$ is common to different stars in a given region (such
as a binary pulsar), all quantities can be computed locally in
terms of the stellar masses. We identify how these relations
depend on the microscopic scalar couplings, agreeing with earlier
workers when comparisons are possible. Explicit analytical
solutions are obtained for the instructive toy model of
constant-density stars, whose properties we compare to more
realistic equations of state for neutron star models. }

\begin{document}

\section{Introduction}

One way in which candidate quantum theories of gravity can differ
from one another is the spectrum of bosons they predict at very
low energies. Such fields, if they exist and are sufficiently
light, can mediate long-range forces that can observably compete
with gravity.

Scalar-tensor theories of gravity are among those that can arise
in this way, with light scalar fields in addition to the usual
metric tensor \cite{chibarev, damourrev, fujiimaeda, faraoni, singhrai, brans}. A variety of scalars commonly arise in fundamental theories, although it is very unusual for them to be light enough to mediate forces over macroscopic distances. They are rarely this light because quantum corrections
famously tend to give scalars masses, even if they would have been
massless at the purely classical level. But in some circumstances
symmetries can protect against masses, such as if the scalar is a
pseudo-Goldstone boson \cite{pGB} for a spontaneously broken approximate
symmetry, or part of the low-energy limit of an extra-dimensional model
\cite{ubernat}.

A good deal of effort has been invested in comparing the
predictions of scalar-tensor theories with observations in various
astrophysical systems \cite{will, efrev, binpult, binpulandgravwav, binpulnew, psaltis, whitedwarf, stellarosc, psaltisbinpul}, in order to improve the
constraints on their existence (or to discover their presence).
Binary pulsars are particularly useful for this purpose, since the
precision of their timing allows accurate measurements
of the relativistic gravitational effects that are generated by
the strong gravitational fields present.

Central to these tests is an understanding of how the stellar
properties depend on the microscopic couplings of the scalar field
to matter. Yet these stellar properties can sometimes be
remarkable in scalar-tensor theories in the presence of
relativistic sources. In particular a phenomenon called
spontaneous scalarization can occur, in which the star above a
critical baryonic mass can locally support a scalar field even if
the scalar-matter coupling vanishes asymptotically far from the
star \cite{damourspsc}. This allows scalar-tensor gravity to
deviate strongly from General Relativity (GR) near the star while
still passing weak-field solar system tests.

Since different fundamental theories predict scalars with
different microscopic couplings, it is useful to be able to survey
how stellar properties depend on these couplings. For this reason
in this paper we re-examine the equilibrium conditions in a star
in scalar-tensor models as a function of scalar couplings. In particular
we do so working as far as possible within analytic approximation
schemes, since these more easily allow the results to be varied
for different kinds of scalar properties. We find semi-analytic
progress is possible using a weak-coupling approximation for the
scalar field near the stellar center. By comparing with numerical
integrations of the equations of stellar structure we find the
domain of validity of these approximations, which allow us to
understand stellar properties fairly well.

As a preliminary to studying the relativistic limit in more
detail, we apply our analysis to the special case of an
incompressible stellar fluid, for which the energy density is
approximately constant. (This can be done consistent with
conservation laws only if the pressure is not also regarded as
being a function of the energy density.) To leading order in the
weak-coupling expansion, the profiles of physical variables of
constant-density stars can be described in closed form in terms of
elementary functions, dilogarithms, and Heun functions.

Although not physically realistic, solutions with this equation of
state can give insight into general features of relativistic
stars, in particular near the maximum mass that can be supported
by gravitational forces. (For GR the maximum mass found under the
assumption of constant density gives an upper bound on the maximum
mass that would be found with more realistic equations of state
\cite{buchdahl}.) We compare the predictions of constant density
in scalar-tensor theory with those of several representative
equations of states for neutron stars.

The rest of the paper is organized as follows. In \S2\ we define
the scalar-tensor theories of interest, and in particular the
parameters describing the couplings of the light scalar to
ordinary matter. Following much of the literature we specialize to
the case where the scalar-matter couplings are only weakly
dependent on the scalar fields themselves --- what we call
quasi-Brans/Dicke (qBD) models --- as well as the constraints on
their couplings that are inferred from solar system tests of
gravity. \S3\ then derives the equations describing hydrostatic
equilibrium for static and spherically-symmetric stars in
scalar-tensor gravity, as well as the matching formulae that
relate the interior and exterior geometries. Some of those general
properties that can be extracted without solving them explicitly
are also discussed, such as whether the pressure need be
monotonic; the kinds of relations they imply amongst quantities
visible to exterior observers; and the non-relativistic limit.
Next, \S4\ provides a discussion of the perturbative solutions to
these equations in the weak-coupling limit, in both the
relativistic and non-relativistic cases. The important distinction
between perturbations in the strength of the coupling measured at
the stellar center, vs its strength at infinity, first arises in
this section. Finally, in section \S5\ the perturbed equations are
solved explicitly to leading nontrivial order for the special case
of incompressible stars, with the results compared to more
realistic equations of state for neutron star models.

\section{Single-field scalar-tensor models}

We start by defining the field equations of the scalar-tensor
systems of interest.

\subsection{Action and field equations}\label{sec:action}

We consider a single light scalar field, and choose its action to
be given by\footnote{Conventions: we use metric signature $(-+++)$
and Weinberg's curvature conventions \cite{Wbg} (differing from
MTW \cite{MTW} only by an overall sign for the Riemann tensor). Units are chosen with $\hbar=c=1$.}
\begin{equation}\label{action}
 S = - \frac{1}{16 \pi G} \int \exd^{4}x \sqrt{-g} \; g^{\mu\nu}
 \bigl( \R_{\mu\nu} + 2 \,\partial_{\mu} \phi\partial_{\nu} \phi
 \bigr) + S_{\rm{m}}[\psi, \tilde{g}_{\mu \nu}] \,,
\end{equation}
where a Weyl re-scaling is performed to go to the Einstein frame
(which eliminates any $\phi$-dependence from the Einstein-Hilbert
term) and the scalar field is redefined to ensure its kinetic term
is minimal (with a conventional factor of 2). Here $G$ is the
Einstein-frame gravitational constant, $g_{\mu \nu}$, is the
Einstein-frame metric whose Ricci tensor is $\R_{\mu\nu}$, and
$S_{\rm m}$ denotes the `matter' action, involving all other
observed fields (collectively denoted here by $\psi$).

There are two physical choices made in writing this action, beyond
the choice of using only a single scalar field.
\begin{itemize}
\item We assume the absence of a scalar potential, which we assume
is small enough to be negligible for the astrophysical
applications of interest. This would necessarily be true if the
scalar is relevant to cosmology, but is also the feature that
quantum contributions make most difficult to achieve in realistic
models (unless there is an approximate symmetry, like shifts $\phi
\to \phi + c$, for constant $c$).
\item We assume the matter action, $S_{\rm m}$, only depends on
$\phi$ and $g_{\mu\nu}$ through the one `Jordan-frame' combination
$\tilde{g}_{\mu \nu} = A^{2}(\phi)g_{\mu \nu}$, where the
conformal factor $A(\phi)$ is a function whose form would be
specified within any particular fundamental theory.
This kind of coupling actually arises in specific models (such as if $\phi$
arises as the breathing mode for the geometry of extra dimensions), and
has the attractive feature that it naturally evades many of the strongest
observational constraints on violations of the equivalence
principle.
\end{itemize}

The field equations obtained by varying (\ref{action}) are
\begin{eqnarray}
\label{fieldeq1}
 \R_{\mu \nu} + 2\partial_{\mu} \phi \partial_{\nu} \phi +
 8 \pi G \left(T_{\mu \nu} - \frac{1}{2} T g_{\mu \nu} \right)
 &=& 0 \\
 \label{fieldeq2}
 \Box \phi + 4 \pi G a(\phi)T &=& 0 \,,
\end{eqnarray}
where $\Box = g^{\mu\nu} \nabla_\mu \nabla_\nu$ is the
d'Alembertian operator built using the Levi-Civita connection of
$g_{\mu \nu}$, while
\begin{equation}
 T^{\mu \nu} = -\frac{2}{\sqrt{-g}}
 \frac{\delta S_{\rm m}}{\delta g_{\mu \nu}}
\end{equation}
is the Einstein-frame energy-momentum tensor, $T=g^{\mu \nu}T_{\mu
\nu}$ is its trace, and $a(\phi) = A'(\phi)/A(\phi)$ defines the
scalar-matter coupling function in terms of the function
$A(\phi)$.

\subsection{Observational constraints}
\label{sec:sctens_constraints}

Because it is $\tilde g_{\mu\nu}$ that appears in the matter
action, it is the geodesic of this Jordan-frame metric along which
the trajectories of matter particles tend to move (in the absence
of other forces). Upon taking the post-Newtonian limit of the
Jordan-frame metric $\tilde{g}_{\mu \nu}$, one finds that the
effective Jordan-frame gravitational constant, measured in asymptotic
Einstein-frame units,\footnote{These are units in which the Einstein-frame
metric is asymptotically Minkowski: diag$(-1,1,1,1)$. If one instead uses units in which the Jordan-frame metric is asymptotically Minkowski, then $\widetilde G = G A^2(\phi_\infty) [1 + a^2(\phi_\infty)] $.} is
\begin{equation}\label{Geff}
 \widetilde{G} = G  \Bigl[ 1+a^{2}(\phi_{\infty}) \Bigr] \,,
\end{equation}
and that the Jordan-frame parameterized post-Newtonian
quantities whose values differ from those of GR are
\begin{equation}\label{ppn}
 \tilde{\beta} = 1+  \frac{a^{2}(\phi_{\infty})b(\phi_{\infty})}
 {2 [ 1+a^{2}(\phi_{\infty}) ]^{2}} \,, \qquad \tilde{\gamma}  =1 -
 \frac{2a^{2}(\phi_{\infty})}{1+a^{2}(\phi_{\infty})} \,,
\end{equation}
where $\phi_{\infty}$ is the asymptotic value of the scalar field
far from the source and $b(\phi) = \exd a(\phi)/\exd\phi$
\cite{damourrev}.

Constraints on these PPN parameters from solar system observations
provide the best bounds on the model, with $|\tilde{\gamma}-1|<
2.3 \times 10^{-5}$ inferred from Cassini tracking, and
$|\tilde{\beta}-1| < 2.3 \times 10^{-4}$ from lunar laser ranging
\cite{will}. This corresponds to the coupling bounds
\begin{equation}\label{ss_bounds}
 a^{2}(\phi_{\infty}^{SS}) < 1.2 \cdot 10^{-5} \,
 \quad \hbox{and} \quad
 a^{2}(\phi_{\infty}^{SS})|b(\phi_{\infty}^{SS})| < 4.6 \cdot
 10^{-4} \,,
\end{equation}
where $\phi_\infty^{SS}$ denotes the asymptotic value of $\phi$ as
one leaves the solar system.

\subsection{(Quasi) Brans/Dicke scalars}
\label{sec:qbd models}

The simplest scalar-tensor theory is Brans-Dicke theory \cite{BD},
for which $a(\phi)=a_{s}$ is constant. In this case -- see
eq.~(\ref{ss_bounds}) -- solar system tests constrain $a_{s} < 3.5
\cdot 10^{-3}$, and so all of the predictions of Brans-Dicke
theory are very close to those of GR.

The next-simplest theory, which we call quasi-Brans/Dicke (qBD)
theory, allows $a(\phi)$ to vary slowly with $\phi$ \cite{qBD}:
\begin{equation}\label{quadmod}
 A(\phi) = \exp(a_{s} \phi + \textstyle{
 \frac{1}{2}} \, b_{s} \phi^{2})\,, \qquad
 a(\phi) = a_{s} + b_{s} \phi\,.
\end{equation}
This introduces a variety of new phenomena because it makes the
strength of the scalar-matter couplings depend on $\phi$, and so
allows them to vary with position and time \cite{PD1, walkingphi}. This
means that couplings in exotic environments (like stellar
interiors) could be stronger than na\"ively expected without
running into conflict with the strong solar-system bounds
mentioned above (this is similar in spirit to, but
different in detail from, evading these bounds through
matter-dependent scalar self-couplings
\cite{chameleon2,chameleon, environscalar}).
Assuming $\phi$ is defined such that $\phi \to 0$
asymptotically far from the Sun, eqs.~\pref{ss_bounds} show that
the strong bound on $a_s$ implies that solar system bounds do not
strongly constrain $b_s$.

The constraints on $b_s$ are comparatively weak, and the best come
from studies of binary pulsars \cite{binarypulsarobs3,
binarypulsarobs1, binarypulsarobs2}. The precise timing
measurements that are possible for binary pulsars allow their
orbits to be accurately measured over long periods of time, and
comparing measurements with the predictions of the qBD model leads
to the constraint $b_{s} \gtrsim -5$
\cite{efrev,binpult,binpulandgravwav,binpulnew}.
Measurements of the redshift of
spectral lines from neutron stars leads to a weaker constraint
$b_{s} \gtrsim -9$ \cite{psaltis}. The main
uncertainties in these bounds come from the poor understanding of
the nuclear equation of state appropriate for neutron star
interiors.

Observations disfavor negative $b_s$ because for $b_{s} \lsim -4$,
compact objects like neutron stars exhibit a phenomenon called
spontaneous scalarization \cite{damourspsc}.
For sufficiently dense objects, whose
precise threshold density depends on $b_{s}$ and the nuclear
equation of state, it is possible for the star to support a
nonzero scalar field even though the scalar coupling vanishes
asymptotically far from the star: $a(\phi_{\infty}) = 0$.
Furthermore it is known that when scalarization takes place it is
the stable solution to the field equations \cite{stab3,chibarev}.
Scalarized neutron stars tend to be disfavored by binary pulsar
observations because scalarization significantly changes the
dynamics and radiation generated by a neutron star.

Collapse processes involving scalarized neutron stars have been
investigated by a number of authors \cite{collapse},
who found that the waveform of the emitted
gravitational radiation depends strongly on $b_{s}$. Moreover,
scalar-tensor gravity allows monopole and dipole radiation, which
are forbidden in GR. Thus, future measurements of gravitational
waves may lead to improved constraints on $b_{s}$ \cite{binpulandgravwav}.

\section{Stellar structure}

One of the potential uncertainties when trying to constrain
scalar-tensor models using astrophysical tests of gravity is the
strength of the scalar field that is supported by objects like the
Sun or a neutron star. For macroscopic astrophysical objects made
up of weakly coupled constituents one's intuition is that the
scalar coupling to the macroscopic object should be proportional
to the scalar coupling to each of its constituents, and we shall
see in this section that this intuition is generally borne out for
weakly coupled non-relativistic systems. We shall also see that it
can fail for relativistic systems, even in the limit of weak
scalar coupling.

In order to do so, we next summarize how the scalar field alters
the physics of stellar interiors, since it is the matching to this
that dictates the properties of the external field configurations
to which external observers have observational access.

\subsection{Equations of hydrostatic equilibrium}\label{sec:st_str}

The equations of hydrostatic equilibrium in scalar-tensor gravity
were first studied in \cite{salmona}, and subsequently in
\cite{nutku,saakmnats,matsuda,yokoi,heintzhill,hillheintz,reizlin,saenz,
bruckman,banerjee,avakyan,zaglauer,damourspsc,bruckman2,salgsud,whinnett,
kozyrev,whintor,salgsud2,yazadjiev,kazanas}.

Following \cite{damourspsc} we model the stellar
interior as a static, spherically-symmetric and perfect fluid in
local thermal equilibrium, locally characterized (in the Jordan
frame) by its pressure, $P$, and mass-energy density, $\rho$.
Time-independence and spherical symmetry allow the use of
Schwarzschild-like coordinates for the Einstein-frame metric
interior to the star:
\begin{equation}\label{sscoords}
 g_{\alpha \beta} \, \exd x^{\alpha} \exd x^{\beta} =
 -e^{\nu(r)} \exd t^{2} +  \frac{\exd r^{2}}{1-2\mu(r)}
 + r^{2} \exd\Omega^{2}\,,
\end{equation}
where $\exd\Omega^2 = \exd \theta^2 + \sin^2\theta \, \exd \phi^2$
denotes the usual round angular metric on the 2-sphere and
$\nu(r)$ and $\mu(r)$ are to-be-determined functions that depend
only on the radial coordinate $r$.

The Jordan-frame energy-momentum tensor for matter is defined by
\begin{equation}
 \tilde{T}^{\alpha \beta} = -\frac{2}{\sqrt{-\tilde{g}}}
 \frac{\delta S_{\rm m}}{\delta \tilde{g}_{\alpha \beta}}\,,
\end{equation}
and is related to the Einstein-frame energy-momentum tensor by
\begin{equation}
 T_{\alpha \beta} = A^{2}(\phi) \, \tilde{T}_{\alpha \beta}\,,
 \qquad
 T_{\alpha}^{\phantom{\alpha} \beta} =
 A^{4}(\phi) \, \tilde{T}_{\alpha}^{\phantom{\alpha} \beta}
 \quad \hbox{and} \quad
 T^{\alpha \beta} = A^{6}(\phi) \, \tilde{T}^{\alpha \beta}\,,
\end{equation}
where indices on $\tilde{T}_{\alpha\beta}$ are raised and lowered
using the Jordan-frame metric, $\tilde{g}_{\alpha\beta}$.

Being a Jordan-frame perfect fluid, the energy-momentum tensor has
the form
\begin{equation}\label{perf_fluid}
 \tilde{T}_{\alpha \beta} = (\rho + P)\tilde{u}_{\alpha}
 \tilde{u}_{\beta} + P \tilde{g}_{\alpha \beta}\,,
\end{equation}
where $P = P(r)$ and $\rho = \rho(r)$, and $\tilde{u}_{\alpha}$ is
the Jordan-frame 4-velocity of the perfect fluid, given in
co-moving coordinates by
\begin{equation}
 \tilde{u}_{\alpha} = e^{\nu/2} \, A(\phi) \, \delta^t_{\alpha}\,,
\end{equation}
so that $\tilde g^{\alpha \beta} \tilde u_\alpha \tilde u_\beta = -1$.

When writing the field equations it is convenient to scale out a
dimensional factor of the central density, $\rho_0 := \rho(0)$,
from the density and pressure,
\begin{equation}
 \Press(r) := \frac{P (r)}{\rho_{0}}
 \quad \hbox{and} \quad
 \Dens(r) := \frac{\rho(r)}{\rho_{0}} \,,
\end{equation}
in terms of which the equation of state is specified by writing
$\Dens(r) = \Dens[\Press (r); \PressC]$. Here
\begin{equation}
 \PressC := \frac{P(0)}{\rho_0} := \frac{P_0}{\rho_0} \,,
\end{equation}
labels the star's central pressure in the same units, and the
$\PressC$-dependence of $\Dens[\Press ; \PressC]$ is meant to
emphasize that the functional form of the equation of state,
$P(\rho)$, in general changes as one changes the central density.
Notice that our notation implies the identity
$\Dens(\PressC;\PressC)=1$.

With these choices the Einstein field equation,
eq.~(\ref{fieldeq1}), in this geometry boils down to the following
three conditions:
\begin{eqnarray}
 \label{eqss1}
 r \mu' + \mu  &=& 4 \pi G \rho_{0} r^{2}  A^{4}(\phi)
 \, \Dens(\Press)  + \frac{r^{2}}{2}(1-2\mu)\phi'^{2}\,, \\
 \label{eqss2}
 \Press ' &=& - [\Dens(\Press) + \Press] \left[
 \frac{ 4 \pi G \rho_{0} r^{2} A^{4}(\phi) \, \Press  + \mu}{r(1-2\mu)} +
 \frac{r}{2} \, \phi'^{2} + a(\phi) \phi' \right]\,, \\
 \label{eqss4}
 \nu' &=& \frac{8 \pi G \rho_{0} r^{2} A^{4}(\phi)
 \, \Press  + 2\mu}{r(1-2\mu)} + r\phi'^{2} \,,
\end{eqnarray}
where primes denote derivatives with respect to $r$. The scalar
wave equation, eq.~(\ref{fieldeq2}), similarly becomes
\begin{equation}
 \label{eqss3}
 \phi'' = \frac{4 \pi G \rho_{0} A^{4}(\phi)}{1-2\mu}
 \Bigl[ a(\phi) [\Dens(\Press) - 3\Press ]
 + r \phi' [\Dens(\Press) - \Press ]  \Bigr]
 - \frac{2(1-\mu)}{r(1-2\mu)} \phi'\,.
\end{equation}
These equations are to be integrated subject to the following
initial conditions at the centre of the star:
\begin{equation}
 \label{ics_1}
 \mu(0)=0\,,\qquad
 \Press (0)=\PressC \,,\qquad
 \phi(0)=\phi_{0}\,,\qquad
 \phi'(0)=0\,.
\end{equation}

Writing eqs.~(\ref{eqss2}) and (\ref{eqss4}) as
\be
 2 \Press' + (\Press + \Dens) \Bigl[ \nu' + 2a(\phi) \phi'
 \Bigr] = 0 \,,
\ee
and integrating once allows $\nu$ to be written in terms of
$\Press $ and $\phi$:
\begin{equation}
 \nu = -2f(\Press ;\PressC) - 2\ln A(\phi) + {\rm const} \,,
\end{equation}
where
\begin{equation}\label{f_defn}
 f(\Press ;\PressC) = \int_{\PressC}^{\Press }
 \frac{\exd\hat\Press}{\hat\Press + \Dens(\hat\Press)} \,.
\end{equation}

Regarding eq.~(\ref{eqss1}) as a linear, first-order differential
equation for the product $r \mu$, allows it to be solved to obtain
\begin{equation}
 \mu = \frac{1}{r} e^{-\int_{0}^{r} \tilde r \phi'^{2} d \tilde r}
 \int_{0}^{r} \hat r^2\left[ \frac{\phi'^{2}}{2} +
 4 \pi G \rho_{0}\, A^{4}(\phi) \, \Dens(\Press) \right]
 e^{\int_{0}^{\hat r} \tilde r \phi'^{2} d \tilde r} \exd \hat r \,,
\end{equation}
where the integration constant is chosen such that (\ref{ics_1})
holds. This expression shows that $\mu$ is always non-negative. It
also shows that $\mu(r)$ should not be interpreted as the
mass-energy inside the ball of radius $r$, unlike in GR.

In principle, eqs.~(\ref{eqss1}) through (\ref{eqss3}) can be
integrated numerically, starting at $r={0}$ and continuing out to
larger $r$. In practice, this system of equations is singular at
$r=0$, so numerical integration must be started at some small
positive $r=r_{0}$. The initial conditions at $r_{0}$ can be
obtained from the power series expansions that are dictated by the
equations of motion themselves:
\begin{eqnarray}
 \mu(r) &=& \frac{4\pi G \rho_{0} A^{4}_{0}}{3} \, r^{2}
 + \mathcal{O}(r^{4}) \,, \nonumber \\
 \Press (r) &=& \PressC + \frac{2\pi G \rho_{0}\,
 A^{4}_{0}}{3} (\PressC + 1) \bigl[ a^{2}_{0} (3\PressC-1)
 -(3\PressC+1) \bigr] r^{2} + \mathcal{O}(r^{4}) \,, \nonumber\\
 \phi(r) &=& \phi_{0} - \frac{2\pi G \rho_{0}\, A^{4}_{0}}{3}
 \, a_{0} (3\PressC-1)r^{2}+ \mathcal{O}(r^{4}) \,,
\end{eqnarray}
where $A_0 = A(\phi_0)$ and $a_0 = a(\phi_0)$.

\subsection{Matching to exterior observables}\label{sec:matching}

The integration within the stellar interior continues until
eventually the pressure $\Press $ vanishes. The value $r = R$
where this happens defines the (Schwarzschild coordinate) radius
of the star, beyond which the appropriate solution instead solves
the `matter-vacuum' field equations with $\rho = P = 0$. Of
course, for generic scalar-tensor theories it might happen that
$\Press$ never actually vanishes, since unlike for GR $\Press $
need not be a monotonically decreasing function. We comment where we
can on the stability of these configurations below, although as we shall
see they are unlikely to happen sufficiently close to the
Brans-Dicke limit and for non-relativistic equations of state.

\subsubsection*{Exterior solutions}

For $r > R$, an exterior solution is required to satisfy the vacuum field
equations,
\begin{equation}
 \R_{\alpha\beta} + 2\,\partial_\alpha \phi \,\partial_\beta \phi = 0
 \quad \hbox{and} \quad
 \Box \phi = 0 \,,
\end{equation}
for which a closed-form static and spherically symmetric solution
may be found by taking $\phi = \phi(r)$ and using the metric
\be \label{sphericalmetric}
    \exd s^2 = - e^{2 u} \, \exd t^2 + e^{-2 u} \, \exd r^2 + e^{2v}
    \, \Bigl[ \exd \theta^2 + \sin^2 \theta \, \exd \varphi^2
    \Bigr] \,,
\ee
where $u = u(r)$ and $v = v(r)$. The field equations for $\phi$,
$u$ and $v$ have the following solutions \cite{extsol}
\be \label{BackgroundSolutions}
    e^{2u} = \left( 1 - \frac{\ell}{r} \right)^x \,, \quad
    e^{2v} = r^2 \left( 1 - \frac{\ell}{r} \right)^{1-x}
    \quad \hbox{and} \quad
    e^{2\phi} = e^{2\phi_\infty}
    \left( 1 - \frac{\ell}{r} \right)^q \,,
\ee
where the real constants $x$ and $q$ satisfy $x^2 + q^2 = 1$ and
are otherwise arbitrary, leaving three free integration constants:
$x$, $\ell$ and $\phi_\infty$.

These three constants are more conveniently rewritten in terms of
the large-$r$ limit of the solution when expressed in the original
Schwarzschild coordinates,
\ba
    \exd s^2 &=& \left( -1 + \frac{2\,GM}{r} + \dots \right) \exd t^2
    + \cdots \nn\\
    e^\phi &=& e^{\phi_\infty} \left( 1 - \frac{GQ}{r}
    + \dots \right) \,,
\ea
where $M$ is the system's ADM mass \cite{ADM} in the Einstein
frame and $Q$ defines its `scalar charge'. For systems where the
scalar coupling, $a(\phi)$, is field-dependent, the asymptotic
value of the scalar field, $\phi_\infty$, may usefully be traded
for the asymptotic value of the scalar coupling strength,
$a_\infty = a(\phi_\infty)$.

Together with the star's radius, we see that from the point of
view of an external observer there are four independent bulk
parameters that characterize any such a star in scalar-tensor
theory: $M$, $Q$, $R$ and $\phi_\infty$. These can be calculated
in principle as functions of the stellar equation of state by
matching the exterior solution to that of the interior at $r = R$,
implying that they are functions of the two parameters, $\phi_0$
and $\PressC$, that define the initial conditions of integration
at the stellar center, eq.~\pref{ics_1}. One of our main goals is
to identify the two relations that must hold amongst the four
external parameters (for any given equation of state),
\be
 \xi_1(R,M,Q,\phi_\infty) = \xi_2(R,M,Q,\phi_\infty) = 0 \,,
\ee
generalizing the familiar mass-radius relation, $R = R(M)$, that
expresses the content of stellar structure within GR. The explicit
form of these constraints for scalar-tensor models is discussed in
section \ref{sec:pertexp} below.

Physically, we expect the value of $\phi_{\infty}$ to depend on
physics external to the star or stellar system of interest,
governed by the local properties of the galaxy in which the stars
are located. $\phi_{\infty}$ could also depend on cosmological
time if $\phi$ is light enough to be evolving over cosmological
time intervals. Thus, only one combination of the four parameters
$M$, $Q$, $R$ and $\phi_\infty$ can vary in practice from star to
star within a specific galactic neighborhood at a given
cosmological epoch. In particular, this means that the scalar
charge $Q$ and mass $M$ of a star are not independent parameters
for any stars within a local neighborhood at a given epoch.

This result greatly simplifies the phenomenological analysis of
binary pulsars. Instead of having two independent parameters
describing each of the two stars in the binary system, there is in
practice only one and so it suffices to describe the observational
constraints as a function of the masses of the two stars, just as
in GR. In principle the dependence on $\phi_\infty$ could
complicate the combining of the implications of many pulsars that
are located far from one another, but a simple estimate shows that
$\phi$ does not vary strongly across the galaxy, so in practice
$\phi_{\infty}$ can be taken to have a common for all stars in the
galaxy. This approximation is implicitly used in
refs.~\cite{efrev,binpult,binpulandgravwav}, when combined
data from several binary pulsars are plotted on one theory-space
exclusion plot in terms only of the masses of the two components.

To estimate the variance of $\phi(r)$ across the galaxy, we may
use the non-relativistic limit for which $g^{\alpha \beta}
T_{\alpha \beta} = A^4(\phi) \tilde g^{\alpha \beta} \tilde
T_{\alpha \beta} = A^4(\phi) (3p - \rho) \simeq - A^4(\phi) \,
\rho$ and so
\be \label{phieqgal}
 \Box \phi = \frac{e^{-\nu/2} \sqrt{1 -2 \mu}}{r^2 }
 \Bigl( e^{\nu/2} \, r^2 \sqrt{1 - 2 \mu} \; \phi'
 \Bigr)' \simeq \frac{1}{r^2} \Bigl( r^2 \, \phi' \Bigr)'
 \simeq 4 \pi G a(\phi) A^4(\phi) \, \rho \,.
\ee
If, on the right-hand-side, we assume $\phi \simeq \phi_g$ to be
approximately constant, then $a(\phi) A^4(\phi) \simeq a_g A_g^4$
is also constant. Then the only $r$-dependence on this side comes
from $\rho$, and we may estimate how large a deviation from
constant $\phi$ is implied by eq.~\pref{phieqgal}. For $r$ large
enough that $\rho$ is dominated by Dark Matter, the density
profile is $\rho \simeq \rho_g \, (\ell/r)^2$ so that orbital
velocities are $r$-independent: $v_g^2 \simeq G M(r)/r \simeq 4
\pi G \rho_g \, \ell^2$. Then eq.~\pref{phieqgal} implies
\be
 \phi(r) \simeq \phi(r') + 4\pi G a_g A_g^4 \rho_g \ell^2 \ln \left(
 \frac{r}{r'} \right) \simeq a_g A_g^4 v_g^2 \ln
 \left( \frac{r}{r'} \right) \,.
\ee
This is clearly small because it depends only logarithmically on
$r$, and both $a_g$ and $v_g$ are small. See Appendix B of \cite{dam2pn} for a related discussion.

\subsubsection*{Matching at $r = R$}

Requiring continuity of the exterior and interior profiles across
$r = R$ implies the three external parameters $M$, $Q$, and
$\phi_{\infty}$ must satisfy \cite{damourspsc}
\begin{eqnarray}
 \label{match1}
 s := \frac{GM}{R} &=& \frac{\mathcal{K}}{2\sqrt{1-2\mu_\star}}
 \exp \biggl[- \frac{\mathcal{K}}{\mathcal{L}} \;
 {\rm arctanh} \left( \frac{\mathcal{L}}
 {\mathcal{J}} \right)\biggr]\,, \\
 \label{match2}
 a_{\ssA} := \frac{Q}{M} &=&  \frac{2R \, \phi_\star'
 (1-2\mu_\star)}{\mathcal{K}}\,,\\
 \label{match3}
 \phi_{\infty} &=& \phi_\star + \frac{2 R \,\phi_\star'
 (1-2\mu_\star)}{\mathcal{L}} \;
 {\rm arctanh} \left( \frac{\mathcal{L}}
 {\mathcal{J}}\right)\,,
\end{eqnarray}
where
\begin{eqnarray}
 \label{matchj}
 \mathcal{J} &:=& 2(1-\mu_\star) + R^{2} \phi_\star'^{2}
 (1-2\mu_\star) \,, \\
 \label{matchk}
 \mathcal{K} &:=& 2\mu_\star + R^{2}\phi_\star'^{2}
 (1-2\mu_\star) \,, \\
 \label{matchl}
 \mathcal{L} &:=& \sqrt{4\mu_\star^{2} +
 4R^{2}\phi_\star'^{2}(1-\mu_\star)(1-2\mu_\star)
 + R^{4} \phi_\star'^{4}(1-2\mu_\star)^{2}} \,,
\end{eqnarray}
and $\mu_\star := \mu(R)$, $\phi_\star := \phi(R)$, and
$\phi_\star' := \phi'(R)$.

It is sometimes useful to decompose the ADM mass as $M = M_\ssB +
\Delta M$, where $\Delta M$ is the gravitational binding energy
and $M_\ssB$ is the baryonic mass, defined by
\begin{equation}\label{barmass}
 M_\ssB := 4 \pi m_{0} \int_{0}^{R} \exd r \;
 \frac{n A^{3}(\phi) \, r^{2}}{\sqrt{1-2\mu}}\,,
\end{equation}
where $n(r)$ is the local baryon-number density, and $m_b$ is the
average mass of an individual baryon. This baryonic mass can be
related to the stellar pressure and density if the star has
constant entropy per nucleon, since in this case it follows from
energy conservation that
\begin{equation}
 \left( \frac{\rho}{n} \right)'
 + P \left(\frac{1}{n}\right)' =0 \,.
\end{equation}
This equation can be used to write $n$ in terms of $\Press $,
\begin{equation}\label{n_eqn}
 n=n_{0} \left[ \frac{\Press
  + \Dens(\Press)}{\PressC + 1} \right] e^{-f(\Press ;\PressC)} \,,
\end{equation}
where $n_{0}=n(0)$ and $f(\Press; \PressC)$ is defined in equation
(\ref{f_defn}). The baryonic mass then becomes
\begin{equation}\label{barmass2}
 M_\ssB = 4 \pi m_{0}n_{0} \int_{0}^{R}
 e^{-f(\Press)} \left[ \frac{\Press  + \Dens(\Press)}{\PressC + 1}
 \right] \frac{A^{3}(\phi) r^{2} \exd r}{\sqrt{1-2\mu}} \,.
\end{equation}

The quantities $s$ and $a_\ssA$ given by the matching equations,
(\ref{match1}) and (\ref{match2}),
are in themselves useful because each has a
physical interpretation. $s$ is called the self-gravity\footnote{Note that
Will \cite{will} uses $s$ to denote the sensitivity of the mass to the
gravitational constant, which he defines as
$-(\partial \log M)/(\partial \log G)$. This definition of $s$ differs from
our definition, but has the same
order of magnitude.}, or
compactness, of the star and it provides a dimensionless measure
of how relativistic its gravitational field is at the stellar
surface, $r = R$. For non-relativistic stars $s \ll 1$, and in
general relativity $s \leq \frac49$ for any star within which the
mass-energy density, $\rho$, is a non-increasing function of $r$,
a result known as Buchdahl's theorem \cite{buchdahl}.\footnote{The
potential generalization of this result to scalar-tensor gravity
is investigated in \cite{buchdahlsctens}.}

Similarly, the quantity $a_{\ssA}$ can often be interpreted as the
effective scalar-matter coupling of the star as seen by the
observer at infinity. This can be understood by considering the
lowest-order non-relativistic interaction energy between two
widely separated stars, A and B, \cite{damourrev}:
\begin{equation}
 U_{\AB} = -\frac{GM_{\ssA}M_{\ssB}}{r_{\AB}}
 -\frac{GQ_{\ssA}Q_{\ssB}}{r_{\AB}}
 \equiv
 - \frac{\widetilde{G}_{\AB}M_{\ssA} M_{\ssB}}{r_{\AB}} \,,
\end{equation}
where
\begin{equation}\label{geff}
 \widetilde{G}_{\AB} = G (1 + a_{\ssA} a_{\ssB})
\end{equation}
is the effective Jordan-frame gravitational constant between the
two stars, and $r_{\AB}$ is the distance between them.
All quantities are measured in units in which
the Einstein-frame metric is asymptotically Minkowski\footnote{As mentioned earlier, to convert to units with an asymptotically Minkowski Jordan-frame metric
multiply $\widetilde{G}_{\AB}$ by $A^{2}(\phi_{\infty})$, multiply $r_{\AB}$ by
$A(\phi_{\infty})$, and divide $M_\ssA$, $M_\ssB$ and $U_{\AB}$ by
$A(\phi_{\infty})$.}.

The connection between $a_\ssA$ and the scalar coupling can also
be seen in another way. Since in the non-relativistic limit we
have $M \simeq M_\ssB$, and since each individual baryon if
separated to infinity would microscopically satisfy $q_b/m_b =
a_\infty = a(\phi_\infty)$, we expect that for macroscopic
non-relativistic systems $Q/M = (N q_b)/(N m_b) = a_\infty$ as
well, so
\begin{equation}\label{limits1}
 \lim_{s \to 0} a_{\ssA} = a(\phi_{\infty})
 \quad \hbox{and} \quad
 \lim_{s \to 0} \widetilde{G}_{\AB} = \widetilde{G} \,,
\end{equation}
where $\tilde{G}$ is the Jordan-frame gravitational constant
defined in equation (\ref{Geff}) \cite{damourrev}. We shall see
below that the prediction $Q/M = a(\phi_\infty)$ is given by a
more detailed matching calculation, where it can be seen to hold
independent of the stellar equation of state in the limit of both
non-relativistic sources and weak scalar coupling.

\subsubsection*{The black hole limit}

For sufficiently massive stars the only stable configuration in GR
is a black hole, for which there is only the exterior solution,
eq.~\pref{BackgroundSolutions} with $x=1$ and $q=0$ \cite{Wbg,MTW}.
In this case the role of matching
to an interior geometry across $r = R$ becomes replaced by the
boundary condition that the geometry not have a physical
singularity at the event horizon, $r = R_{bh}$, making this the
natural equivalent of the stellar radius, $R$, for a black hole.
Since for spherically symmetric stars in GR we have $R_{bh} = 2\,
GM$, we see that for black holes the mass-radius condition
predicted is particularly simple: $s = GM/R_{bh} = \frac12 >
\frac49$.

For a black hole coupled to scalar fields the exterior solution,
\pref{BackgroundSolutions}, always has a curvature singularity at
$r = \ell$ whenever $q \ne 0$ \cite{damourrev}. 
This shows that only $q = 0$
describes a legitimate black hole, for which $\phi = \phi_\infty$
is constant and arbitrary (this is unchanged if $a_\infty =
a(\phi_\infty) \ne 0$, because there is no matter exterior to the
black hole with which to couple). The metric is then given by the
Schwarzschild geometry, just as for GR, implying $s = \frac12$ and
$Q = 0$ (for all $\phi_\infty$) even when scalars are present.
This is a special case of the no-hair theorems, originally formulated for 
gravity in \cite{nohairorig} and extended to scalar fields in
\cite{nohairold}. 

\subsubsection*{Stability}
\label{sec:stability}

\FIGURE[r]{\epsfig{file=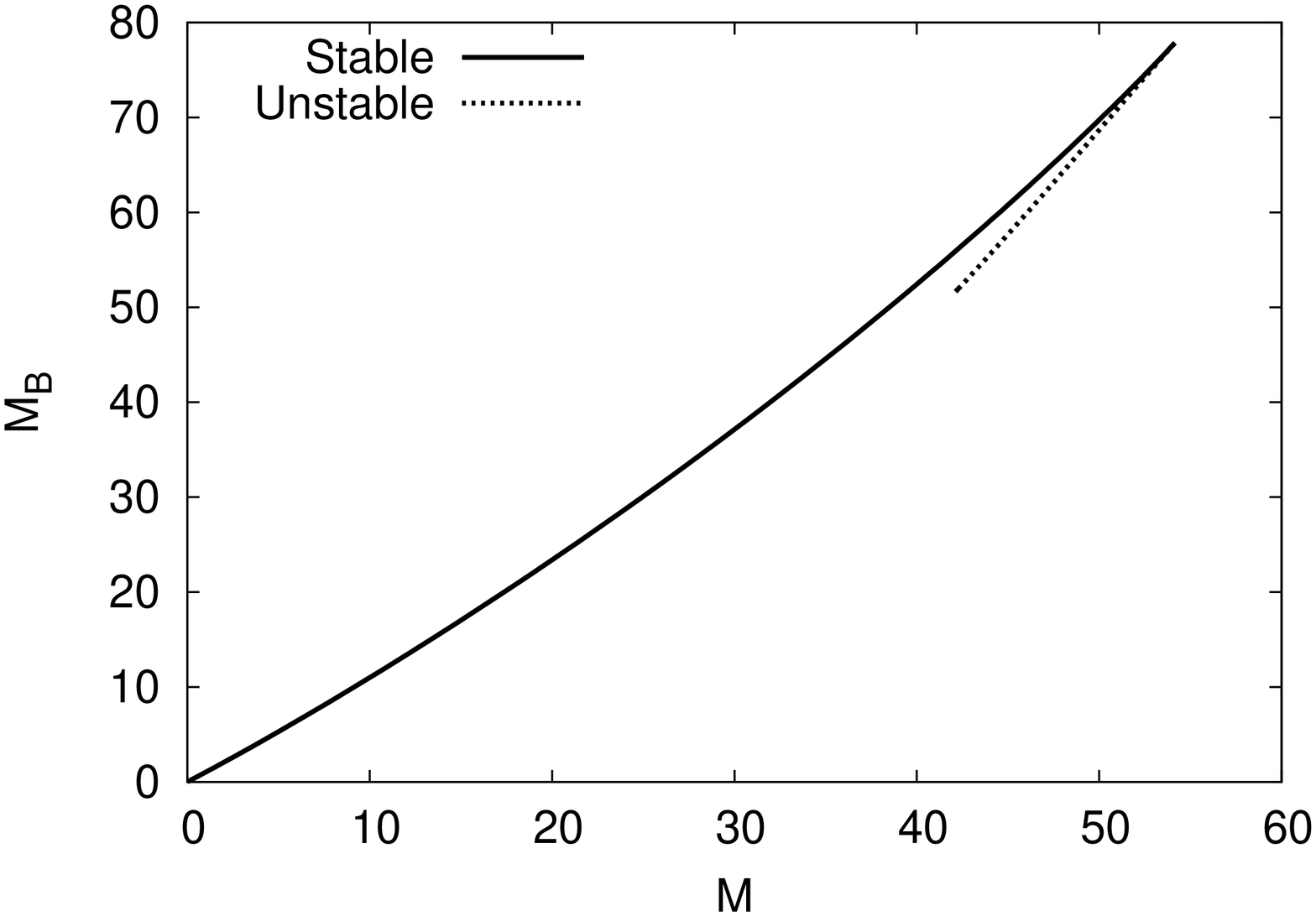,angle=270,width=7.5cm}
\label{fig:stab_example}
\caption{$M_{\ssB}$ versus $M$ for constant-density stars in
Brans-Dicke theory, with $\phi_{\infty}=0$ and $a_{s}^{2}=0.2$.}
}

Once a solution of equations (\ref{eqss1})-(\ref{eqss3}) is obtained, 
and the external parameters (\ref{match1})-(\ref{match3}) are calculated, 
it must be checked whether the solution is stable against perturbations.
To do this properly requires working with the time-dependent equations 
of stellar structure. We here instead follow \cite{damourspsc,stab3,chibarev}
and perform a simplified analysis.

The main idea builds on the observation that (for general relativity) 
the equations of hydrostatic equilibrium are equivalent to
the problem of extremizing $M$, while keeping $M_{\ssB}$ fixed 
\cite{Wbg}. The idea used in \cite{damourspsc,stab3,chibarev} is to assume
the same is true for scalar-tensor theories (with $\phi_{\infty}$ 
fixed as well as $M_\ssB$) although we have been unable to
prove this so far in scalar-tensor gravity.

For given values of $M_{\ssB}$ and $\phi_{\infty}$, one generically
finds that there can exist multiple stellar configurations with 
different values of $M$. This is illustrated by the plot of
$M_{\ssB}$ versus $M$ given in figure \ref{fig:stab_example}
for a one-dimensional family of stellar configurations
that share a common value for $\phi_{\infty}$. Whenever two
or more values of $M$ are are obtained for a given $\phi_\infty$
we take the configuration with the lowest value of $M$ to be stable, 
and the others to be unstable.

%\clearpage

\subsection{Stellar structure for quasi-Brans/Dicke scalars}
\label{sec:st_str_quadmod}

Of course all of the predictions for quantities like $Q/M$ depend
in principle on the details of the coupling function, $a(\phi)$,
in addition to depending on the stellar equation of state. In this
section we specialize the equations of the previous section to the
quasi-Brans/Dicke (qBD) model of eq.~(\ref{quadmod}), since this
boils the dependence on $a(\phi) = a_s + b_s \phi$ down to the
dependence on the two parameters $a_s$ and $b_{s}$.

Because the gravity-scalar part of the theory is invariant under
constant shifts, $\phi \to \phi + c$, this transformation can be
used to set $a_{s}$ to zero with no loss of generality, provided
that $b_{s} \neq 0$. With this choice, the matter coupling can be
seen to have a reflection symmetry $\phi \to -\phi$. We henceforth
choose this convention for $a_s$, making it sufficient to follow
the dependence of observables on $b_s$.

Consider now integrating the field equations starting from $r =
0$. The information that used to reside in $a_s$ now resides in
the value of the field (or coupling) at the stellar origin, $a_0
:= a(\phi_{0}) = b_s \phi_0$. Suppose first that the initial
value, $\phi_0$, chosen for $\phi$ at the stellar center is such
that the coupling vanishes there, $a(\phi_{0}) = 0$. In this case
the reflection symmetry implies $\phi(r) \equiv 0$ for all $r$ is
a solution to the field equations, and so $a(\phi) = a_0 = 0$,
everywhere within the stellar interior. (That this is a solution
can be seen by direct inspection of eq.~\pref{eqss3}.) In this
case the equations of stellar structure, (\ref{eqss1}) through
(\ref{eqss3}), uniquely reduce to those of GR.

If instead $a_0 = a(\phi_{0}) = b_s \phi_0 \neq 0$, then, after
the change of variables
\begin{equation}
 \label{chvar1}
 \varphi = (\phi - \phi_{0})/a_{0} \,, \qquad
 u = 8 \pi G \rho_{0} A^{4}_{0} \, r^{2} \,,
\end{equation}
%
%\pagebreak

\noindent
the equations of stellar structure simplify to
\begin{eqnarray}\label{eqssnew1}
 \dot{\mu} &=& -\frac{\mu}{2u}
 + \frac{\Dens(\Press) \, e^{4a_0^2 \varphi \, (1
 +b_{s} \varphi / 2)}}{4} + a_0^2 u(1-2\mu)\dot{\varphi}^{2}\,, \\
 \label{eqssnew2}
 \dot{\Press } &=& - [\Dens(\Press) + \Press\,]
 \left[ \frac{\mu}{2u(1-2\mu)} +
 \frac{\Press  e^{4a_0^2 \varphi(1+b_{s} \varphi / 2)}}{4(1-2\mu)}
 +a_0^2 \dot{\varphi}(1+u\dot{\varphi}+b_{s}\varphi) \right]\,, \\
 \label{eqssnew3}
 \ddot{\varphi} &=& -\frac{(3-4\mu)\dot{\varphi}}{2u(1-2\mu)}
 + \frac{e^{4a_0^2\varphi(1+b_{s}\varphi / 2)}}{8u(1-2\mu)}
 \Bigl[ (1+b_{s}\varphi)[\Dens(\Press)-3\Press\,]
 + 2[\Dens(\Press)-\Press \,] u\dot{\varphi} \Bigr]\,,
\end{eqnarray}
where dots denote derivatives with respect to $u$. These equations
show that $a_0 = a(\phi_0) = a(r = 0)$ only enters physical
observables through its square: $a^2_{0}$. In terms of these
variables the initial conditions are
\begin{equation}
 \label{quadmod_ics}
 \mu(0)=0 \,, \qquad
 \Press (0)=\PressC \,, \qquad
 \varphi(0)=0 \,, \qquad
 \dot{\varphi}(0)=\frac{1}{12} \, (1-3\PressC) \,.
\end{equation}
The profiles $\mu(u)$, $\Press (u)$ and $\varphi(u)$ obtained by
integrating these equations depend only on the three parameters
$a_0^2$, $b_{s}$, $\PressC$, as well as on the choice of equation
of state $\Dens(\Press)$.

The baryonic mass, eq.~(\ref{barmass2}), then becomes
\begin{eqnarray}
 M_\ssB &=& \frac{2 \pi m_{0} n_{0}}{(8 \pi G
 \rho_{0})^{3/2}A^{3}_{0}} \int_{0}^{U} \exd u \;
 \frac{\Press  +  \Dens(\Press)}{\PressC + 1}
 \, e^{-f(\Press ;\PressC)} e^{3  a_0^2 \varphi (1+b_{s}\varphi/2)}
 \sqrt{\frac{u}{1-2\mu}}  \,,
 \nonumber \\
 \label{barmass_quadmod} & := & \frac{2 \pi m_{0} n_{0}}{(8 \pi
 G \rho_{0})^{3/2}A^{3}_{0}} \;
 \mathcal{M}(a_0^2,b_{s},\PressC) \,,
\end{eqnarray}
where
\begin{equation}\label{U_defn}
 U(a_0^2,b_{s},\PressC) =  8 \pi G \rho_{0} A^{4}_{0} \, R^{2}
\end{equation}
is the value $U = u(R)$ corresponding to the boundary of the star.
For a given equation of state, $U$ depends only on $a_0^2$,
$b_{s}$ and $\PressC$, because it can be computed by finding the
point at which $\Press (u)$ vanishes.

Note that the quantities $\mathcal{J}$, $\mathcal{K}$, and
$\mathcal{L}$, defined in equations (\ref{matchj}) through
(\ref{matchl}), depend only on $\mu_\star = \mu(U)$ and
$R^{2}\phi'^{2}(R)=4 a_0^2 U^{2}\dot{\varphi}^{2}(U)$. Therefore,
the compactness $s$ --- given by equation (\ref{match1}) --- also
depends only on $a_0^2$, $b_{s}$ and $\PressC$. The matching
conditions at the stellar surface, eqs.~(\ref{match2}) --
(\ref{match3}), can be re-written using these variables as
\begin{equation}\label{f}
 \frac{\phi_{\infty}-\phi_{0}}{a_{0}} = \varphi(U) + \frac{4U
 \dot{\varphi}(U) (1-2\mu_\star)}{\mathcal{L}}
 \, {\rm arctanh} \left(
 \frac{\mathcal{L}}{\mathcal{J}}\right) :=
 \mathcal{F}(a_0^2,b_{s},\PressC) \,,
\end{equation}
and
\begin{equation}\label{aa}
 a_{\ssA} = \frac{Q}{M} =  \frac{4a_{0} \, U \dot{\varphi}(U)
 (1-2\mu)}{\mathcal{K}} := a_{0}
 \A(a_0^2,b_{s},\PressC) \,,
\end{equation}
which define the quantities $\F$ and $\A$.

%\subsubsection*{Stability}
%
%In terms of the functions $\mathcal{M},s,\mathcal{A},\mathcal{F}$, the
%condition for a turning point is
%\begin{equation}\label{stab_tp}
%1 + b_{s} \mathcal{F} + 2 a_{0}^{2} b_{s}
%\frac{\partial \mathcal{F}}{\partial a_{0}^{2}} =
%a_{0}^{2} \frac{\partial \mathcal{F} / \partial \PressC}
%{\partial \mathcal{M} / \partial \PressC}
%\left(2b_{s}
%\frac{\partial \mathcal{M}}{\partial a_{0}^{2}}-3\mathcal{M} \right) \,,
%\end{equation}
%and the condition for instability in the neighbourhood of a turning point is
%\begin{eqnarray}
%\left[ b_{s}
%\left( \mathcal{A} + 2a_{0}^{2} \frac{\partial \mathcal{A}}{\partial a_{0}^{2}} \right)
%- a_{0}^{2} \frac{\partial \mathcal{A}/\partial \PressC}
%{\partial \mathcal{M} / \partial \PressC}
%\left( 2b_{s} \frac{\partial \mathcal{M}}{\partial a_{0}^{2}}-3\mathcal{M}\right)
%\right] \times
%\nonumber
%\\
%\label{stab_instab}
%\left[
%1 + b_{s} \mathcal{F} + 2 a_{0}^{2} b_{s}
%\frac{\partial \mathcal{F}}{\partial a_{0}^{2}} -
%a_{0}^{2} \frac{\partial \mathcal{F} / \partial \PressC}
%{\partial \mathcal{M} / \partial \PressC}
%\left(2b_{s}
%\frac{\partial \mathcal{M}}{\partial a_{0}^{2}}-3\mathcal{M} \right)
%\right] > 0 \,.
%\end{eqnarray}

\subsubsection*{Properties of the pressure profile}
\label{sec:phys_real}

This section briefly pauses to investigate whether the pressure,
$\Press$, decreases monotonically, and whether the equation
$\Press (u) = 0$ must have a solution.

It is clear that (unlike for GR) $\exd P/\exd r$ can be positive
for sufficiently large couplings and pressures. This follows from
evaluating equations (\ref{eqssnew2}) and (\ref{quadmod_ics}) at
the stellar center,
\begin{equation}
 \dot{\Press }(0) =  - \frac{1+\PressC}{12} \Bigl[ 1+3\PressC +
 a_0^2(1-3\PressC) \Bigr] \,,
\end{equation}
which (for $\PressC \ge 0$) is strictly non-positive unless
$\PressC = P_0/\rho_0 > \frac13$ and
\begin{equation}\label{monpresineq1}
 a_0^2 > {a^2_0}_{\rm crit} :=
 \frac{3\PressC+1}{3\PressC-1} \,.
\end{equation}
In general, to determine whether the pressure profile is monotonically
decreasing everywhere,
it is not sufficient to look only at $\dot{\Press}(0)$. In the qBD theory
with $b_{s}>0$, it can happen that $\dot{\Press}(0) >0$,
$\dot{\Press}(u_{\star})=0$ for some $u_{\star} > 0$, and
$\Press(U) = 0$ for some $U>u_{\star}$. Such a star has its maximum pressure
in between the centre and the surface.
Similarly, if $b_{s}<0$, it can happen that $\dot{\Press}(0)<0$,
$\dot{\Press}(u_{\star})=0$ for some $u_{\star}>0$,
$\Press(u) > 0$ for all $u \geq 0$, and $\Press(u) \to \infty$ as
$u \to \infty$. Such a solution describes an object of infinite extent,
and seems unphysical.

By contrast, for Brans Dicke theory our numerical calculations 
suggest that this kind of complicated behaviour does not take 
place, and the pressure profile
$\Press(u)$ cannot have any local extrema. We have been unable to
prove this analytically, but we have shown that if
$a^2_0={a^2_0}_{\rm crit}$, then $\Pi(u)=\eta$ for all $u$,
for all equations of state. This constant-pressure
solution appears to be the boundary
between the solutions with $\dot{\Press} > 0$, and those with
$\dot{\Press} < 0$. For Brans Dicke theory Salmona 
\cite{salmona} has shown that if
$\Press / \varrho < 1/3$ everywhere, then $\Press$ decreases
monotonically everywhere.

If we assume that this simple behaviour of the pressure profile in
Brans-Dicke theory is correct, then
imposing certain requirements on the properties of the solutions of
the equations of hydrostatic equilibrium leads to constraints on
$a_{0}^{2}$. If we require that the pressure is
decreasing for all $\PressC$, then it follows that $a_{0}^{2} < 1$.

A more conservative constraint on $a_0^2$ can be obtained by using
some information about the equation of state. For these purposes
it suffices to consider only neutron stars, because only these are
relativistic enough to have $\PressC > 1/3$. But neutron-star
interiors can be modelled as relativistic polytropes
\cite{astrophys}, with
\begin{equation}
 \frac{\rho}{m_{b}} = n + \frac{\kappa \, \hat{n}}{\gamma-1}
 \left( \frac{n}{\hat{n}} \right)^{\gamma}
 \quad \hbox{and} \quad
 \frac{P}{m_{b}} =  \kappa \, \hat{n} \left( \frac{n}{\hat{n}}
 \right)^{\gamma}\,,
\end{equation}
where $n$ is the baryon number density, $\hat{n}=0.1 \; {\rm
fm}^{-3}$ is a typical nuclear density, $m_{b} = 1.66 \cdot
10^{-24} \, {\rm g}$ is the mass of an average baryon, and
$\kappa$ and $\gamma$ are constants ($\gamma$ is called the
polytropic index). Notice that these choices imply the central
density and pressure are related by $\rho_0/m_b = n_0 + \kappa \,
\hat n (n_0/\hat n)^\gamma/(\gamma - 1)$, and so
\be
 1 = \frac{m_b n_0}{\rho_0} + \frac{\PressC}{\gamma - 1} \,,
\ee
implying the maximum value of $\PressC$ that is possible is
$\PressC^{\rm max} = \gamma - 1$. The functions $\Dens(\Press)$
and $f(\Press)$ are similarly given by
\begin{eqnarray}
 \Dens(\Press ;\PressC) &=&
 \frac{\Press }{\gamma-1} + \left( \frac{\Press }{
 \PressC}\right)^{1/\gamma} \left(1-\frac{
 \PressC}{\gamma-1} \right) \,, \\
 f(\Press ;\PressC) &=& \ln \left[
 \frac{\gamma - 1 - \PressC + \gamma \Press
 (\PressC/\Press )^{1/\gamma}}
 {(\PressC+1)(\gamma - 1)} \right] \,.
\end{eqnarray}

%Using these in the equation for $\dot p$ shows that a necessary
%condition for $\dot p < 0$ (and so for relativistic polytropes
%with index $\gamma$ to exist) in Brans-Dicke theory is
Requiring that the pressure is decreasing for all relativistic polytropes
yields the constraint
\begin{equation}\label{monpresineq2}
 a_0^2 < \frac{3\gamma -2}{3\gamma - 4} \,.
\end{equation}
For the neutron-star equations of state EOS II and EOS A of
ref.~\cite{damourspsc}, these bounds evaluate to $|a_{0}| < 1.29$
and $|a_{0}| < 1.26$, respectively. Although nowhere near as good
as the solar system bounds of eqs.~(\ref{ss_bounds}), they are
complementary because they apply to the coupling deep within a
neutron star interior, and rely only on general considerations,
rather than precise observations.

\subsection{Non-relativistic limit}
\label{sec:nonrellim}

In most stars, relativistic effects are negligible, allowing us to
take $\Press = P/\rho_0 \ll 1$ and $\mu \ll 1$. In this case, the
equations of stellar structure, eqs.~(\ref{eqss1}) through
(\ref{eqss3}), simplify to
\begin{eqnarray} \label{newt1}
 r\mu' + \mu &\simeq& 4 \pi G r^{2}A^{4}(\phi)\rho +
 \frac{r^{2} \phi'^{2}}{2} \ , \\
 \label{newt2}
 P' &\simeq& -\rho \left[ \frac{\mu}{r} + \frac{r\phi'^{2}}{2} +
 a(\phi)\phi' \right] \,, \\
 \label{newt3}
 \phi'' &\simeq& 4 \pi G A^{4}(\phi) \rho
 \Bigl[ a(\phi) + r\phi' \Bigr] - \frac{2\phi'}{r}\,, \\
 \label{newt4}
 \nu' &\simeq& \frac{2\mu}{r} + r\phi'^{2}\,,
\end{eqnarray}
where the energy density, $\rho$, is equivalent to the mass
density in the non-relativsitic limit.

The matching conditions, eqs.~(\ref{match1}) through
(\ref{match3}), similarly simplify to
\begin{eqnarray} \label{newt_match1}
 \frac{GM}{R} &=& \mu(R) + \frac{[R \, \phi'(R)]^{2}}{2}\,, \\
 \label{newt_match2}
 \frac{GQ}{R} &=& R \, \phi'(R)\,, \\
 \label{newt_match3}
 \phi_{\infty} &=& \phi(R) + R \, \phi'(R)\,.
\end{eqnarray}
Notice that these last two matching conditions quite generally
imply
\begin{equation}
\phi(R) = \phi_{\infty} - \frac{GQ}{R} \,,
\end{equation}
as is appropriate for the non-relativistic limit of the known
external solutions.

\subsubsection*{Newtonian polytropes}

The equation of state that is of most interest for Newtonian
systems is that of a polytrope,
\begin{equation}\label{eq:polytrope}
 P = K \rho^{1+1/\chi}\ ,
\end{equation}
where $\chi$ is the polytropic index (a constant that need not be
an integer), and $K$ is a constant. We briefly specialize to this
equation of state here for later use in subsequent sections.

Specializing eqs.~(\ref{newt1}) -- (\ref{newt4}) to this equation
of state, (\ref{eq:polytrope}), the above equations simplify after
changing to dimensionless variables
\begin{equation}\label{newtpoly_chvar}
 r = r_s w \,,\qquad \rho=\rho_{0} \,\theta^{\chi} \,,
\end{equation}
where $\rho_{0} = \rho(0)$, and $r_s$ is a length scale that is to
be specified later. Then equations (\ref{newt1}) -- (\ref{newt4})
become
\begin{eqnarray}
\label{eq:poly1}
 \ddot{\phi} &=& -\frac{2 \dot{\phi}}{w} +
 C A^{4}(\phi) \theta^{\chi}
 \Bigl[ a(\phi) + w \dot{\phi} \Bigr]\, \\
 \label{eq:poly2}
 \frac{\exd}{\exd w} \left[ \zeta w^{2} \dot{\theta}
 +  \frac{1}{2}w^{3}
 \dot{\phi}^{2} + w^{2} a(\phi) \dot{\phi} \right]
 &=& -C A^{4}(\phi) w^{2} \, \theta^{\chi}
 - \frac{w^{2}}{2} \, \dot{\phi}^{2}\ ,
 \\
 \label{eq:poly3}
 \mu &=& - w\left[ \zeta \dot{\theta}
 + \frac{w}{2} \, \dot{\phi}^{2} +
 a(\phi) \dot{\phi} \right]\ ,
\end{eqnarray}
where dots now denote $\exd/\exd w$, and the dimensionful
parameters are all rolled into the new dimensionless constants,
\begin{equation}\label{eq:a_polytrope}
 \zeta := K(\chi+1) \rho_{0}^{1/\chi}\ ,\ \
 C := 4 \pi G r_s^{2} \rho_{0}\ .
\end{equation}
In these variables the initial conditions are
\begin{eqnarray}\label{eq:newtpoly_ic1}
 &&\theta(0)=1 \,, \qquad \dot{\theta}(0)=0\ , \\
 \label{eq:newtpoly_ic2}
 \hbox{and} \quad
 &&\phi(0)=\phi_{0} \,, \quad\;
 \dot{\phi}(0) = 0\ .
\end{eqnarray}

We now specialize to the quasi-Brans/Dicke models of
eq.~(\ref{quadmod}), for which we had
\begin{equation} \label{chvar2}
 a_0^2 = a^{2}(\phi_{0})=(a_{s} +  b_{s} \phi_{0})^{2}
 \quad \hbox{and} \quad
 \varphi = \frac{\phi - \phi_{0}}{a_{0}} \ .
\end{equation}
If we choose
\begin{equation}\label{newtpoly_s}
 r_s = \frac{1}{A^{2}_{0}} \sqrt\frac{\zeta}{4 \pi G \rho_{0}}
 = \frac{1}{A^{2}_{0}} \sqrt{\frac{K(\chi+1)}{4\pi G}}
 \rho_{0}^{(1-\chi)/2\chi} \,,
\end{equation}
so that
\begin{equation}
 \zeta=C A^{4}_{0}  \,,
\end{equation}
then equations (\ref{eq:poly1}) -- (\ref{eq:poly3}) become
\begin{eqnarray} \label{eq:poly_mod1_1}
 - \frac{\exd}{\exd w} (w^{2} \dot{\theta}) - \frac{a_0^2
 \, b_{s}}{\zeta} \, w^{2} \dot{\varphi}^{2} &=&
 w^{2} e^{4 a_0^2 \varphi (1 + b_{s} \varphi / 2)}
 \theta^{\chi} \Bigl[ 1 + a_0^2 \left( 1 + w\dot{\varphi}
 + b_{s} \varphi \right)^{2} \Bigr]  \\
 \label{eq:poly_mod1_2}
 \frac{\exd}{\exd w}(w^{2} \dot{\varphi}) &=&  \zeta
 w^{2} e^{4 a_0^2 \varphi (1 + b_{s} \varphi / 2)}
 \theta^{\chi} \left( 1 + w \dot{\varphi}
 +  b_{s} \varphi \right) \\
 \label{eq:poly_mod1_3}
 \hbox{and} \quad
 \mu &=& -w \left(\zeta \dot{\theta} +
 \frac{1}{2}a_0^2 w \dot{\varphi}^{2} + a_0^2(1+b_{s}
 \varphi)\dot{\varphi}\right)\ ,
\end{eqnarray}
and the initial conditions of eq.~\pref{eq:newtpoly_ic1} for
$\varphi$ now are:
\begin{equation}
 \varphi(0)=0\ ,\ \ \ \dot{\varphi}(0)=0\ .
\end{equation}

This system implies the solutions have the following power-series
expansions near $w = {0}$:
\begin{eqnarray}
 \theta(w) &=& 1 - \frac{1}{6}(1+a_0^2)w^{2} + \mathcal{O}(w^{4})\,\\
 \varphi(w) &=& \frac{\zeta}{6} \, w^{2} + \mathcal{O}(w^{4})\ .
\end{eqnarray}

When written in terms of the variables $\theta$, $\varphi$, and
$w$, the matching equations, (\ref{newt_match1}) --
(\ref{newt_match3}), become
\begin{eqnarray}\label{poly_match1}
 \frac{GM}{R} &=& - W \Bigl(\zeta \dot{\theta}(W) +
 a_0^2 [1+b_{s}\varphi(W)] \dot{\varphi}(W) \Bigr)  \\
 \label{poly_match2}
 \frac{GQ}{R} &=& a_{0} W \dot{\varphi}(W) \\
 \label{poly_match3}
 \hbox{and} \quad
 \phi_{\infty} &=& \phi_{0} + a_{0} [ \varphi(W)
 + W\dot{\varphi}(W) ] \,,
\end{eqnarray}
where $W = w(R) = R/r_s$ denotes the stellar boundary.

\section{Solutions for weak central coupling}

Most of what is known about the solutions to the equilibrium
equations derived above is based on integrating them numerically,
revealing several surprising features such as the phenomenon of
spontaneous scalarization \cite{damourspsc}. But the regime of
most practical interest is weak coupling, $a_0^2 \ll 1$, and since
interesting phenomena like scalarization are already present in
this limit, it is worth exploiting the simplicity of the
weak-coupling regime at the outset, both to simplify the numerics
required and (in some cases, see below) to allow analytic
solutions to be obtained.

Our goal in this section is to systematically expand in powers of
the scalar coupling at the stellar centre, $a_0^2 = a^2(\phi_0) =
a^2(r = 0) \ll 1$. By comparing these perturbative results with
direct numerical integrations, we show that for small $a_0^2$
their domain of validity typically covers the entire stellar
interior.

Our motivation for pursuing the simplifications introduced by this
expansion is the ease of generalizing to different kinds of scalar
couplings and to different equations of state. However in this
paper we confine our attention to the well-studied qBD case,
$a(\phi) = a_s + b_s \phi$, in order to better compare with known
results.

\subsection{The weak-central-coupling expansion}

The weak central-coupling expansion is clearest in the case of qBD
models, for which the entire coupling function, $a(\phi)$, is
determined by the two parameters $a_0 = a(\phi_0)$ and $b_s$. In
this case the weak central-coupling solutions are obtained by
expanding the differential equations in powers of $a_0^2$. For
Brans-Dicke theory ($b_s = 0$) the coupling is constant and known
to be small, $a_0^2 = a_s^2 \lsim 1.2 \times 10^{-5}$.

For qBD theories the lowest-order expansion of the equilibrium
equations, \pref{eqssnew1} through \pref{eqssnew3} gives
\begin{eqnarray}\label{eqssnew1p}
 \dot{\mu} + \frac{\mu}{2u} - \frac{\Dens(\Press)}{4} &\simeq&
 a_0^2 \left[ \Dens(\Press) \varphi \, \left( 1
 + \frac{b_{s} \varphi }{ 2} \right) +
 u(1-2\mu)\dot{\varphi}^{2} \right] + \O(a_0^4)  \\
 \label{eqssnew2p}
 \dot{\Press } + \frac{[\Dens(\Press) + \Press\,]
 (2\mu + u \Press)}{4u(1-2\mu)} &\simeq& - a_0^2
 [\Dens(\Press) + \Press\,]
 \left[ \frac{\Press \varphi(1+b_{s} \varphi / 2)}{1-2\mu}
 + \dot{\varphi}(1+u\dot{\varphi}+b_{s}\varphi) \right]
 + \O(a_0^4)  \\
 \label{eqssnew3p}
 \ddot{\varphi} + \frac{(3-4\mu)\dot{\varphi}}{2u(1-2\mu)} &-&
  \frac{1}{8u(1-2\mu)}
 \Bigl[ (1+b_{s}\varphi)[\Dens(\Press)-3\Press\,]
 + 2[\Dens(\Press)-\Press \,] u\dot{\varphi} \Bigr] \\
 &\simeq& \frac{a_0^2\varphi(1+b_{s}\varphi / 2)}{2u(1-2\mu)}
 \Bigl[ (1+b_{s}\varphi)[\Dens(\Press)-3\Press\,]
 + 2[\Dens(\Press)-\Press \,] u\dot{\varphi} \Bigr]
 + \O(a_0^4) \,,\nn
\end{eqnarray}
where dots denote derivatives with respect to $u = 8 \pi G
\rho_{0} A^{4}_{0} \, r^{2}$. Notice that the leading contribution
to the $\varphi$ equation depends on the self-coupling $b_s$, even
if $a_0^2 \to 0$. The boundary conditions are as before: $\mu(0)=
\varphi_0 = 0$, $\Press (0) = \PressC$ and $\dot{\varphi}(0) =
\frac{1}{12} \, (1-3\PressC)$ (and so $\dot \varphi(0) > 0$
provided $\PressC < \frac13$).

We seek interior profiles $\mu(u)$, $\Press (u)$ and $\varphi(u)$
obtained by integrating these equations subject to the series {\em
ans\"atze},
\begin{equation}
 \mu(u) = \sum_{i=0}^\infty \mu_{(i)}(u) \, a_0^{2i} \,, \quad
 \Press(u) = \sum_{i=0}^\infty \Press_{(i)}(u) \, a_0^{2i}
 \quad \hbox{and} \quad
 \varphi(u) = \sum_{i=0}^\infty \varphi_{(i)}(u) \, a_0^{2i} \,,
\end{equation}
with the leading expressions for $\Press_0(u)$ and $\mu_0(u)$
agreeing with the results from GR. In particular, because $a_0^2$
is small, the pressure profile decreases monotonically, ensuring
the existence of a solution for $R$ of $p(R) = 0$. Because of the
explicit factor of $a_0$ appearing in the definition $\varphi :=
(\phi - \phi_0)/a_0$, given a solution for $\mu$, $\Press$ and
$\varphi$ correct to order $a_0^{2k}$, we have a solution for
$\phi$ that is valid to order $a^{2k+1}_{0}$.
Such solutions are obtained explicity for $k=0$ and $k=1$ for incompressible
stars in sections \ref{sec:fo_sol} and \ref{sec:chi_corrections} below.

\subsection{Perturbative relations amongst observables}
\label{sec:pertexp}

This same $a_0^2$ expansion is inherited by the expressions
relating the external physical parameters, $M$, $R$, $Q$ and
$\phi_\infty$, by virtue of the matching conditions at $r = R$.
Although it is difficult to characterize these constraints
analytically in the general case, an expansion in powers of $a_0^2
= a^{2}(\phi_{0})$ allows some progress to be made. To this end
write
\begin{eqnarray} \label{Fnexpns}
 U(a_0,b_s,\PressC) &\simeq& U_{(0)}(\PressC) + U_{(1)}
 (b_s, \PressC) \, a_0^2 + \O(a_0^4) \nn\\
 \mathcal{F}(a_0, b_s, \PressC) &\simeq& \mathcal{F}_{(0)}(b_s,
 \PressC)
 + \mathcal{F}_{(1)}(b_s,\PressC) \, a_0^2 + \O(a_0^4) \nn\\
 \A(a_0, b_s, \PressC) &\simeq& \mathcal{A}_{(0)}(b_s, \PressC)
 + \mathcal{A}_{(1)}(b_s,\PressC) \, a_0^2 + \O(a_0^4) \\
 s(a_0, b_s, \PressC) &\simeq& s_{(0)}(\PressC) + s_{(1)}(b_s,\PressC) \, a_0^2 + \O(a_0^4) \nn\\
 \mathcal{M}(a_0, b_s, \PressC) &\simeq& \mathcal{M}_{(0)}(\PressC)
 + \mathcal{M}_{(1)}(b_s, \PressC) \, a_0^2 + \O(a_0^4) \,.\nn
\end{eqnarray}
where $s = GM/R$, $\F = (\phi_\infty - \phi_0)/a_0$, $\A = Q/M a_0 = a_\ssA/a_0$ and so on.

We seek two constraints among the four quantities $M$, $Q$, $R$
and $\phi_\infty$, and it is convenient to write the first of
these as a relationship between $M$, $R$ and $\phi_{\infty}$, and
the second as a relationship between $a_{\ssA} = Q/M$, $s = GM/R$
and $a_\infty = a(\phi_{\infty})$. The convenience of this choice
comes from the GR limit, for which the first constraint becomes
the usual $M$-$R$ relation, and the second constraint degenerates
into something vacuous: $0 = 0$.

An important point about these constraints is that the dependence
of observables like $M$, $Q$ and $R$ (or $U$) on $\phi_\infty$ ---
or $a_\infty = a(\phi_\infty)$ --- arises completely through their
dependence on $a_0 = a(\phi_0)$. Obtaining this dependence
therefore requires a relation between the scalar field at the
origin and infinity: $\phi_{0}(\phi_{\infty})$. This is
accomplished by using the function $\mathcal{F}$, whose definition
-- see eq.~(\ref{f}) -- states $\phi_\infty = \phi_0 + a_0 \, \F$,
and so
\begin{eqnarray} \label{infty_0_rel_A}
 A(\phi_{\infty}) &=& A(\phi_{0}) \exp \left[ a_0^2
 \left( 1 + \frac{b_{s} \mathcal{F}}{2} \right)
 \mathcal{F} \right] \,, \\
 \label{infty_0_rel_alpha}
 a(\phi_{\infty}) &=& a(\phi_{0})
 (1+b_{s} \mathcal{F}) \,.
\end{eqnarray}

It is tempting to ask at this point whether we are working too
hard. In particular, since it is $a_\infty$ and not $a_0$ that
directly controls the strengths of interactions that we see, being
asymptotic observers, perhaps we could avoid the exercise of
trading $a_0$ for $a_\infty$ by directly expanding the field
equations in powers of $a_\infty$ rather than $a_0$. The reason we
do not do so --- and indeed the point of expanding in powers of
$a_0$ --- is that the mapping defined by
eq.~\pref{infty_0_rel_alpha} between $a_0$ and $a_\infty$ is in
general not one-to-one. This is the lesson of scalarization, which
relies on $a_\infty = 0$ corresponding to {\em several} choices:
$a_0 = 0$ and $a_0 \ne 0$. It is the option of having a second
choice that allows the star to support a scalar field ($Q \ne 0$)
despite the vanishing of $a_\infty$. It is the fact that
integration of the field equations makes stellar properties
single-valued in $a_0$ that makes this the natural expansion
parameter. The existence of several branches to the function
$a_0(a_\infty)$ means that stellar properties need not also be
analytic in $a_\infty$. We describe the relevance of this to
scalarization in more detail below.

\subsubsection*{The generalized mass-radius relation}

To obtain the leading form of the constraint generalizing the
$M(R)$ relation of GR, set $a_0^2=0$ in equations (\ref{eqssnew1})
-- (\ref{eqssnew2}). The result implies that $\mu_{(0)}$ and
$\Press_{(0)}$ do not depend on $b_{s}$. Consequently
$U_{(0)}(\PressC)$, which is defined by $\Press_{(0)}(U_{(0)},
\PressC) = 0$, also cannot depend on $b_{s}$ --- a fact already
indicated in eqs.~\pref{Fnexpns}. The same is then true for the
compactness,
\begin{equation} \label{GRcompcond}
 s(U,\phi_\infty;b_s) = \frac{GM}{R} \simeq s_{(0)}[
 U_{(0)}, \PressC(U_{(0)})] + \O(a_0^2) \,,
\end{equation}
implying this constraint goes over to the GR limit to leading
order in $a_0^2$, even though the profile for $\varphi(u)$ need
not be trivial (for nonzero $b_s$). Thus is reproduced the usual
$M$ vs $R$ (or $U$) relation of GR.

\FIGURE[ht]{ \epsfig{file=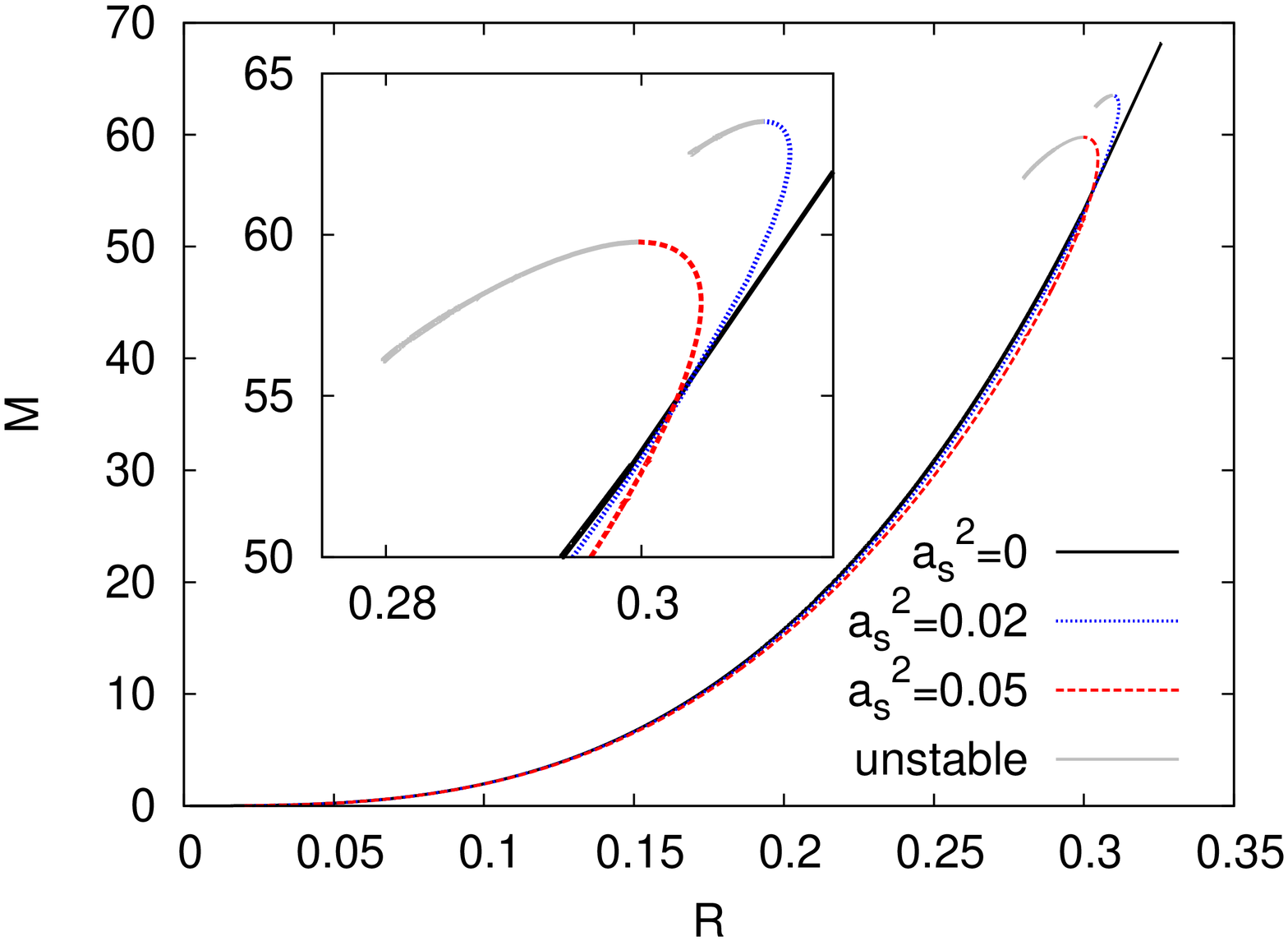,angle=270,width=0.8\hsize}
\\
\epsfig{file=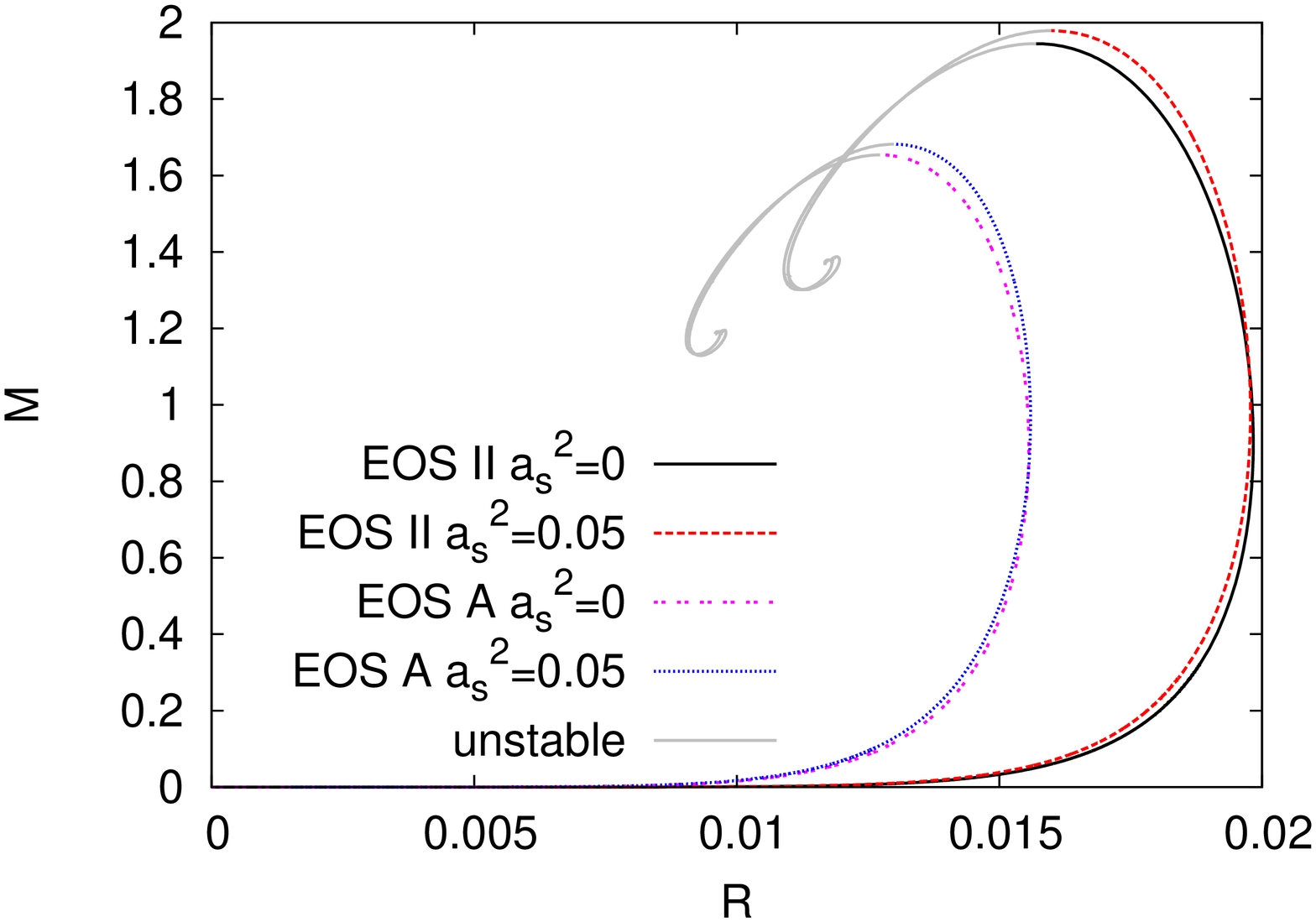,angle=270,width=0.8\hsize}
\caption{$M$ vs $R$ in Brans-Dicke theory for various equations of state
and values of $a_{s}$. The central value, $\PressC =
P_0/\rho_0$ varies along each curve. The starting point
of each curve ($\PressC \to 0$) is at $M=R=0$, and the endpoint of each
curve corresponds to the ultra-relativistic limit where $\PressC
\to \PressCmax$. The values of $\phi_{0}$ are chosen
such that $\phi_{\infty}$ is constant along each curve. The stellar
configurations become unstable after the first turning point where
$dM/dR=0$.
Top: Incompressible stars, for which $\PressCmax=\infty$.
Notice that the curves with non-zero
scalar coupling have smaller maximum radii and masses.
Bottom: Relativistic polytrope models of neutron stars, as defined
in \cite{damourspsc}, for which $\PressCmax$ is finite (colour online).
} \label{Fig1} }

Figure \ref{Fig1} illustrates how the generalized mass-radius
relation depends on $a_{s}$ in Brans-Dicke theory ({\em i.e.} $b_s
= 0$) for incompressible stars (discussed in
more detail in the next section) and relativistic polytrope models
of neutron stars.
Each curve traces the
relationship between $M$ and $R$ as $\PressC$ is varied,
beginning at $M=R=0$ where $\PressC \to 0$, and terminating at the point where $\PressC \to \PressCmax$.
For incompressible stars,
$\PressCmax$ is infinite. In general relativity, $M \propto R^{3}$,
and the maximum $M$ and $R$ that can be supported against gravitational
collapse are attained in the ultra-relativistic limit. However, once the
scalar-matter coupling is turned on, the maximum values of $M$ and
$R$ are attained at a finite value of $\PressC$.
As $a_{s}$ increases, the maximum values of $M$ and $R$
decrease.

The equations of state EOS II and EOS A are defined in reference
\cite{damourspsc}. They are relativistic polytropes, with a maximum
central pressure of $\PressCmax=\gamma-1$, where $\gamma$ is the
polytropic index. Their $M-R$ curves are more complicated than
those of incompressible stars, and turning on a weak scalar-matter coupling
slightly shifts these curves.

Numerically carrying out the stability
analysis described in section \ref{sec:matching}
shows that in all cases, the stellar configurations
become unstable after the first turning point where
$dM/dR=0$. Thus, the scalar field destabilizes ultra-relativistic
incompressible stars.

For non-relativistic systems the matching condition simplifies to
eq.~\pref{newt_match1},
\begin{equation}
 s_{(0)} = \mu_{(0)}(U_{(0)};\PressC) \ll 1\,,
\end{equation}
and although numerical methods are usually required to follow the
dependence on $\PressC(R)$, more explicit statements about the
scalar corrections to this relation are possible for specific
choices of equation of state. The examples of Newtonian polytropes
and incompressible stars are considered more explicitly below.

\subsubsection*{Scalar-coupling constraint}

For a given equation of state the second observable constraint,
eq.~(\ref{aa}), gives an expression for $a_{\ssA} = Q/M$ in terms
of $a_{0}$, $b_{s}$ and $\PressC$. Equation
(\ref{infty_0_rel_alpha}) can be used to relate $a(\phi_{\infty})$
with $a(\phi_{0})$, and on expansion yields
\begin{equation}\label{spsc_eqn}
 a(\phi_{\infty}) \simeq (1+b_{s}\mathcal{F}_{(0)}) a_{0} + b_{s}
 \mathcal{F}_{(1)} a^{3}_{0} + \mathcal{O}(a^{5}_{0}) \,.
\end{equation}
Therefore,
\begin{equation}\label{constraint2_fo}
 a_{\ssA} = \frac{Q}{M} = a_0 \, \A(a_0^2, b_s, \PressC)
 \simeq \frac{a(\phi_{\infty}) \A_{(0)}}
 {1+b_{s}\mathcal{F}_{(0)}} + \mathcal{O}(a^{3}_{0}) \,.
\end{equation}

This expression diverges when $1 + b_{s}\mathcal{F}_{(0)} \to 0$. In
this limit, one must include the $\mathcal{O}(a_{0}^{3})$ terms
in order to obtain a meaningful result. This will be described in the
section below.

\FIGURE[ht]{ \epsfig{file=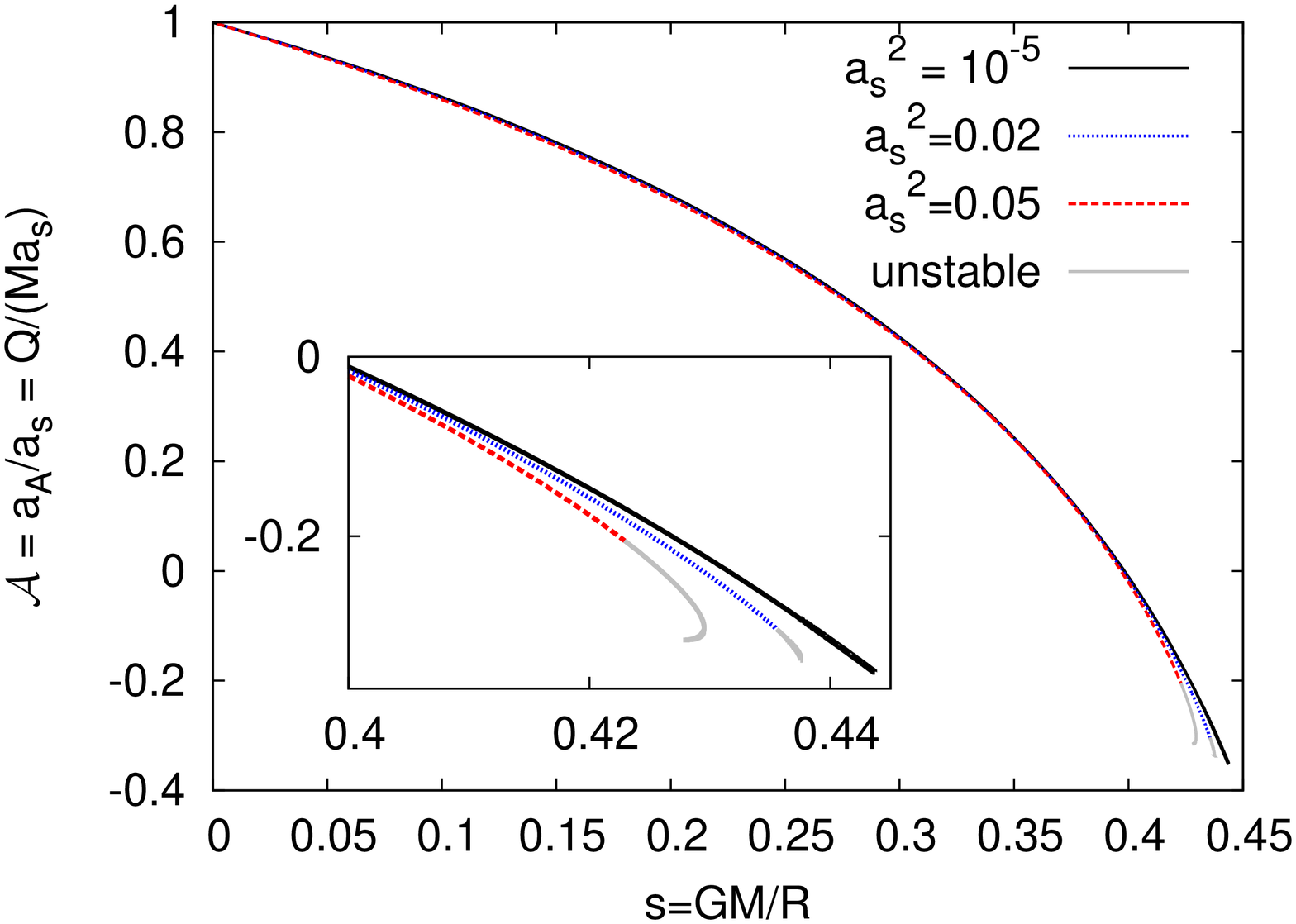,angle=270,width=0.9\hsize}
\\
\epsfig{file=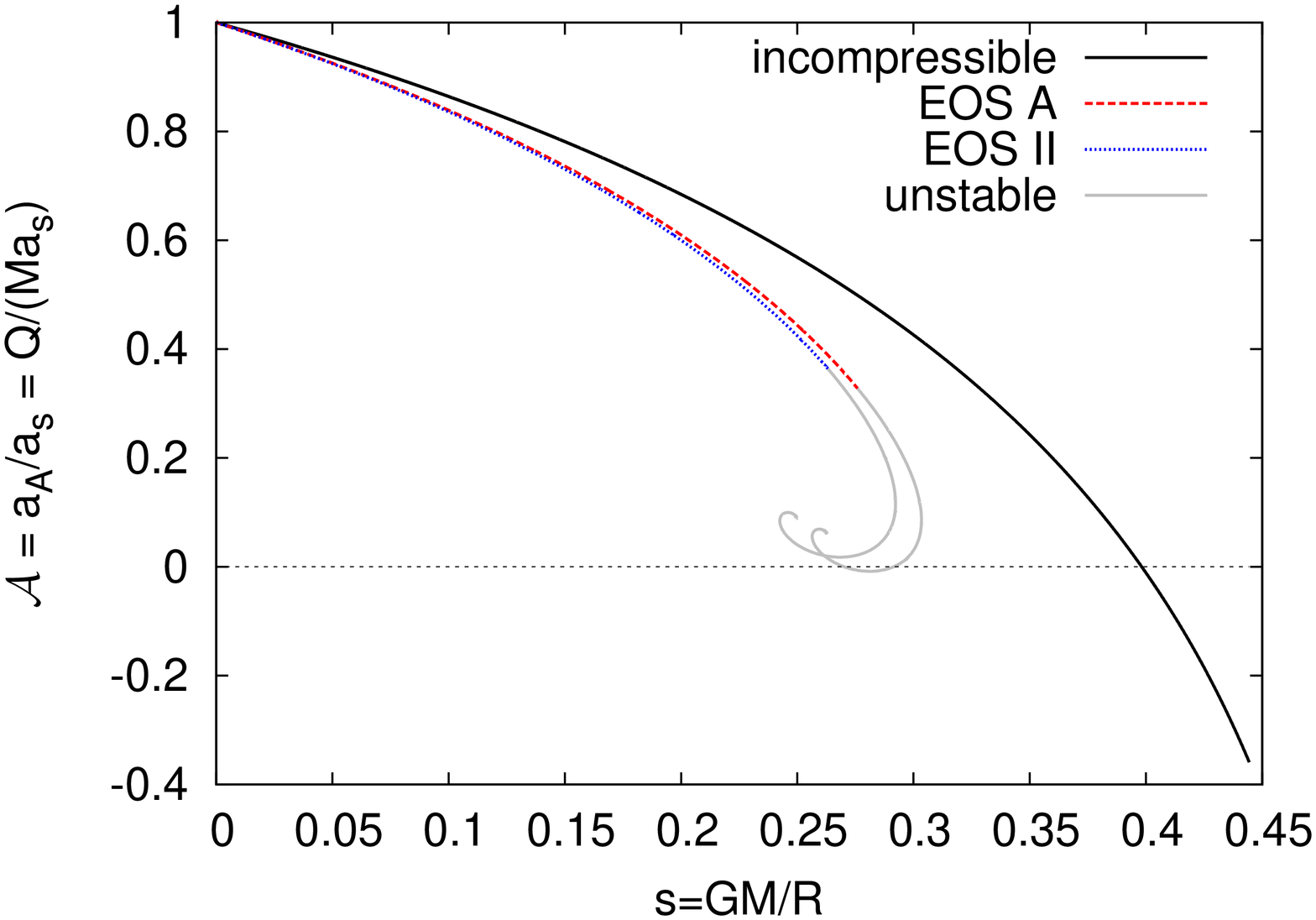,angle=270,width=0.9\hsize}
\caption{Top: A comparison of $\A = a_{\ssA}/a_{s} = Q/(Ma_{s})$
vs $s=GM/R$ for incompressible stars in Brans-Dicke theory, for
various values of $a_{s}$. The curves start at $s=0$, $\mathcal{A}=1$ where
$p_{0} \to 0$, and terminate where $\PressC \to
\infty$. Bottom: The same quantity comparing incompressible stars
with two kinds of neutron-star equations of state (relativistic
polytropes) given in \cite{damourspsc}, using Brans-Dicke theory
to leading order in $a_{s}$.} \label{Fig2} \label{Fig3}}

The dependence of $a_\ssA/a_0$ on $s$ is shown in Figure
\ref{Fig2} for several choices of $a_0 = a_s$ in the special case
of Brans-Dicke theory ($b_s = 0$), using for illustration an
incompressible star equation of state (see \S5, below).
Each curve can be parametrized by $\PressC$. The starting point is
at $s=0$, $\mathcal{A}=1$ when $\PressC \to 0$, and the endpoint
is reached in the limit $\PressC \to \PressCmax$.
%$\PressC$
%can be regarded as being a parameter along these curves, which
%terminate at the point corresponding to $\PressC \to \infty$.
These curves show that the small-$a_0$ limit works well in this
case even out to very relativistic stars. Notice that $a_\ssA$ is
in general smaller than $a_0$, with the suppression increasing for
more relativistic stars.

Fig.~\pref{Fig3} also compares the amount of this suppression for
several other choices for the equation of state, indicating that
the suppression for relativistic polytropes (more about which
later) is under-estimated for incompressible stars, although they
all agree in the non-relativistic limit (for which $s_{(0)} \ll
1$), in agreement with intuition.

For incompressible stars $a_\ssA$ eventually becomes negative. 
As seen in the insert in Fig.~\pref{Fig3}, this change in sign 
takes place before the onset of instability, 
so there exist stable incompressible stars with $a_\ssA < 0$. 
The scalar interaction between two stars $A$ and $B$ is attractive if 
$a_\ssA a_\ssB >0$, and repulsive if $a_\ssA a_\ssB < 0$. If at least one of 
$A$, $B$  
is an incompressible star, then both cases are possible.
%indicating a repulsive rather than an attractive scalar
%interaction.
However, negative values of $a_\ssA$ are not seen for the more
realistic equations of state, although these do approach $a_\ssA =
0$ when extremely relativistic. This should be compared with the
corresponding universal result, $Q = 0$, found above for a static,
spherically symmetric black hole.  
%All of the solutions with 
%negative $a_\ssA a_\ssB$ we have found are unstable, according to
%the criterion given above in \S\ref{sec:stability}. 

\subsubsection*{Spontaneous Scalarization}

In the above section, the scalar-coupling constraint was expanded in powers of $a_{0}$, and it was found that at leading order, $a_{\ssA}$ is a single-valued function of $a_{\infty}$. In this section, we take the expansion to next order, and demonstrate that $a_{\ssA}$ becomes a multi-valued function $a_{\infty}$. As a consequence, $a_0$ and $a_\ssA$ can both be nonzero even when $a_\infty$ vanishes. The phenomenon where $Q/M$ is nonzero even though $a_\infty = 0$ is called spontaneous scalarization. To simplify notation, write
\begin{eqnarray}
\label{ainfty_exp}
 a_{\infty} = d_{1}a_{0} + d_{2}a_{0}^{3} + \mathcal{O}(a_{0}^{5}) \,,
 \\
 \label{aa_exp}
 a_{\ssA} = e_{1}a_{0} + e_{2}a_{0}^{3} + \mathcal{O}(a_{0}^{5})\,.
\end{eqnarray}
Dropping terms of order $a_{0}^{5}$ and inverting equation (\ref{ainfty_exp}) yields
\begin{equation}
 a_{0} = \omega C_{+} + \bar{\omega} C_{-} \,,
\end{equation}
where
\begin{equation}
 C_{\pm} = \sqrt[3]
 {
 \frac{a_{\infty}}{2d_{2}}
 \pm
 \sqrt{D}
 } \,,
 \qquad
 D=\left(\frac{a_{\infty}}{2d_{2}}\right)^{2}
 +\left(\frac{d_{1}}
 {3d_{2}}\right)^{3} \,,
\end{equation}
and $\omega = 1, -e^{-i\pi/3}, -e^{i\pi/3}$. Thus,
\begin{equation}
 a_{\ssA} = (\omega C_{+} + \bar{\omega}C_{-})e_{1} +
 (\omega C_{+} + \bar{\omega}C_{-})^{3}e_{2}
 + \mathcal{O}(a_{0}^{5}) \,.
\end{equation}
If $D>0$, then the $\omega=1$ solution is real, and the other two solutions are complex. Thus, specification of $a_{\infty}$ determines a unique stellar configuration.

If $D<0$, then all solutions are real. Thus, specification of $a_{\infty}$ determines three stellar configurations.

In the limit $a_{\infty} \to 0$, the $\omega=1$ solution vanishes,
and the other two solutions become
\begin{equation}
 \label{sc_cond1}
 a_{0} = \pm  \sqrt{-\frac{d_{1}}{d_{2}}} \,.
\end{equation}
Therefore, whenever the quantity inside the square root is positive, there exist stellar configurations with $a_{\infty}=0$ and
\begin{equation}
 a_{\ssA} =
 \pm \left\{ \left(-\frac{d_{1}}
 {d_{2}}\right)^{1/2} e_{1}
 + \left(-\frac{d_{1}}
 {d_{2}}\right)^{3/2} e_{2} \right\}
 + \mathcal{O}(a_{0}^{5})
 \,.
\end{equation}
This is precisely the phenomenon of spontaneous scalarization.

The coefficients $d_{i}$ and $e_{i}$ are functions of $b_{s}$ and $\PressC$, and they also depend on the equation of state. The solution obtained in section \ref{sec:fo_sol} for incompressible stars can be used to calculate
\begin{eqnarray}
 \label{d1_eqn}
 d_{1} &=& {\rm HeunG}(\tilde{a},\tilde{q};\tilde{\alpha},\tilde{\beta},
 \tilde{\gamma},\tilde{\delta} ; Z)
 - \frac{1+\PressC}{1+3\PressC}\log \left( \frac{1+\PressC}{1+3\PressC} \right)
 {\rm HeunG'}(\tilde{a},\tilde{q};\tilde{\alpha},\tilde{\beta},
 \tilde{\gamma},\tilde{\delta} ; Z) \,,
 \\
 \label{e1_eqn}
 e_{1} &=& \frac{1+\PressC}{b_{s}(1+3\PressC)}
 {\rm HeunG'}(\tilde{a},\tilde{q};\tilde{\alpha},\tilde{\beta},\tilde{\gamma},
 \tilde{\delta} ; Z) \,,
\end{eqnarray}
where the arguments inside the Heun functions are given in equations (\ref{heunpars1})-(\ref{heunpars2}), and $Z=\PressC/(3\PressC+1)$. Note that in the limit $\PressC \to 0$, we have $d_{1} \to 1$ and $e_{1} \to 1$. In principle, the next coefficients $d_{2}$ and $e_{2}$ can be calculated using the solution found in section \ref{sec:chi_corrections}. However, the resulting expressions are very complicated.

It follows from equation (\ref{sc_cond1}) that the onset of scalarization implies $d_{1}=0$. Thus, all the points $(b_{s},\PressC)$ at which scalarization starts can be found by doing a root search of equation (\ref{d1_eqn}).

In figure (\ref{fig:d1_bneg}),
equation (\ref{d1_eqn}) is plotted versus
$\PressC$, for various negative values of
$b_{s}$. In all cases, $d_{1}$ is a convex
function of $\PressC$, with one global minimum.
If $b_{s} \in (-4.329,0)$, then $d_{1}$ never crosses
zero, and there is no scalarization.
If $b_{s} \in (-\infty,-4.329)$, then $d_{1}$ has two crossings of zero,
which correspond to the scalarization which has been extensively studied
in the literature.

In figure (\ref{fig:d1_bpos}), equation (\ref{d1_eqn}) is plotted
versus $\PressC$, for various positive values of $b_{s}$. In all cases,
$d_{1}$ is an oscillatory function of $\PressC$, and there are
multiple regions of scalarization. We have verified numerically that
scalarization does actually occur when $d_{1}<0$. However, application of
of the stability criterion of \S\ref{sec:stability} shows that these 
scalarized stars are unstable whenever $b_s > 0$.

%We have found evidence that incompressible
%stars can also scalarize for $b_{s}>0$,
%and are currently investigating whether
%these solutions are stable, and have physical relevance.

\FIGURE[r]{\epsfig{file=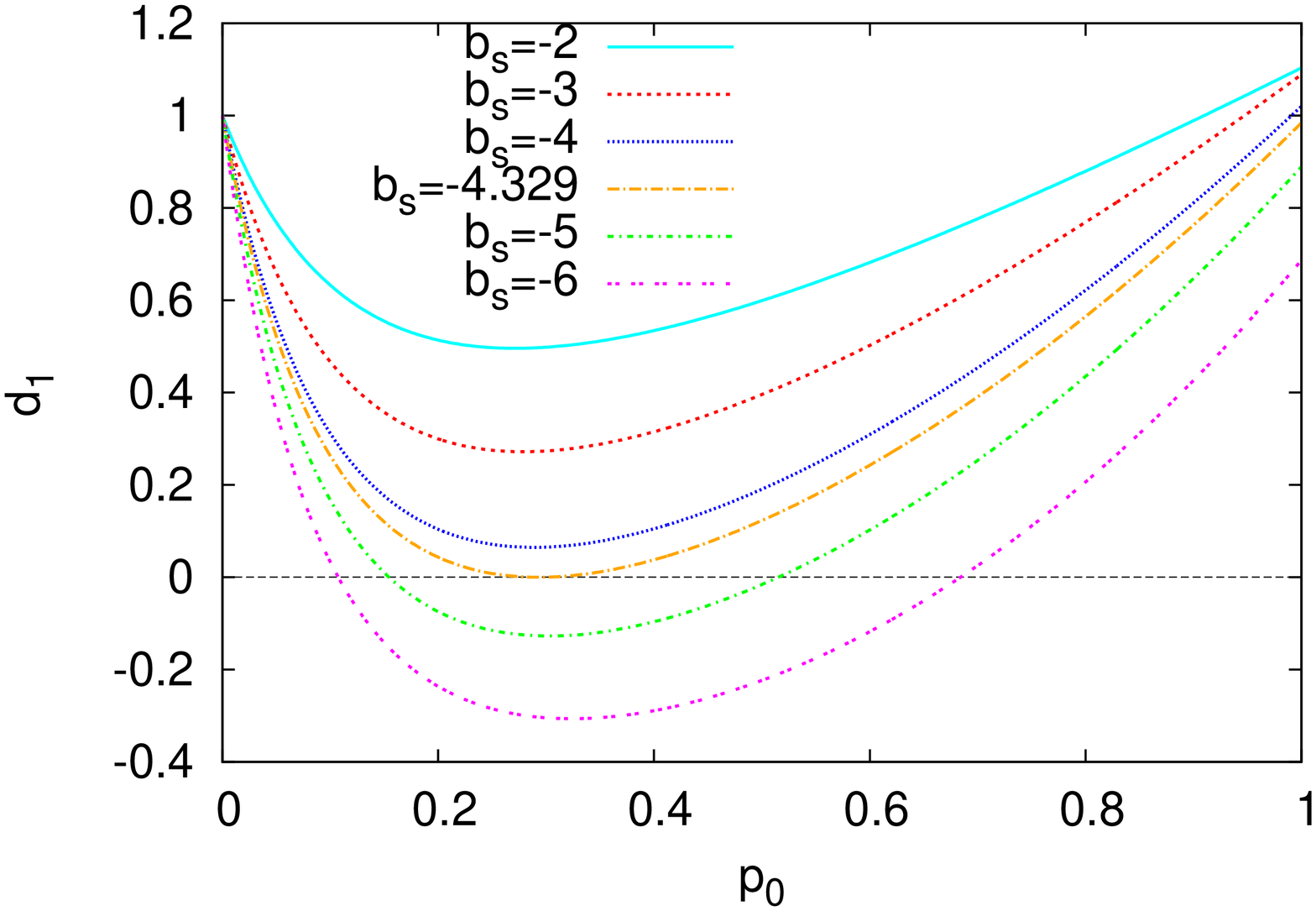,angle=270,width=0.8\hsize}
\label{fig:d1_bneg}
\caption{The coefficient $d_{1}$ plotted versus $\PressC$, for constant-density stars, for several choices of $b_{s}<0$. Scalarization becomes possible for
$b_{s} < -4.329$.}
}

\FIGURE[r]{\epsfig{file=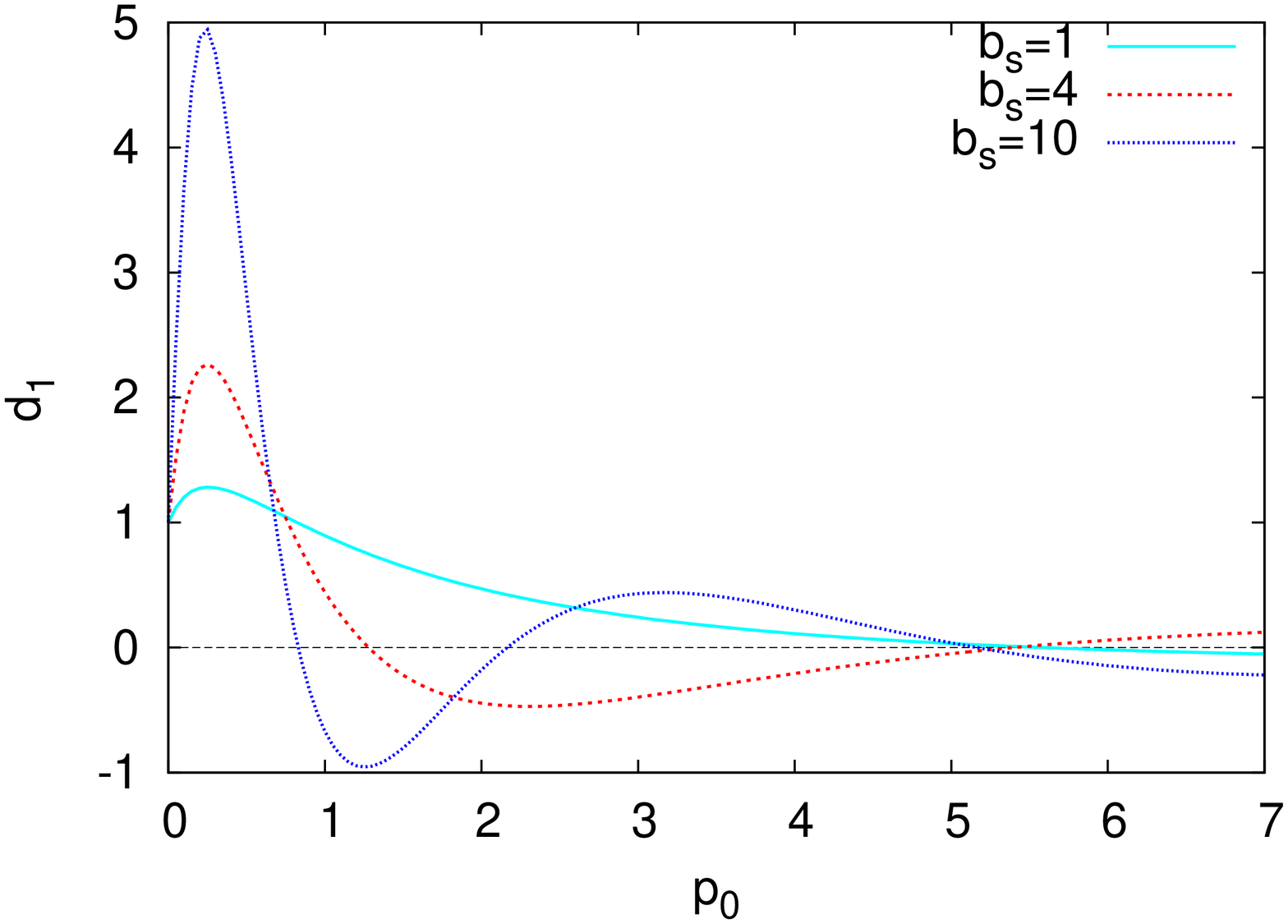,angle=270,width=0.8\hsize}
\label{fig:d1_bpos}
\caption{The coefficient $d_{1}$ plotted versus $\PressC$, for constant-density stars, for several choices of $b_{s}>0$. There are multiple regions of
scalarization.}
}

\clearpage

\subsection{Non-relativistic polytropes}

As an example for which the scalar-field dependence of the above
constraints can be more explicitly explored, consider the case of
Newtonian polytropes discussed \S\ref{sec:nonrellim} with equation
of state $P = K \rho^{1+1/\chi}$. In this case for weak central
coupling the scalar field, $\varphi$, and the dimensionless
density, $\theta = (\rho/\rho_0)^{1/\chi}$, can be expanded in
$a_0^2$, $b_s$, as well as the parameter $\zeta = (1 + \chi) K
\rho_0^{1/\chi} = (1 + \chi) P_0/\rho_0 = (1 + \chi) \PressC$:
\begin{eqnarray}
 \theta &=& \sum_{i} \theta_{(i)} a_0^{2i} = \sum_{i,j}
 \theta_{(i,j)} \, a_0^{2i} \, b_{s}^{j} = \sum_{i,j,k}
 \theta_{(i,j,k)} \, a_0^{2i} \, b_{s}^{j} \, \zeta^{k}\ , \\
 \varphi &=& \sum_{i} \varphi_{(i)} a_0^{2i} = \sum_{i,j}
 \varphi_{(i,j)} \, a_0^{2i} \, b_{s}^{j} = \sum_{i,j,k}
 \varphi_{(i,j,k)} \, a_0^{2i} \, b_{s}^{j} \, \zeta^{k}\ .
\end{eqnarray}
We show below that $\zeta \ll 1$ for the polytropes of practical
interest, such as white dwarfs and main-sequence stars.

In the limit $a_0^2=0$, eqs.~(\ref{eq:poly_mod1_1}) --
(\ref{eq:poly_mod1_2}) become
\begin{eqnarray} \label{eq:le0_1}
 {\theta}'' &=& - \frac{2 {\theta}'}{w} - \theta^{\chi}\ ,
 \\ \label{eq:le0_2}
 {\varphi}'' &=& - \frac{2 {\varphi}'}{w} + \zeta
 \theta^{\chi} [1 + w {\varphi}' + b_{s} \varphi ]\ .
\end{eqnarray}
where for later notational convenience we use primes in this
section to denote differentiation with respect to $w$. Equation
(\ref{eq:le0_1}) is called the Lane-Emden equation, and its
solutions are well-studied because of its important role in the
theory of stellar structure. It can be solved analytically when
$\chi=0,1,5$ \cite{chandbook}. The solutions of the Lane-Emden
equation with initial conditions (\ref{eq:newtpoly_ic1}) are
called Lane-Emden functions, and are denoted by
$\Theta_{\chi}(w)$. Thus,
\begin{equation}
 \theta_{(0)}(w) = \Theta_{\chi}(w)\ .
\end{equation}

If we now take $b_{s}=0$ then equation (\ref{eq:le0_2}) can be
solved for $\varphi$ in terms of $\theta$, giving
\begin{eqnarray}
 {\varphi'_{(0,0)}} &=& - \frac{\zeta}{w^{2}}
 e^{-\zeta (w\Theta_{\chi})'}
 \int^w (\hat w^{2}{\Theta}_{\chi}')'
 e^{\zeta(\hat w \Theta_{\chi})'}
 \exd \hat w\ \\
 &=&
 -\frac{1}{w} + \frac{1}{w^{2}} e^{-\zeta (w \Theta_{\chi})'}
 \int^w e^{\zeta (\hat w \Theta_{\chi})'} \exd \hat w \,.
\end{eqnarray}
Expanding this in powers of $\zeta$ then yields
\begin{equation}
 \varphi_{(0,0)} \simeq \zeta(1-\Theta_{\chi}) +
 \mathcal{O}(\zeta^{2})\,.
\end{equation}
and so
\begin{equation}
 \varphi_{(0,0,0)} = 0
 \quad \hbox{and} \quad
 \varphi_{(0,0,1)} = 1 - \Theta_{\chi}\ .
\end{equation}

Expanding the matching equations, eqs.~(\ref{poly_match1}) --
(\ref{poly_match3}) in powers of $a_0^2$, $b_{s}$ and $\zeta$
similarly yields
\begin{eqnarray}
 \label{poly_match1_pert} \frac{GM}{R} &\simeq& -\zeta W
 {\Theta'}_{\chi}(W) + \mathcal{O}(a_0^2) \,,
 \\ \label{poly_match2_pert}
 \frac{Q}{Ma_{0}} &\simeq& 1 + \mathcal{O}(a_0^2)
 + \mathcal{O}(b_{s}) + \mathcal{O}(\zeta) \,,
 \\ \label{poly_match3_pert}
 \frac{\phi_{\infty}-\phi_{0}}{a_{0}} &\simeq& \zeta
 \Bigl[ 1-\Theta_{\chi}(W) - W {\Theta'}_{\chi}(W) \Bigr]
 + \mathcal{O}(a_0^2) + \mathcal{O}(b_{s})
 + \mathcal{O}(\zeta^{2}) \,.
\end{eqnarray}
Notice in particular that eq.~(\ref{poly_match2_pert}) implies
that
\begin{equation}
 a_{A} \simeq a_{0} \Bigl[ 1 + \mathcal{O}(a^{2}_{0})+
 \mathcal{O}(b_{s}) + \mathcal{O}(\zeta) \Bigr] \,,
\end{equation}
which is consistent with the limit $a_{A} \to a(\phi_{\infty})$
for weakly self-gravitating stars. Similarly, equation
(\ref{poly_match1_pert}) can be re-written as
\begin{eqnarray} \label{poly_mass}
 GM &=& -\zeta r_s W^2 \Theta'_\chi + \O(a_0^2) \nn\\
 &=& - W^{2} {\Theta}'_{\chi}(W) \; \left[
 \frac{\rho_{0}^{(3-\chi)/2\chi}}{A^{2}_{0}}
 \sqrt{\frac{K^{3}(\chi+1)^{3}}{4 \pi G}} \right]
 + \mathcal{O}(a_0^2) \,.
\end{eqnarray}
which is independent of $\rho_{0}$, as advertised, when $\chi = 3$.

To get an idea about the validity of the $\zeta$ expansion, we
close this section by estimating its size for the examples of
white dwarfs and main-sequence polytropes.

\subsubsection*{White dwarfs}

A white dwarf can be modeled by a degenerate fermion gas which
satisfies in the ultra-relativistic limit the polytropic equation
of state, eq.~(\ref{eq:polytrope}), with
\begin{equation}
     K = \frac{3^{1/3}\pi^{2/3}}{4} \left( \frac{Y_{e}}{m_{b}}
     \right)^{4/3}
     \quad \hbox{and} \quad
     \chi=3 \,.
\end{equation}
Here $Y_{e}$ is the number of electrons per nucleon and $m_{b}$ is
the average nucleon mass \cite{astrophys}. Although the fermion
gas is highly relativistic, the gravitational field generated by
it is not strong, allowing use of the non-relativistic expressions
developed above.

The typical central densities of white dwarfs are $\rho_{0} \sim
(10^{7} \ldots 10^{14} ) \; {\rm kg}\, {\rm m}^{-3}$
\cite{astrophys}, corresponding to $10^{-5} \lsim \zeta \lsim
10^{-2}$ for $Y_{e} = \frac12$. This shows that a perturbative
expansion in $\zeta$ is likely a good approximation for white
dwarfs.

Since the ultra-relativistic limit of the fermion gas is used, the
mass calculated below using the matching conditions --- {\em i.e.}
equation (\ref{poly_mass}) --- is greater than the actual mass of
the white dwarf. It is instead to be regarded as the Chandrasekhar
limit: an upper bound on the mass of white dwarfs. Because $\chi =
3$ its value turns out to be independent of $\rho_{0}$, and for
$Y_{e} = \frac{1}{2}$, its value is approximately $1.4
M_{\odot}$ \cite{chandmass}.

\subsubsection*{Main-sequence models}

In the Eddington stellar model, a star is regarded as an ideal gas
whose energy is transported by radiation, and it is assumed that
the gas makes up a fixed fraction of the total pressure, $\beta :=
{P_{\rm gas}}/{P} = {\rm const}$. Here $P_{\rm gas}$ is the
pressure of the ideal gas, and the total pressure is $P = P_{\rm
gas} + P_{\rm rad}$, where $P_{\rm rad}$ is the radiation
pressure. The Eddington model leads to a polytropic equation of
state with
\begin{equation}
     K = \left[ \frac{45}{\pi^2 k_\ssB^4}  \left(\frac{R_g}{\mu}\right)^{4}
     \frac{1-\beta}{\beta^{4}}\right]^{1/3} \quad \hbox{and}
     \quad \chi = 3 \,,
\end{equation}
where $R_g$ is the universal gas constant, $\mu$ is the molar mass
of the ideal gas and $k_\ssB$ is Boltzmann's constant
\cite{astrophys2}. Main-sequence stars can be approximately
described by the Eddington standard model, even though convection
also plays a role in heat transfer for more realistic models.

For a more accurate single-polytrope model of the Sun, $\chi=3.35$
and $\zeta \sim 10^{-5}$ \cite{solarmodel}, so the perturbative
expansion in $\zeta$ remains a good approximation.

\section{Incompressible stars}

We now specialize the discussions of the previous sections to the
special case of an incompressible star, for which the stellar
density, $\rho$, is constant. Since constant density can only be
consistent with the pressure gradients required for hydrostatic
equilibrium if $\Press \ne \Press(\rho)$, we no longer impose this
kind of equation of state. (It is not necessary in any case, since
the closure of the field equations is now accomplished by the
incompressibility condition, $\rho(r) \equiv \rho_0$.)

The purpose of this exercise is to have a toy example for which
all of the above manipulations can be simply carried through
explicitly in closed form. Performing the same exercise for GR
provides an interesting example that displays the main features of
relativistic structure, including the existence of a maximum
compactness for a star, $s_{(0)} \le \frac49$, that can be
supported against gravitational collapse. The maximum that is
found for incompressible stars turns out to provide an upper bound
to the maximum compactness that can be achieved with other
equations of state.

Several earlier works have numerically investigated
incompressible stars in scalar-tensor gravity
\cite{saakmnats,hillheintz,avakyan},
but our quasi-analytical treatment of these stars is new.

\subsection{Incompressible stars with quasi-Brans/Dicke scalars}
\label{sec:constrho}

We cut right to the chase and specialize directly to qBD scalars,
for which $a(\phi) = a_s + b_s \phi$, since this case is broad
enough to be of wide interest, but restricted enough to be
explored in detail.

\subsubsection*{Field equations}

Taking the equation of state to be $\rho = \rho_0$, or
$\Dens(\Press ) = 1$, equations (\ref{eqssnew1}) --
(\ref{eqssnew3}) become
\begin{eqnarray}\label{constrho1}
 \dot{\mu} &=& -\frac{\mu}{2u} + \frac14 \, e^{4a_0^2 \varphi(1+b_{s}
 \varphi / 2)} + a_0^2 \, u(1-2\mu)\dot{\varphi}^{2}
 \\ \label{constrho2}
 \dot{\Press } &=& - (1+\Press )\left[
 \frac{\mu}{2u(1-2\mu)} + \frac{\Press }{4(1-2\mu)} \,
 e^{4a_0^2 \varphi(1+b_{s} \varphi / 2)} +a_0^2 \,
 \dot{\varphi}(1+u\dot{\varphi}+b_{s}\varphi) \right]
 \\ \label{constrho3}
 \ddot{\varphi} &=& -\frac{(3-4\mu)\dot{\varphi}}{2u(1-2\mu)} +
 \frac{e^{4a_0^2\varphi(1+b_{s}\varphi / 2)}} {8u(1-2\mu)} \Bigl[
 (1+b_{s}\varphi)(1-3\Press ) + 2(1-\Press )u\dot{\varphi}\Bigr]\,.
\end{eqnarray}
Similarly, the function $f(\Press )$ defined in eq.~(\ref{f_defn})
becomes
\begin{equation}
 f(\Press ) = \ln \left( \frac{1+\Press }{1+\PressC}\right) \,,
\end{equation}
and so the baryon number density, $n(r)$, computed from
eq.~(\ref{n_eqn}) is also constant, $n=n_{0}$. The function
$\mathcal{M}$ defined in eq.~\pref{barmass_quadmod} similarly
becomes
\begin{eqnarray}
 \mathcal{M} &=& \int_{0}^{U} \exd u \sqrt{\frac{u}{1-2\mu}}
 \exp\left[ 3a_0^2 \varphi \left(1+ \frac{b_{s}\varphi}{2}
 \right) \right]  \,.
\end{eqnarray}

\subsection{Perturbative solutions: leading order}\label{sec:fo_sol}

Because the GR problem can be explicitly integrated for
incompressible stars, equations (\ref{constrho1}) --
(\ref{constrho3}) can be solved analytically when $a_0^2 = 0$. The
zeroth-order profiles, $\mu_{(0)}$ and $\Press_{(0)}$, are given
by \cite{GRincomp, carroll}
\begin{eqnarray}
 \mu_{(0)}(u) &=& \frac u6 \,, \\
 \Press_{(0)}(u) &=& \frac{(1+3\PressC)\sqrt{1-u/3}-(1+\PressC)}
 {3(1+\PressC) - (1+3\PressC)\sqrt{1-u/3}} \,,
\end{eqnarray}
as a function of the central pressure $\PressC$. We return to
computing the profile, $\varphi(u)$, below.

At leading order the stellar radius is determined as the zero of
$\Press_{(0)}$, which vanishes at $u = U_{(0)}$ where
\begin{equation}
 U_{(0)}= \frac{12\PressC(1+2\PressC)}{(1+3\PressC)^{2}} \,,
\end{equation}
corresponding to the radius $r = R_{(0)}$ with
\begin{equation}\label{radius}
 R_{(0)} = \frac{2}{A^{2}_{0}(1+3\PressC)}
 \sqrt{\frac{3\PressC(1+2\PressC)}{8\pi G \rho_{0}}}  \,.
\end{equation}
The leading components of the functions relevant to matching to
the exterior solutions --- {\em i.e.} $\mathcal{F}$, $\A$, $s$ and
$\mathcal{M}$ --- are
\begin{eqnarray} \label{match0_constrho}
 \mathcal{F}_{(0)} &=& \varphi_{\star(0)}
 -12\left(\frac{1+\PressC}{1+3\PressC}\right)^{2}
 \ln \left( \frac{1+\PressC}{1+3\PressC} \right)
 \dot{\varphi}_{\star(0)} \,,  \\ \label{match1_constrho}
 \A_{(0)} &=& 12\left( \frac{1+\PressC}{1+3\PressC}\right)^{2}
 \dot{\varphi}_{\star(0)}\,,
 \\ \label{match2_constrho}
 s_{(0)} &=& \frac{2\PressC(1+2\PressC)}{(1+3\PressC)^{2}}\,, \\
 \mathcal{M}_{(0)} &=& 3\sqrt{3}
 \left\{ \arccos \left( \frac{1+\PressC}{1+3\PressC}\right)
 - 2 \frac{(1+\PressC) \sqrt{\PressC(1+2\PressC)}}{(1+3\PressC)^{2}} \right\} \,,
\end{eqnarray}
where, as before, the subscript `$\star$' denotes evaluation at $u
= U$, so $\varphi_{\star(0)} := \varphi_{(0)}(U_{(0)})$.

Equation (\ref{match2_constrho}) implies that the compactness is
an increasing function of $\PressC$, which vanishes when
$\PressC=0$ and asymptotes to $\frac 49$ as $\PressC \to \infty$.
Thus we reproduce the GR result $0 \leq s_{(0)} \leq \frac49$ for
constant-density stars. This prediction gets modified once
$O(a_0^2)$ corrections are included, however, as is discussed in
detail in the next section.

The mass-radius constraint, eq.~(\ref{GRcompcond}), in this case
becomes
\begin{equation}
 s = \frac{GM}{R} = \frac{4\pi G\rho_{0}}{3} A^{4}_{\infty}  R^{2}
 + \mathcal{O}(a^{2}_{0}) \,,
\end{equation}
which states $M \propto R^3$, as might be expected for constant
density. This behaviour is seen explicitly in the $a_s = 0$ curve
in Fig.~\pref{Fig1}.

The scalar-coupling constraint, eq.~(\ref{constraint2_fo}),
similarly becomes
\begin{eqnarray} \label{qm_constraint_2_alt}
 a_{\ssA} &=&
 \frac{12 a_{\infty}(1-2s)\dot{\varphi}_{\star(0)}}
 {1+b_{s} \varphi_{\star(0)} -
 6b_{s}(1-2s) \ln(1-2s) \dot{\varphi}_{\star(0)}} +
 \mathcal{O}(a^{3}_{0})\,,
\end{eqnarray}
where
\begin{equation}
 \PressC = \frac{1-3s - \sqrt{1-2s}}{9s-4}
 + \mathcal{O}(a^{2}_{0})
\end{equation}
should be substituted into $\varphi_{\star{(0)}}$ and $\dot
\varphi_{\star{(0)}}$ on the right-hand side. Since further
progress requires knowing the scalar profile, we next turn to
solving its field equation.

\subsubsection*{Scalar profile in Brans-Dicke theory
(when $b_{s}=0$)}\label{sec:solb0}

If $b_{s}=0$, then the model reduces to Brans-Dicke theory, and
$a(\phi) \equiv a_{0} = a_s$. Equation (\ref{constrho3}) with
$a_0^2 = 0$ becomes a first-order linear differential equation for
$\dot{\varphi}$, which can be solved analytically:
\begin{equation}\label{phidot_beta0}
 \dot{\varphi}_{(0)}(u) = \frac{ 27(1+\PressC)
 \left(\frac{\arcsin\sqrt{u/3}}{\sqrt{u/3}} - \sqrt{1-u/3}\right)
 -4(1+3\PressC)u}
 {12u\sqrt{1- u/3}(3(1+\PressC) - (1+3\PressC)\sqrt{1-u/3})}\,.
\end{equation}
Integrating this expression once more with respect to $u$ then
gives
\begin{eqnarray} \label{phi_beta0}
 \varphi_{(0)} &=& \frac{1}{8(2+3\PressC)}\biggl[
 (9\PressC+5)(3\PressC-1) \ln
 \left( 3(1+\PressC)-(1+3\PressC)\sqrt{1-u/3}\right)
 \nonumber \\
 &&\hphantom{\frac{1}{8(2+3\PressC)}\biggl[}
 -9(1+\PressC)\left( 1+3\PressC + 3(1+\PressC)\sqrt{1-u/3}\right)
 \frac{\arcsin{\sqrt{u/3}}}{\sqrt{u/3}} \biggr]
 \nonumber \\
 && + \frac{9(1+3\PressC)^{2}(1+\PressC)}{16(2
 +3\PressC)^{3/2}} \biggl[
 {\rm Li}_{2} (\lambda_{-}) - {\rm Li}_{2} (\lambda_{+})
 + i\ln \left( \frac{1-\lambda_{-}}{1-\lambda_{+}}\right)
 \arcsin{\sqrt{u/3}}\biggr]\,,
\end{eqnarray}
where
\begin{equation}
 \lambda_{\pm} := \frac{(1+3\PressC)(\sqrt{1-u/3}+i\sqrt{u/3})}
 {3(1+\PressC) \pm 2\sqrt{2+3\PressC}}
 \equiv |\lambda_{\pm}| e^{i \arcsin \sqrt{u/3}}
 \,,
\end{equation}
and
\begin{equation}
 {\rm Li}_{2}(z) = \sum_{k=1}^{\infty}\frac{z^{k}}{k^{2}}
\end{equation}
is the dilogarithm function.

By using the identity \cite{gradshteyn}
\begin{equation}
 \sum_{k=1}^{\infty}p^{k} \sin (kx) =
 \frac{p \sin x}{1 - 2p \cos x + p^{2}} \,,
\end{equation}
it can be shown that the imaginary part of (\ref{phi_beta0}) is
constant, and can thus be absorbed into an integration constant.
The final normalized and manifestly-real expression for
$\varphi_{(0)}$ is then
\begin{eqnarray}
 \label{phi_beta0_final}
 \varphi_{(0)}(u) &=& \frac{1}{8(2+3\PressC)}\biggl[
 (9\PressC+5)(3\PressC-1) \log
 \left( \frac{3}{2}(1+\PressC)-\frac{1}{2}(1+3\PressC)
 \sqrt{1-u/3}\right)
 \nonumber \\
 &&\hphantom{\frac{1}{8(2+3\PressC)}\biggl[}
 -9(1+\PressC)\left( 1+3\PressC + 3(1+\PressC)\sqrt{1-u/3}\right)
 \frac{\arcsin{\sqrt{u/3}}}{\sqrt{u/3}} \biggr]
 \nonumber \\
 && + \frac{9(1+3\PressC)^{2}(1+\PressC)}{16(2+3
 \PressC)^{3/2}} \biggl[
 \arctan \left(
 \frac{2\sqrt{(2+3\PressC)u/3}}{1+3\PressC
 -3(1+\PressC)\sqrt{1-u/3}}
 \right)
 \arcsin \sqrt{u/3}
 \nonumber
 \\
 &&\hphantom{+ \frac{9(1+3\PressC)^{2}(1+
 \PressC)}{16(2+3\PressC)^{3/2}} \biggl[}
 +\Re \left[ {\rm Li}_{2} (\lambda_{-}) -
 {\rm Li}_{2} (|\lambda_{-}|)
 - {\rm Li}_{2} (\lambda_{+}) + {\rm Li}_{2} (|\lambda_{+}|) \right]
 \biggr]
 \nonumber
 \\
 &&+\frac{9}{4}(1+\PressC) \,.
\end{eqnarray}

If $\PressC > \frac{1}{\sqrt{3}}$, then the inverse tangent in the
above expression changes branch at the critical value
\begin{equation}
 u_{\rm crit} = \frac{4(2+3\PressC)}{3(1+\PressC)^{2}} < U_{(0)}\,,
\end{equation}
so that $\varphi_{(0)}$ is a continuous function of $u$.

Substituting equation (\ref{phidot_beta0}) into (\ref{match1_constrho})
yields
\begin{equation}
 \A_{(0)} = \frac{9(1+\PressC) (1+3\PressC)^{2}}{16 [\PressC(1+2\PressC)]^{3/2}}
 \arccos \left( \frac{1+\PressC}{1+3\PressC} \right)
 -\frac{41\PressC^{2} + 34\PressC + 9}{8\PressC(1+2\PressC)} \,,
\end{equation}
so the scalar-coupling constraint,
eq.~(\ref{qm_constraint_2_alt}), can be explicitly evaluated,
\begin{eqnarray}\label{bd_constraint2}
 a_{\ssA} &=&
 a_{0} \left(
 \frac{5}{2} - \frac{9}{4s}
 - \frac{9\sqrt{1-2s}
 [27s-14-9\sqrt{1-2s}(2-3s)]}
 {8[s(5-9s+3\sqrt{1-2s})]^{3/2}} \arccos \sqrt{1-2s}
 \right)
 \nonumber
 \\
 &&
 + \mathcal{O}(a_{0}^{3})
 \\
 &=& a_{0}\left(1 - \frac{6}{5}s + \mathcal{O}(s^{2})\right)
 + \mathcal{O}(a_{0}^{3})
 \,.
\end{eqnarray}
This confirms that although $a_\ssA \to a_0$ in the
non-relativistic limit $s \to 0$ --- consistent with
eq.~(\ref{limits1}) --- it is in general depressed relative to
$a_0$ for relativistic systems, even when additional powers of
$a_0^2$ are neglected.

Notice that in the opposite limit we have
\begin{eqnarray}
 \lim_{s \to 4/9} a_{\ssA} &=&
 a_{0} \left( \frac{81}{64}\sqrt{2} \arccos ( 1/3 ) -
 \frac{41}{16} \right) + \mathcal{O}(a_{0}^{3})
 \nonumber
 \\ & \simeq & -0.359 a_{0} + \mathcal{O}(a_{0}^{3}) \,.
\end{eqnarray}
Equation (\ref{bd_constraint2}) is plotted in Figure \ref{Fig3},
showing that for incompressible stars $a_{\ssA}$ passes through
zero, changing sign at $s \sim 0.398$. This is compared in the
same figure to the corresponding curves for neutron stars modeled
by relativistic polytropes, using the equations of state EOS A and
EOS II defined in \cite{damourspsc}. For small $s$ all three
curves agree reasonably well, but differ for large $s$. For large
$s$ the neutron star curves significantly deviate from the
constant-density curve, as might be expected given that
relativistic polytropes have a maximum value $\PressC^{\max} =
\gamma -1$, where $\gamma$ is the polytropic index, while
constant-density stars have no such maximum value for $\PressC$.

Substituting equations (\ref{phi_beta0_final}) and
(\ref{phidot_beta0}) into (\ref{match0_constrho}) yields
\begin{eqnarray}\label{f_beta0}
 \mathcal{F}_{(0)} &=& \frac{9}{4}(1+\PressC)
 - \frac{41\PressC^{2} + 34 \PressC + 9}{8 \PressC
 (1+2\PressC)} \log(1+3\PressC)
 \nonumber \\
 && + \frac{9(1+\PressC)(1+3\PressC
 +3 \PressC^{2})}{4(2+3\PressC)\sqrt{\PressC(1+2\PressC)}}
 \left( \frac{1+\PressC}{\sqrt{\PressC(1+2\PressC)}} \log(1+\PressC) -
 \arccos \left(\frac{1+\PressC}{1+3\PressC} \right)\right)
 \nonumber \\
 && - \frac{9(1+3\PressC)^{2}(1+\PressC)}{16(2+3\PressC)^{3/2}} \biggl\{
 \arccos \left(\frac{1+\PressC}{1+3\PressC}\right) \left[
 \left(\frac{2+3\PressC}{\PressC(1+2\PressC)}\right)^{3/2}
 \log \left( \frac{1+\PressC}{1+3\PressC} \right) \right.
 \nonumber  \\
 && \left.\hphantom{- \frac{9(1+3\PressC)^{2}(1+\PressC)}{16
 (2+3\PressC)^{3/2}} \biggl\{}
 \hphantom{\arccos \left(\frac{1+\PressC}{1+3\PressC}\right) \biggl[}
 + \widetilde{\arctan} \left( \frac{2\sqrt{\PressC(1+
 2\PressC)(2+3\PressC)}}{1-3\PressC^{2}}
 \right)\right]
 \nonumber\\
 && \hphantom{- \frac{9(1+3\PressC)^{2}(1+\PressC)}{16(2+
 3\PressC)^{3/2}} \biggl\{}
 -\Re [ {\rm Li}_{2} (\Lambda_{-}) - {\rm Li}_{2} (|\lambda_{-}|)
 - {\rm Li}_{2} (\Lambda_{+}) + {\rm Li}_{2} (|\lambda_{+}|) ]
 \biggr\} \,,
\end{eqnarray}

where

\begin{equation}
\Lambda_{\pm} = \frac{1+\PressC + 2i \sqrt{\PressC(1+2\PressC)}}
{3(1+\PressC) \pm 2 \sqrt{2+3\PressC}}
\end{equation}

is the value of $\lambda_{\pm}$ when $u=U_{(0)}$, and

\begin{equation}
\widetilde{\arctan}\, X =
\begin{cases}
\arctan \, X - \pi & \text{if } 0 < \PressC < \frac{1}{\sqrt{3}}
\,,
\\
\arctan \, X & \text{if } \PressC > \frac{1}{\sqrt{3}} \,.
\end{cases}
\end{equation}

Equation (\ref{f_beta0}) is plotted in Figure \ref{Fig4}, and is
compared to the corresponding curves for neutron stars. Again, the
curves are close for small $s$, and diverge for large $s$.
$\mathcal{F}_{(0)}(s)$ is positive and increasing on the interval
$0 < s < 0.25$; positive and decreasing when $0.25 < s < 0.36$;
and negative and decreasing for $0.36 < s < \frac49$. The maximum
value, attained at $s \sim 0.25$, is $\mathcal{F}_{(0),{\rm max}}
\sim 0.27$. As $s \to \frac49$, $\mathcal{F}_{(0)}$ tends to
$-\infty$.

\FIGURE[ht]{ \epsfig{file=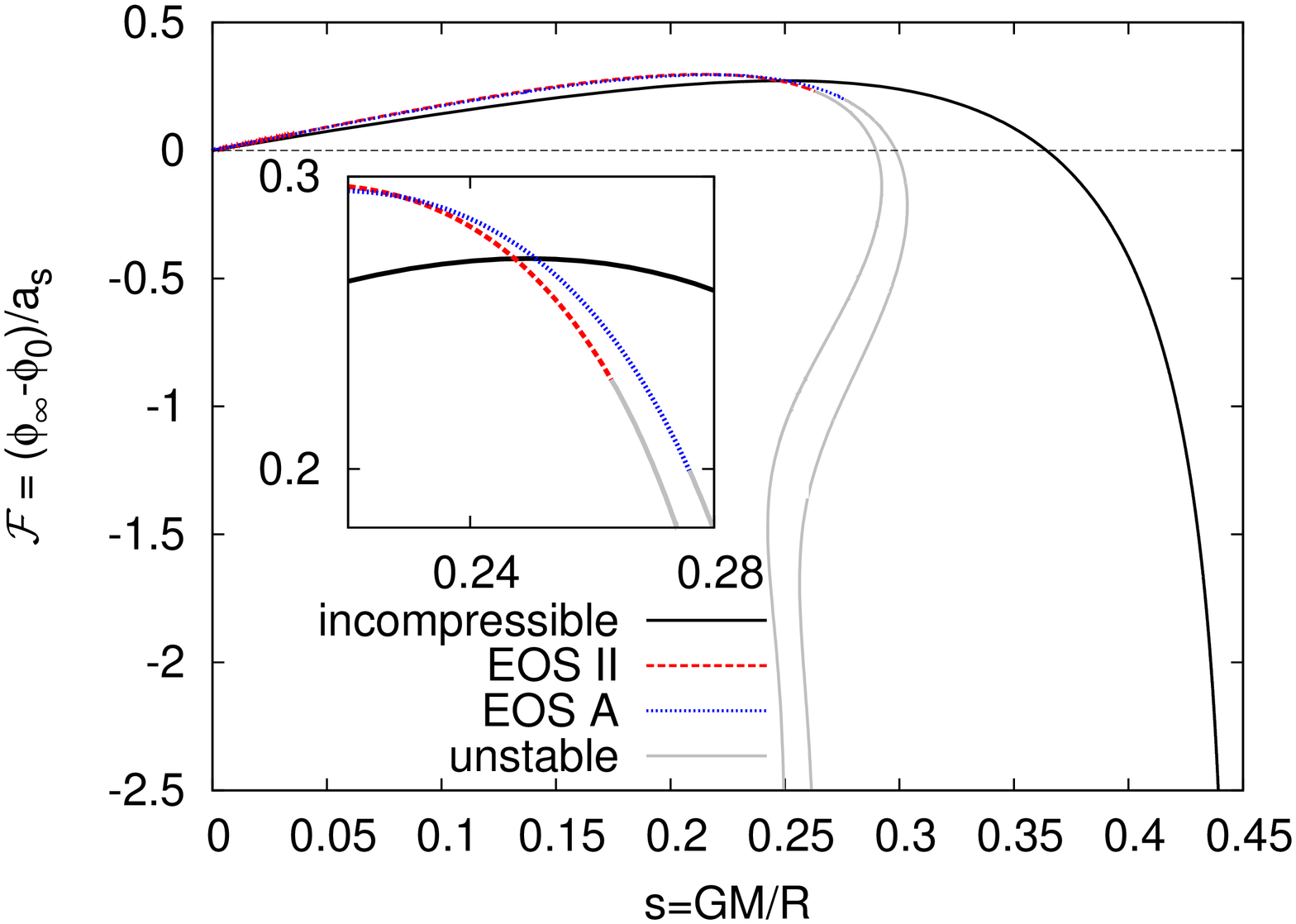,angle=270,width=0.9\hsize}
\caption{$\mathcal{F}=(\phi_{\infty}-\phi_{0})/a_{0}$ vs $s=GM/R$
for various stars in Brans-Dicke theory, in the limit $a_{0} \to
0$.} \label{Fig4} }

\subsubsection*{Solution for $b_{s} \neq 0$}\label{sec:solbn0}

Next suppose $b_{s} \neq 0$. To calculate $\varphi_{(0)}$, change
variables from $(\varphi_{(0)},u)$ to $(\psi, z)$ with
\begin{equation} \label{chvar_bn0}
 \psi = 1+b_{s}\varphi_{(0)} \,,
 \qquad z = \frac12 \left( 1- \sqrt{1- \frac{u}{3}} \right) \,.
\end{equation}
Initial conditions for $\psi(z)$ are then
\begin{equation}\label{heun_ic}
 \psi(0)=1\,,\qquad \frac{
 \exd \psi}{\exd z}\biggr|_{z=0} =
 b_{s}(1-3\PressC)\,.
\end{equation}
Equation (\ref{constrho3}) with $a_0^2=0$ becomes
\begin{equation}\label{heuneq}
 \frac{\exd^{2}\psi}{\exd z^{2}} + \left( \frac{\tilde{\gamma}}{z} +
 \frac{\tilde{\delta}}{z-1} + \frac{\tilde{\epsilon}}{z-\tilde{a}}
 \right) \frac{\exd\psi}{\exd z} + \frac{\tilde{\alpha}
 \tilde{\beta} \, z -\tilde{q}}{z(z-1)(z-\tilde{a})} \psi = 0\,,
\end{equation}
where
\begin{equation} \label{heunpars1}
 \tilde{a} = -\frac{1}{1+3\PressC}\,,\qquad
 \tilde{q} = \frac{3b_{s}}{2} \left( \frac{3\PressC -1}{3\PressC +
 1} \right) \,,
 \qquad
 \tilde{\gamma} = \tilde{\delta} =
 \textstyle{\frac{3}{2}}\,,\qquad \tilde{\epsilon} = 1\,,
\end{equation}
\begin{equation} \label{heunpars2}
 \tilde{\alpha} = \frac{3}{2} \left( 1- \sqrt{1-
 \frac{8b_{s}}{3}} \right) \,,\qquad
 \tilde{\beta} = \frac{3}{2} \left( 1+
 \sqrt{1- \frac{8b_{s}}{3}} \right)\,,
\end{equation}
and
\begin{equation}
 \tilde{\gamma} + \tilde{\delta} + \tilde{\epsilon}
 = \tilde{\alpha} + \tilde{\beta} + 1\,.
\end{equation}

Equation (\ref{heuneq}) is called Heun's equation \cite{heun}, and
is a linear second-order differential equation with singularities
at $z=0,1,\tilde{a},\infty$. It is a natural generalization of the
hypergeometric equation to the situation having four regular
singular points. The solution which satisfies initial conditions
(\ref{heun_ic}) is the local Frobenius solution about $z=0$ with
exponent $0$, and is given by the power series
\begin{equation}\label{heunseries}
 \psi(z) = {\rm HeunG}(\tilde{a},\tilde{q};
 \tilde{\alpha},\tilde{\beta},
 \tilde{\gamma},\tilde{\delta};z) =
 \sum_{r=0}^{\infty} c_{r}z^{r}\,,
\end{equation}
where the first two coefficients are given by
\begin{equation} \label{recrel1}
 c_{0}=1\,,\qquad
 c_{1}=\frac{\tilde{q}}{\tilde{a}\tilde{\gamma}}\,,
\end{equation}
and the higher coefficients are found by solving the three-term
recurrence relation
\begin{eqnarray} \label{recrel2}
 && (r-1+\tilde{\alpha})(r-1+\tilde{\beta})
 \; c_{r-1} \nonumber\\
 && \qquad\qquad - [r(r-1+\tilde{\gamma})(1+\tilde{a})
 +r(\tilde{a}\tilde{\delta}
 +\tilde{\epsilon})+\tilde{q}] \; c_{r}  \nonumber\\
 && \qquad\qquad\qquad\qquad
 + \tilde{a}(r+1)(r+\tilde{\gamma}) \; c_{r+1}=0\,.
\end{eqnarray}

In terms of $b_{s}$ and $\PressC$, the recursion relation for the
coefficients $c_{r}$ become
\begin{equation}
 c_{0} = 1\,,\qquad
 c_{1} = b_{s}(1-3\PressC)\,,
\end{equation}
\begin{eqnarray}
 \label{recrel_new}
 && 2(1+3\PressC)(r^{2}+r-2+6b_{s}) \; c_{r-1}
 \nonumber\\
 && \qquad\qquad -\left[ r(6\PressC r + 9\PressC
 -1)+3b_{s}(3\PressC-1)\right] \; c_{r}
 \nonumber\\
 && \qquad\qquad\qquad\qquad
 -(r+1)(2r+3) \; c_{r+1}=0\,.
\end{eqnarray}
This implies the coefficients $c_{r}$ of the power series
(\ref{heunseries}) can be written explicitly as a polynomial of
degree $r$ in $b_s$,
\begin{equation}
 c_{r} = \sum_{i=0}^{r} a_{i}^{(r)}b_{s}^{i}\,,
\end{equation}
where $a_{i}^{(r)}$ is itself a polynomial in $\PressC$ of degree
$r$.

The solutions for $a_i^r$ and $c_r$ are found explicitly in the
Appendix, where it is also shown that the coefficient of the
largest power of $b_s$ has a particularly simple form:
\begin{equation}\label{coef11}
 a_{r}^{(r)} = \frac{[6(1-3\PressC)]^{r}}{(2r+1)!}\,.
\end{equation}
Because $b_s$ is relatively poorly constrained, it can be larger
than unity so far as phenomenology is concerned. In this case
eq.~\pref{coef11} can be used to obtain an approximation for
$\varphi(r)$ for large $b_s$. This gives (see Appendix for
details)
\begin{eqnarray}
 \label{g0} \A_{(0)} &=& \frac{1+\PressC}{\PressC} \left\{
 \frac{1}{2b_{s}} \left( \cosh \sqrt{T} - \frac{\sinh
 \sqrt{T}}{\sqrt{T}}\right) + \sum_{k=1}^{\infty}\sum_{j=0}^{2k-1}
 \frac{P_{k,j}(\PressC)f_{k,j+1}(T)}
 {b_{s}^{k+1}[6(1-3\PressC)]^{2k}} \right\} \,,
 \\
 \label{f0} \mathcal{F}_{(0)} &=& \frac{1}{b_{s}} \left( \left[ 1 +
 \frac{1+\PressC}{2\PressC}L \right] \frac{\sinh\sqrt{T}}{\sqrt{T}}
 - \frac{1+\PressC}{2\PressC}L \cosh \sqrt{T} - 1\right)
 \nonumber\\
 && + \sum_{k=1}^{\infty}\sum_{j=0}^{2k-1} \left( f_{k,j}(T) -
 \frac{1+\PressC}{\PressC}Lf_{k,j+1}(T)\right)
 \frac{P_{k,j}(\PressC)}{b_{s}^{k+1}[6(1-3\PressC)]^{2k}} \,,
\end{eqnarray}
where $T=6b_{s}\PressC(1-3\PressC)/(1+3\PressC)$ and
$L=\log(1+\PressC)-\log(1+3\PressC)$. If $b_{s}>0$, then $T \leq
(6-4\sqrt{2})b_{s} \sim 0.34 b_{s}$. If $b_{s}<0$, then $T \geq
-(6-4\sqrt{2})|b_{s}| \sim -0.34 |b_{s}|$. The above expressions
(\ref{g0}) and (\ref{f0}) can be used to calculate the second
constraint (\ref{constraint2_fo}).

\subsubsection*{Compactness vs central density}\label{sec:compactness}

\FIGURE[ht]{ \epsfig{file=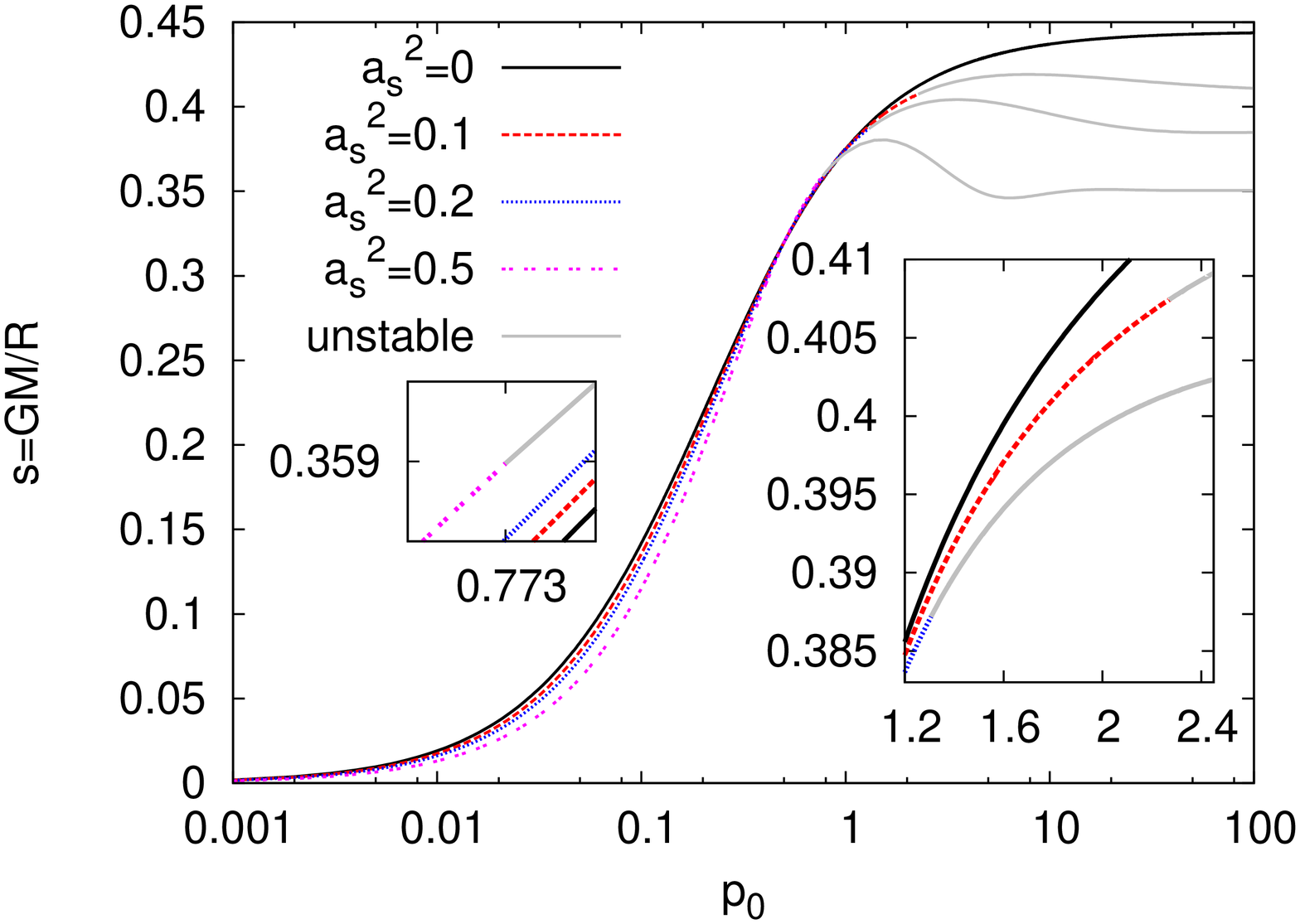,angle=270,width=0.9\hsize}
\caption{The compactness $s=GM/R$ plotted versus $\PressC =
P_{0}/\rho_{0}$ for constant-density stars in Brans-Dicke theory,
for various values of $a_0^2$.} \label{Fig5} }

Equation (\ref{match2_constrho}) describes how the compactness
depends on $\PressC$ in GR. In order to find how scalar-matter
couplings modify this behaviour, it is necessary to solve the
equations of stellar structure to first order in $a_0^2$, and
calculate $s_{(1)}$.

Figure \ref{Fig5} plots the compactness vs $\PressC$ in
Brans-Dicke theory, for various values of the Brans-Dicke coupling
$a_{0} = a_s$. Notice that the compactness eventually stops
growing with $\PressC$, approaching instead an asymptotic value as
$\PressC \to \infty$. In
GR, this asymptotic value is $GM/R=4/9$, which is the maximum allowed
by Buchdahl's theorem. As $a_{0}$ increases, this asymptotic value
decreases. This is consistent
with the results of \cite{hillheintz}.

\subsection{Perturbative solutions: next-to-leading corrections}
\label{sec:chi_corrections}

In this section, $\mathcal{O}(a_0^2)$ corrections are calculated.
The defining equation for $U$ is
$\Press (U)=0$. Expanding it in powers of $a_0^2$ yields
\begin{equation}
 U_{(1)} = \frac{12(1+\PressC)^{2}}{(1+3\PressC)^{2}} \;
 \Press_{(1)}(U_{(0)}) \,.
\end{equation}
Solving equations (\ref{constrho1}) -- (\ref{constrho3})
perturbatively in $a_0^2$ yields
\begin{eqnarray}
 \label{mu_corrn} \mu_{(1)}(u) &=& \frac{1}{\sqrt{u}}\int_{0}^{u}
 \exd \hat u \sqrt{\hat u} \left[ \varphi_{(0)} \left(
 1+\frac{b_{s} \varphi_{(0)}}{2} \right)
 +\hat u(1- \hat u/3)(\dot{\varphi}_{(0)})^{2}\right] \,, \\
 \label{pi_corrn} \Press_{1}(u) &=&
 \frac{2(1+\PressC)\varphi_{(0)}
 \left( 1+ {b_{s}\varphi_{(0)}}/{2} \right)}
 {(1+3\PressC)\sqrt{1-u/3}-3(1+\PressC)} \nonumber \\
 &&+ \frac{2(1+\PressC)\mu_{(1)}(u)}{\sqrt{1-u/3}} \cdot
 \frac{(1+\PressC)(3-2u)\sqrt{1-u/3}-(1+3\PressC)}
 {((1+3\PressC)\sqrt{1-u/3}-3(1+\PressC))^{2}}
 \nonumber \\
 &&+ \frac{2(1+\PressC)\sqrt{1-u/3}}{6((1+3\PressC)
 \sqrt{1-u/3}-3(1+\PressC))^{2}}
 \int_{0}^{u} \frac{\exd \hat u}{(1- \hat u/3)^{3/2}} \;
 J(\hat u) \,,
\end{eqnarray}
where the function $J(u)$ appearing in equation (\ref{pi_corrn})
is given by
\begin{eqnarray}
 J(u) &=& 2 \Bigl[ 6(1+3\PressC) \sqrt{1-u/3}
 -(1+\PressC)(2u^{2}-9u+18) \Bigr]
 u(1-u/3)(\dot{\varphi}_{(0)})^{2}
 \nonumber \\
 &&- (1+\PressC) (4u^{2}-18u+9)
 \varphi_{(0)} \left( 1 + \frac{b_{s}
 \varphi_{(0)}}{2} \right) \,.
\end{eqnarray}

The perturbation to the scalar profile is similarly
\be
 \varphi_{(1)} = \Phi_{1}\psi+ \Phi_{2} \tilde{\psi} \,.
 \label{phi_corrn}
\ee
where the functions $\psi$ and $\tilde{\psi}$ are local Frobenius
solutions of equation (\ref{heuneq}) (with parameters
(\ref{heunpars1}) -- (\ref{heunpars2})) about $z=0$ with exponents
$0$ and $-\frac{1}{2}$, respectively. They are given by
\begin{eqnarray}
 \psi
 %&=&
 %1 + b_{s} \varphi_{(0)} \nonumber \\
 &=& {\rm HeunG} \biggl( \frac{-1}{1+3\PressC} , \frac{3}{2} b_{s}
 \cdot \frac{3\PressC-1}{3\PressC+1} ; \nonumber \\
 &&\hphantom{{\rm HeunG} \biggl(} \frac{3}{2}(1-\sqrt{1-8b_{s}/3}),
 \frac{3}{2}(1+\sqrt{1-8b_{s}/3}), \frac{3}{2}, \frac{3}{2};
 z\biggr) \,, \\
 \tilde{\psi} &=& \frac{1}{\sqrt{z}} \; {\rm HeunG} \biggl(
 \frac{-1}{1+3\PressC},
 \frac{6b_{s}(3\PressC-1)+1-6\PressC}{4(1+3\PressC)}; \nonumber
 \\
 &&\hphantom{ \frac{1}{\sqrt{z}} {\rm HeunG} \biggl(}
 1+\frac{3}{2}\sqrt{1-8b_{s}/3}, 1-\frac{3}{2} \sqrt{1-8b_{s}/3} ,
 \frac{1}{2}, \frac{3}{2}; z \biggr) \,,
\end{eqnarray}
where $z=(1-\sqrt{1-u/3})/2$. The coefficients $\Phi_{1}$ and
$\Phi_{2}$ are given by
\begin{eqnarray}
 \Phi_{1} &=& 288\int_{0}^{z}\tilde{\psi} (F \Press _{(1)} +
 G\mu_{(1)} + H) (1-2z)^{2} [z(1-z)]^{3/2}
 [1+(1+3\PressC)z] \, \exd z \,, \\
 \Phi_{2} &=& -288\int_{0}^{z} \psi (F \Press _{(1)} + G\mu_{(1)} +
 H) (1-2z)^{2} [z(1-z)]^{3/2} [1+(1+3\PressC)z] \, \exd z \,,
\end{eqnarray}
where
\begin{eqnarray}
 F &=& -\frac {1+b_{s} \varphi_{(0)} + 8z(1-z)\dot{\varphi}_{(0)}}
 {32z(1-z)(1-2z)^{2}} \,, \\
 G &=& \frac{(4(1+3\PressC)z+1-3\PressC)(1+b_{s}\varphi_{(0)})}
 {48z(1-z)(1-2z)^{4}(1+(1+3\PressC)z)} \nonumber \\
 && \qquad \qquad -
 \frac{[12(1+3\PressC)z^{3}-6(1+7\PressC)z^{2}
 +(9\PressC-5)z+1]\dot{\varphi}_{(0)}}
 {12z(1-z)(1-2z)^{4}(1+(1+3\PressC)z)} \,, \\
 H &=& \frac{(4(1+3\PressC)z + 1-3\PressC)
 \varphi_{(0)}(1+b_{s}\varphi_{(0)})(1+b_{s}\varphi_{(0)}/2)}
 {24z(1-z)(1-2z)^{2}(1+(1+3\PressC)z)} \nonumber \\
 && \qquad \qquad + \frac{(2(1+3\PressC)z+1-\PressC)
 \varphi_{(0)}(1+b_{s}\varphi_{(0)}/2)\dot{\varphi}_{(0)}}
 {(1-2z)^{2}(1+(1+3\PressC)z)} \,.
\end{eqnarray}

\FIGURE[ht]{
\epsfig{file=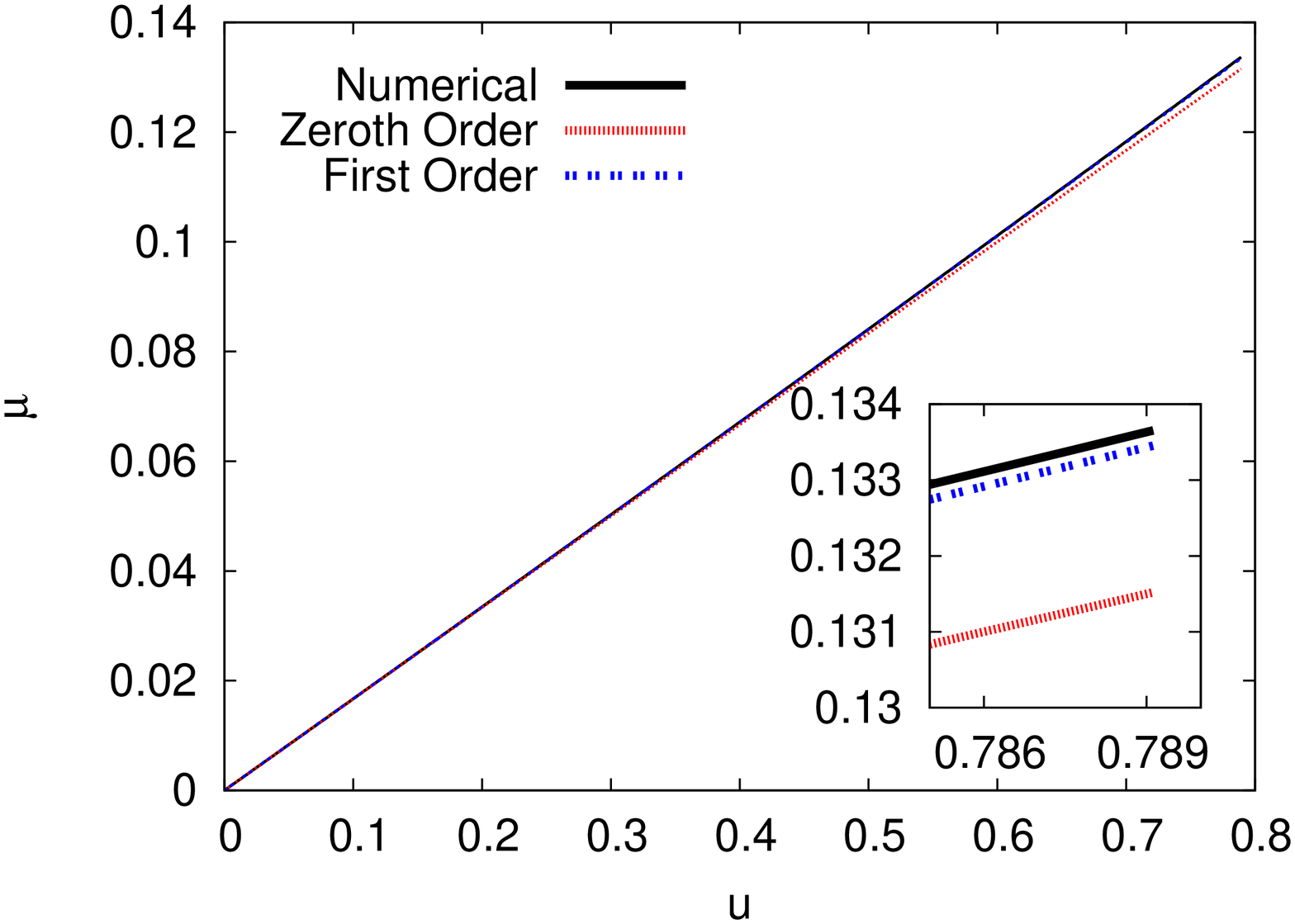,angle=270,width=0.4\hsize}
\epsfig{file=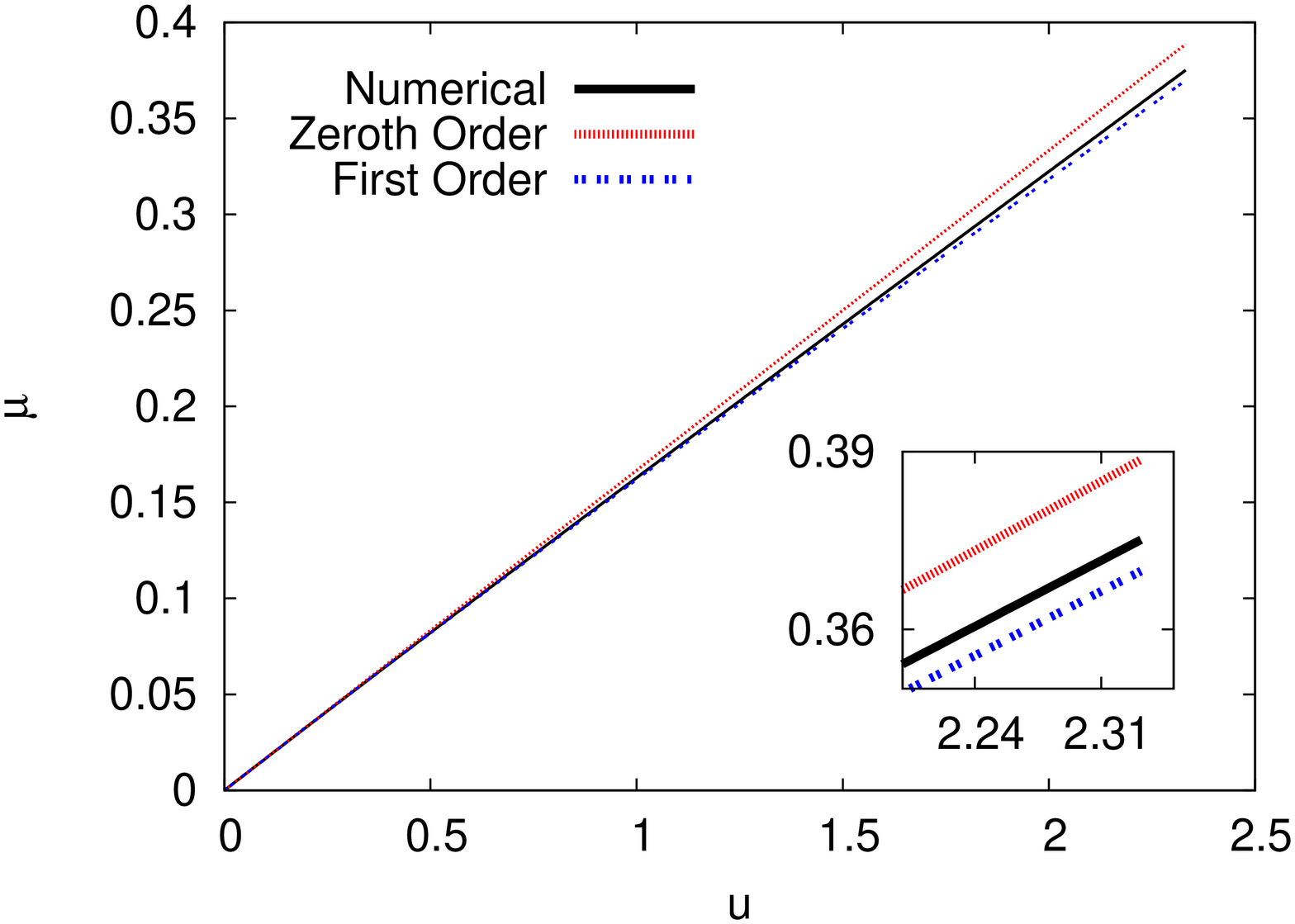,angle=270,width=0.4\hsize}
\\
\epsfig{file=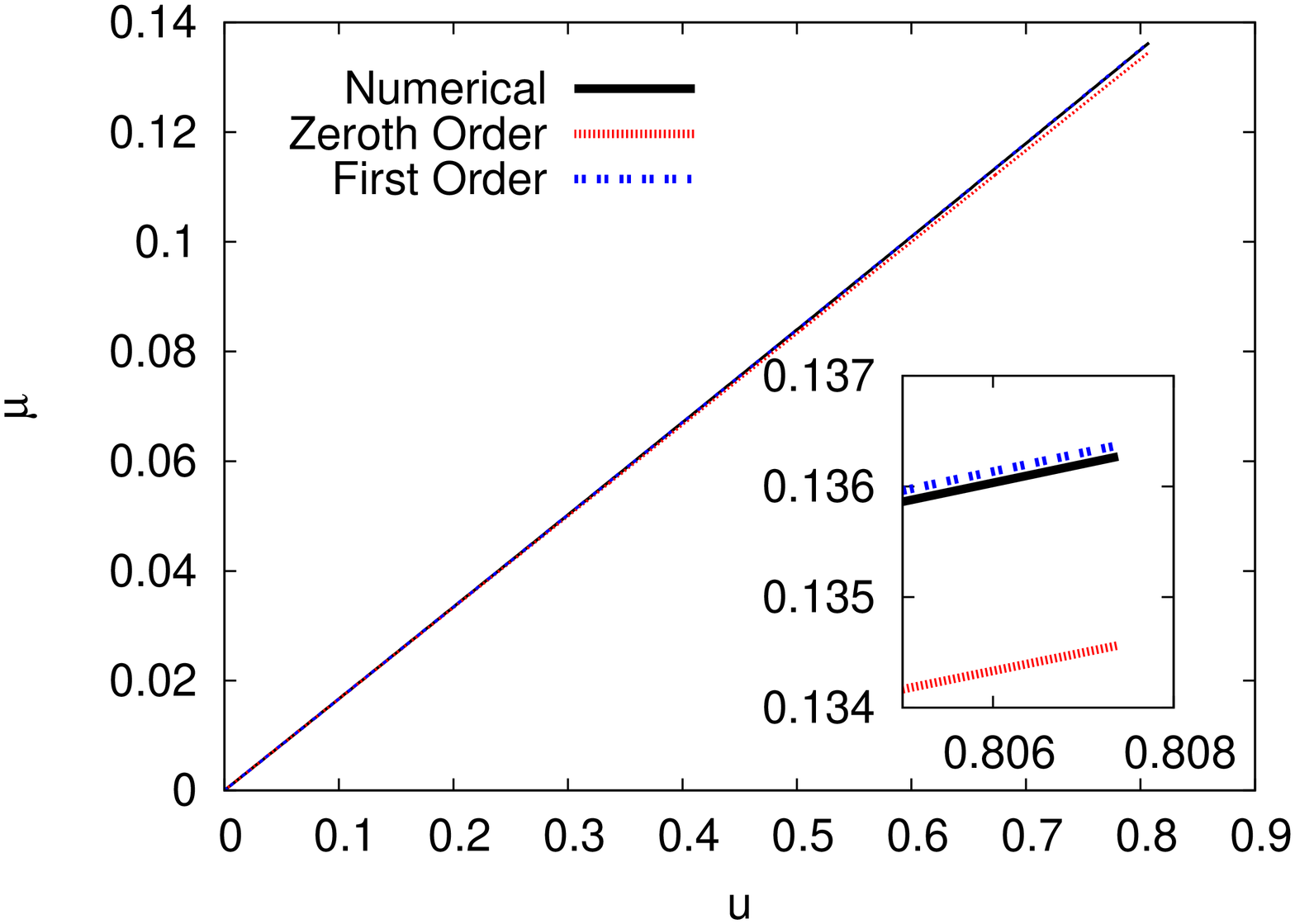,angle=270,width=0.4\hsize}
\epsfig{file=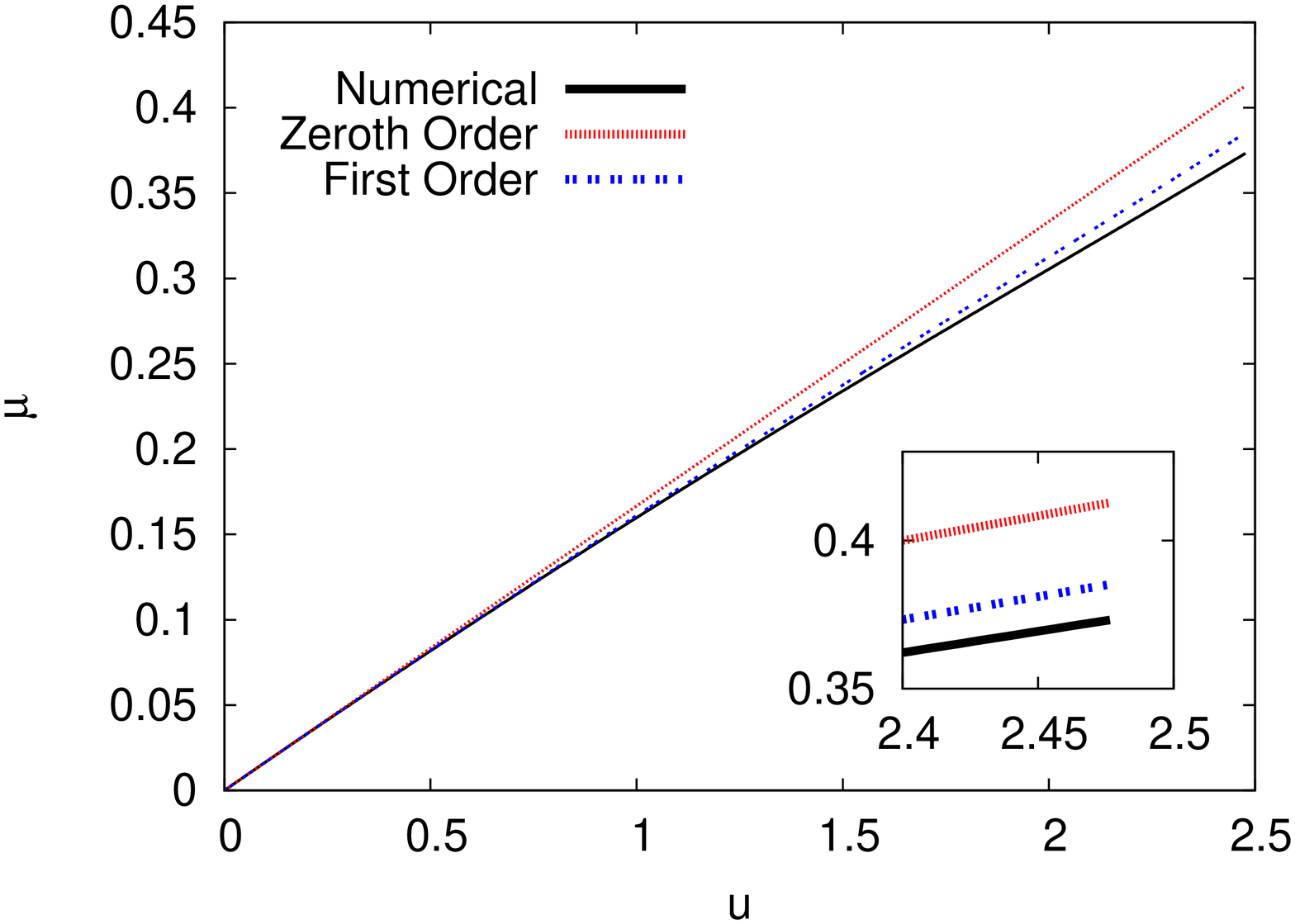,angle=270,width=0.4\hsize}
\caption{Comparison of $\mu$ vs $u$ calculated perturbatively and
numerically for an incompressible star in quasi-Brans/Dicke theory
with $a_{0}^{2}=0.1$ and $b_{s}=4$, $\PressC=0.1$ (top left); $b_s
= 4$, $\PressC=1$ (top right); $b_{s}=-4$, $\PressC=0.1$ (bottom
left); and $b_s = -4$, $\PressC=1$ (bottom right). All curves
terminate at the stellar exterior, $u=U$.} \label{Fig6}
\label{Fig7} }

The first-order corrections to the external parameters are given by
\begin{eqnarray}
 \mathcal{F}_{(1)} &=& \varphi_{(1)} - 12 e^{2L} L
 \dot{\varphi}_{(1)} -
 \frac{3(1+\PressC)^{2}}{2\PressC(1+2\PressC)}
 L(1+b_{s}\varphi_{(0)})\Press _{(1)} \nonumber
 \\
 &&+36e^{2L} \left(
 1+\frac{1+4\PressC+5\PressC^{2}}{2\PressC(1+2\PressC)}L \right)
 \left( \Press _{(1)}\dot{\varphi}_{(0)} + 48
 e^{2L}\frac{\PressC(1+2\PressC)}{(1+3\PressC)^{2}}
 (\dot{\varphi}_{(0)})^{3} \right) \nonumber
 \\
 && +
 12\left(1+\frac{(1+3\PressC)^{2}}{2\PressC(1+2\PressC)}L\right)
 \dot{\varphi}_{(0)}\mu_{(1)} \,,
 \\
 \A_{(1)} &=& 12e^{2L} \dot{\varphi}_{(1)} +
 \frac{3(1+\PressC)^{2}}{2\PressC(1+2\PressC)}
 (1+b_{s}\varphi_{(0)})\Press _{(1)} - 6
 \frac{(1+3\PressC)^{2}}{\PressC(1+2\PressC)}
 \dot{\varphi}_{(0)}\mu_{(1)} \nonumber
 \\
 && - 1728 e^{4L}\frac{\PressC(1+2\PressC)}{(1+3\PressC)^{2}}
 (\dot{\varphi}_{(0)})^{3} - 18 e^{2L}
 \frac{(5\PressC^{2}+4\PressC+1)}{\PressC(1+2\PressC)}
 \dot{\varphi}_{(0)}\Press _{(1)} \,,
 \\
 s_{(1)} &=& \mu_{(1)} +2e^{2L}\Press _{(1)} -144 L
 \frac{\PressC(1+2\PressC)}{(1+3\PressC)^{2}}e^{4L}
 (\dot{\varphi}_{(0)})^{2} \,,
 \\
 \mathcal{M}_{(1)} &=&
 \frac{24(1+\PressC)\sqrt{3\PressC(1+2\PressC)}} {(1+3\PressC)^{2}}
 \Press _{(1)} +
 \frac{12\sqrt{3\PressC(1+2\PressC)}}{1+\PressC}\mu_{(1)} \nonumber
 \\
 && -3 \int_{0}^{U_{(0)}} du \sqrt{\frac{u}{1-u/3}}
 [\varphi_{(0)}(1+b_{s}\varphi_{(0)}/2) +
 2u(1-u/3)(\dot{\varphi}_{(0)})^{2}] \,,
\end{eqnarray}
where $L=\log(1+\PressC)-\log(1+3\PressC)$, and the profiles are
all to be evaluated at $U_{(0)}$.

If $b_{s} \neq 0$, then the relation
$a_{\ssA} = \partial \log M / \partial \phi_{\infty}$ \cite{damourrev}
can be used to simplify $\M_{(1)}$:
\begin{eqnarray}
 \mathcal{M}_{(1)} &=& \frac{9\sqrt{3}}{2b_{s}} \arccos \left(
 \frac{1+\PressC}{1+3\PressC}\right)
 - \frac{\sqrt{3\PressC(1+2\PressC)}(41\PressC^{2}+34\PressC+9)}
 {b_{s}(1+\PressC)(1+3\PressC)^{2}}
 \nonumber
 \\
 && - \frac{96(1+\PressC)\sqrt{3\PressC^{3}(1+2\PressC)^{3}}}
 {b_{s}(1+3\PressC)^{4}}\; \dot{\varphi}_{(0)}(1+b_{s}\varphi_{(0)})
 + \frac{8\sqrt{3\PressC(1+2\PressC)}}{1+\PressC} \;  \mu_{(1)}
 \nonumber
 \\
 && +
 \frac{24(1+\PressC)\sqrt{3\PressC(1+2\PressC)}}{(1+
 3\PressC)^{2}} \; \Press_{(1)} \,.
\end{eqnarray}

\FIGURE[ht]{
\epsfig{file=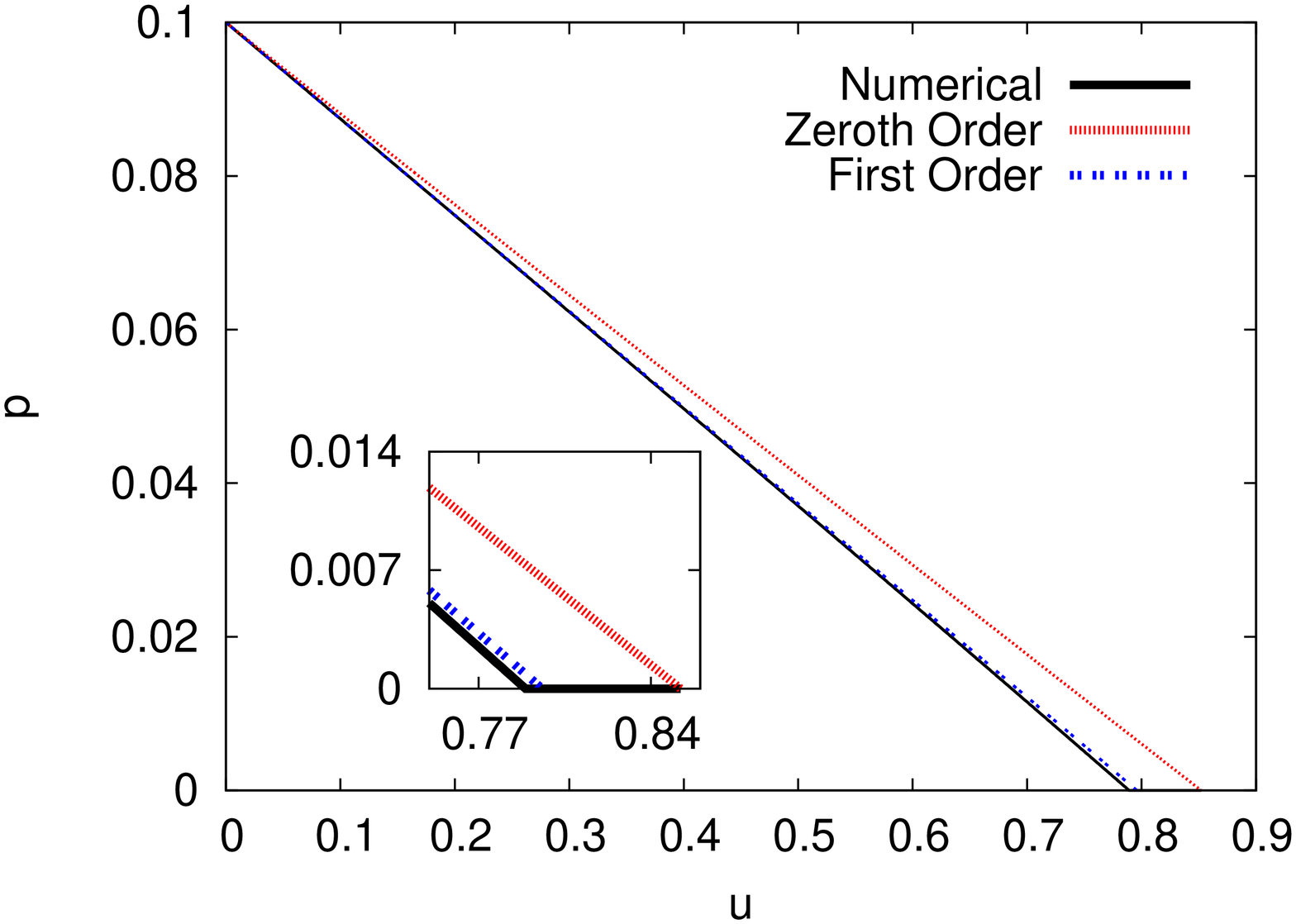,angle=270,width=0.4\hsize}
\epsfig{file=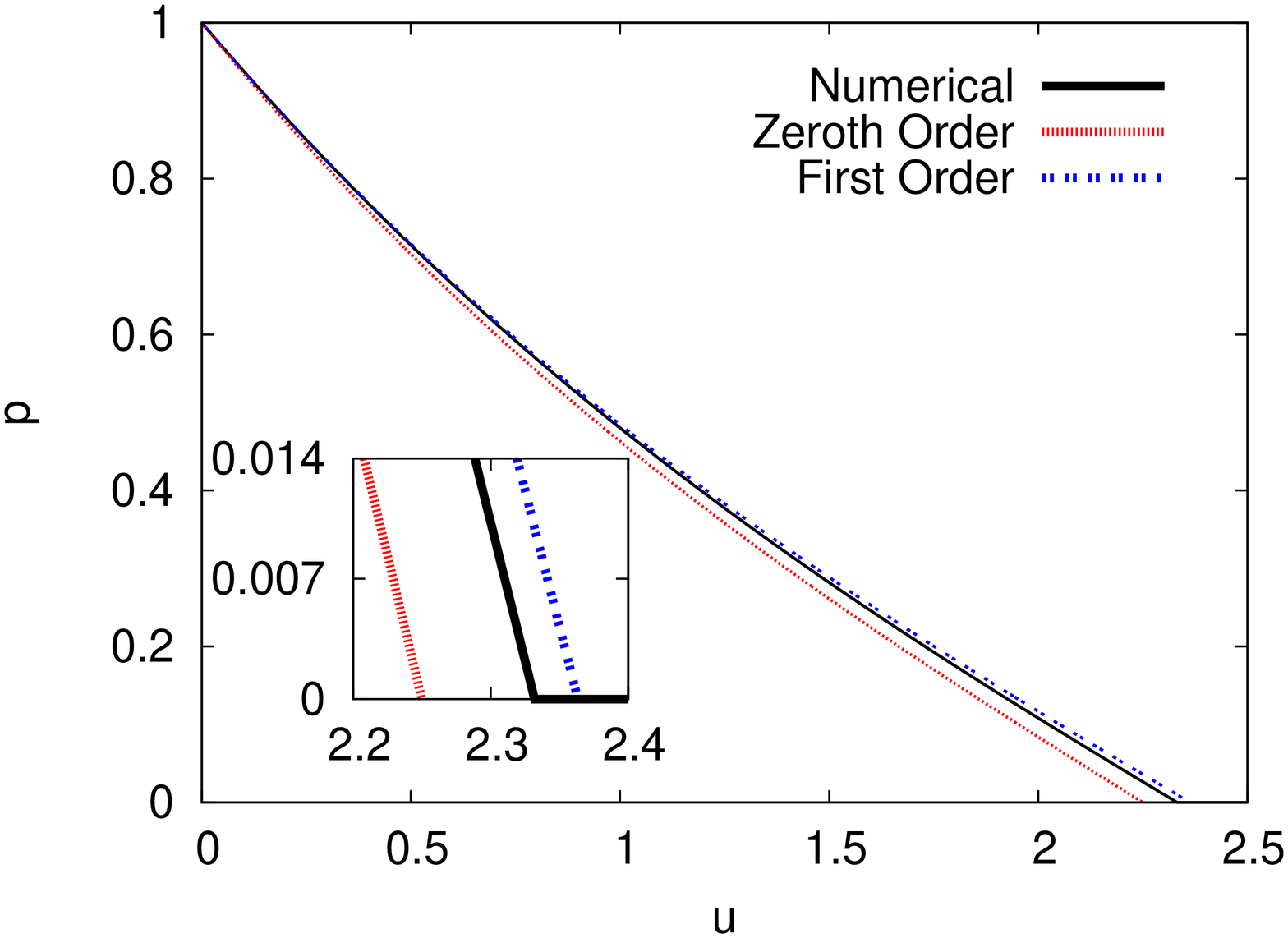,angle=270,width=0.4\hsize}
\epsfig{file=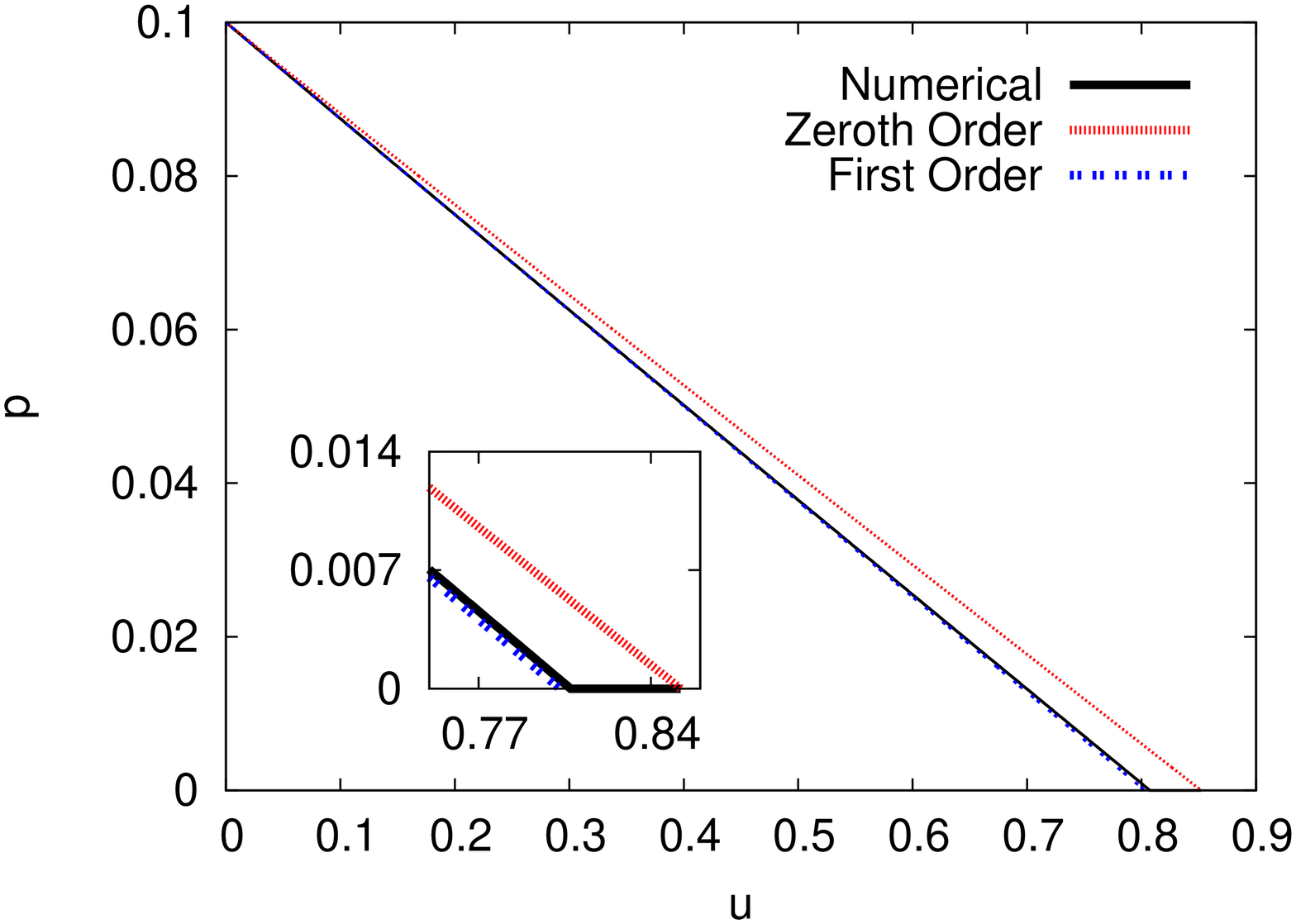,angle=270,width=0.4\hsize}
\epsfig{file=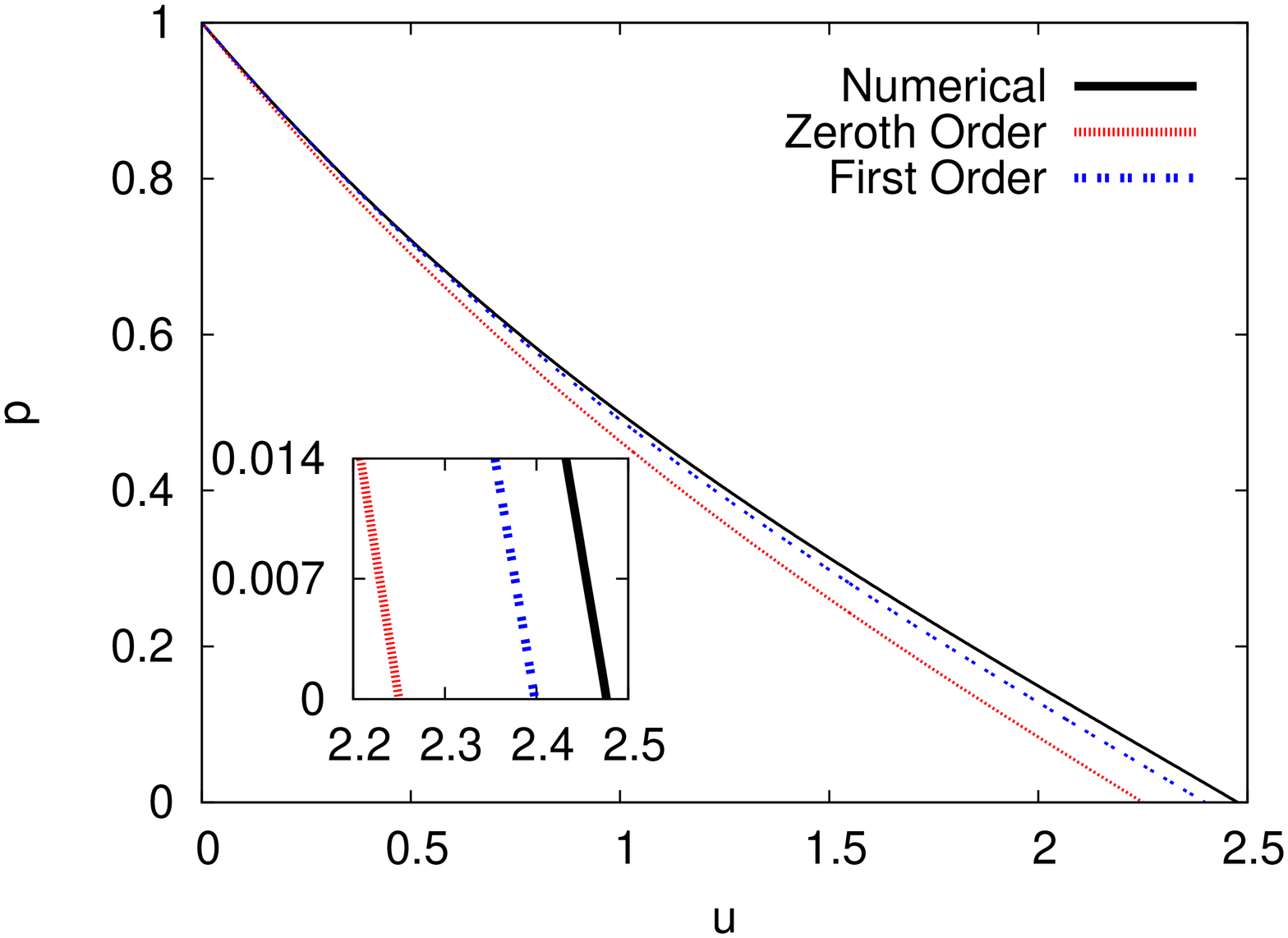,angle=270,width=0.4\hsize}
\caption{Comparison of $\Press$ vs $u$ calculated perturbatively
and numerically for an incompressible star in quasi-Brans/Dicke
theory with $a_{0}^{2}=0.1$ and $b_{s}=4$, $\PressC=0.1$ (top
left); $b_s = 4$, $\PressC=1$ (top right); $b_{s}=-4$,
$\PressC=0.1$ (bottom left); and $b_s = -4$, $\PressC=1$ (bottom
right). All curves terminate at the stellar exterior, $u=U$.}
\label{Fig8} \label{Fig9} }

\subsection{Comparing perturbative solutions with numerics}
\label{sec:num_an_cmp}

Part of the utility of analyzing the incompressible star in such
detail is that such explicit expressions for the perturbative
solutions allow a detailed comparison with direct numerical
integrations. This helps indicate the domain of validity of the
perturbative results.

First, we look at the profiles for the physical variables
$\mu(u)$, $\Press(u)$, and $\varphi(u)$ across the interior of the
star. The quantities $\mu$, $\Press$ and $\varphi$ are
respectively plotted versus $u$ in Figures \pref{Fig6},
\pref{Fig8} and \pref{Fig10}, for $a_{0}^2 = 0.1$ and several
choices for $b_{s}$, and $\PressC$. The line labelled ``Zeroth
Order'' plots the zeroeth-order result, {\em e.g.} $\mu_{(0)}$,
while the line labelled ``First Order'' includes also the first
correction, {\em e.g.} $\mu_{(0)} + a_{0}^{2}\mu_{(1)}$. Notice
that the curves all lie close to one other for small $u$, but
begin to separate at the stellar exterior, $u \to U$. Furthermore,
the separation is largest for the more relativistic stars, for
which $\PressC$ is larger. However in all cases displayed the
perturbative results capture the full numerics quite well
throughout the entire star, with the strongest deviations
happening for $\varphi(u)$ when $b_s < 0$.

\FIGURE[ht]{
\epsfig{file=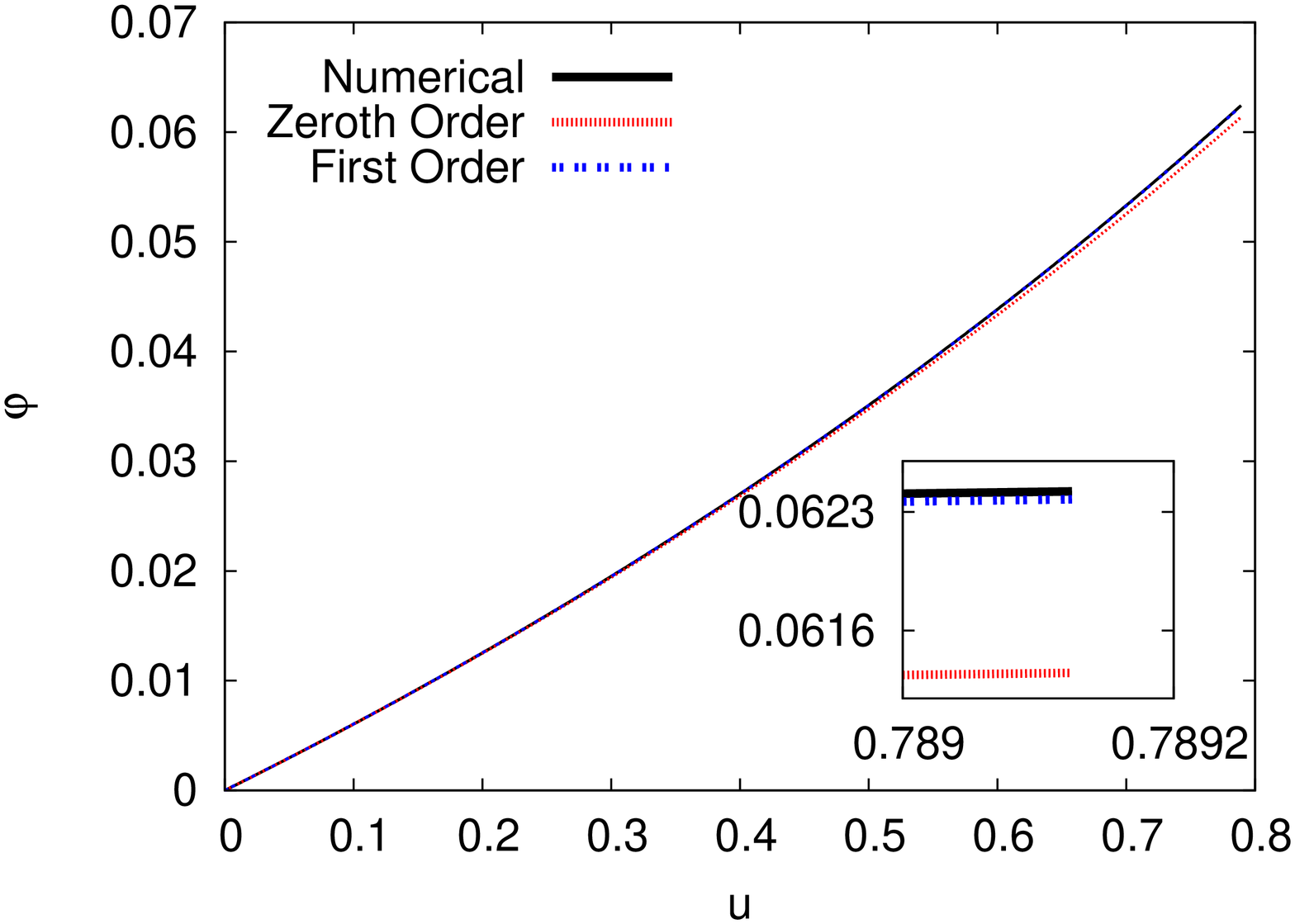,angle=270,width=0.4\hsize}
\epsfig{file=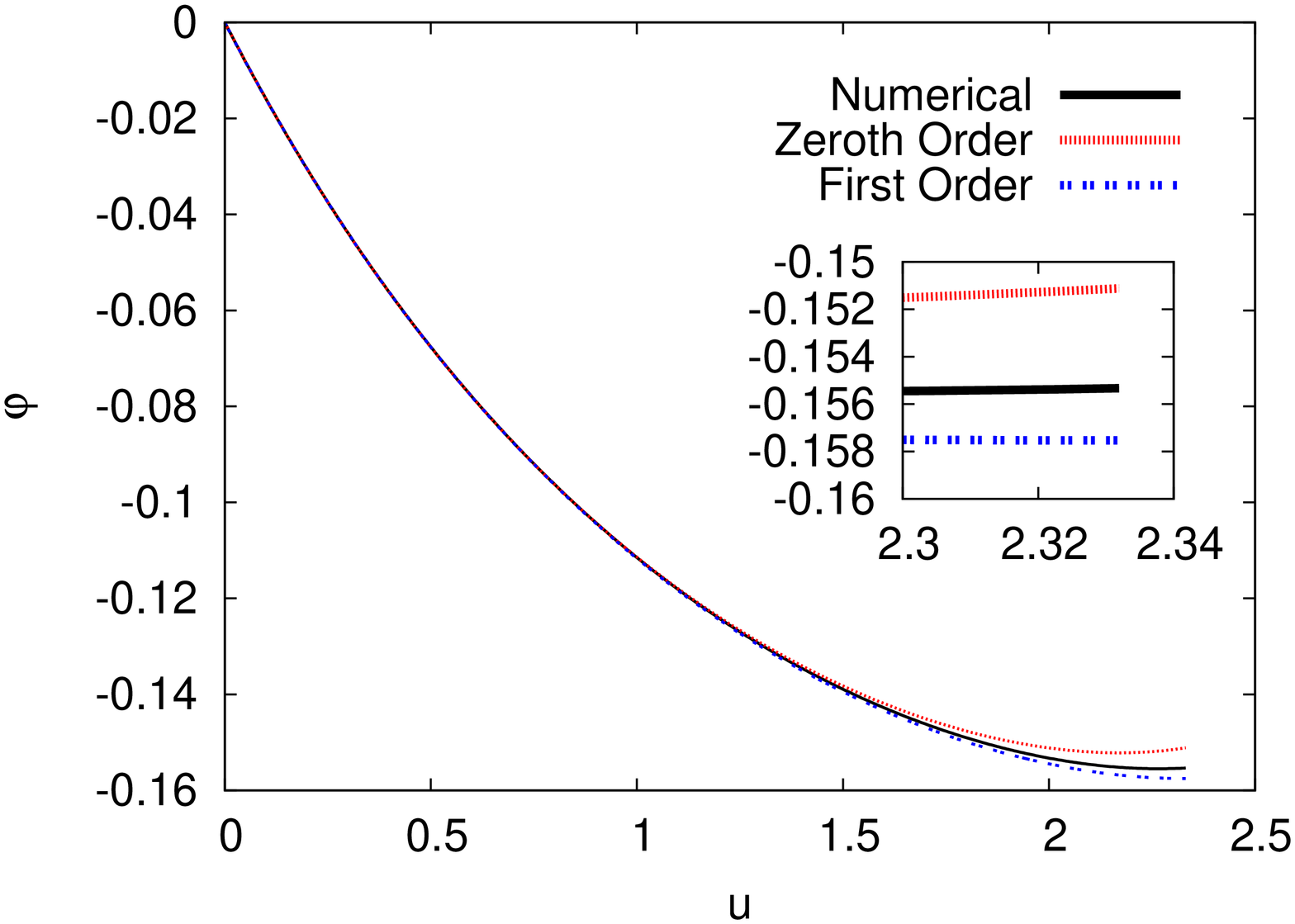,angle=270,width=0.4\hsize}
\epsfig{file=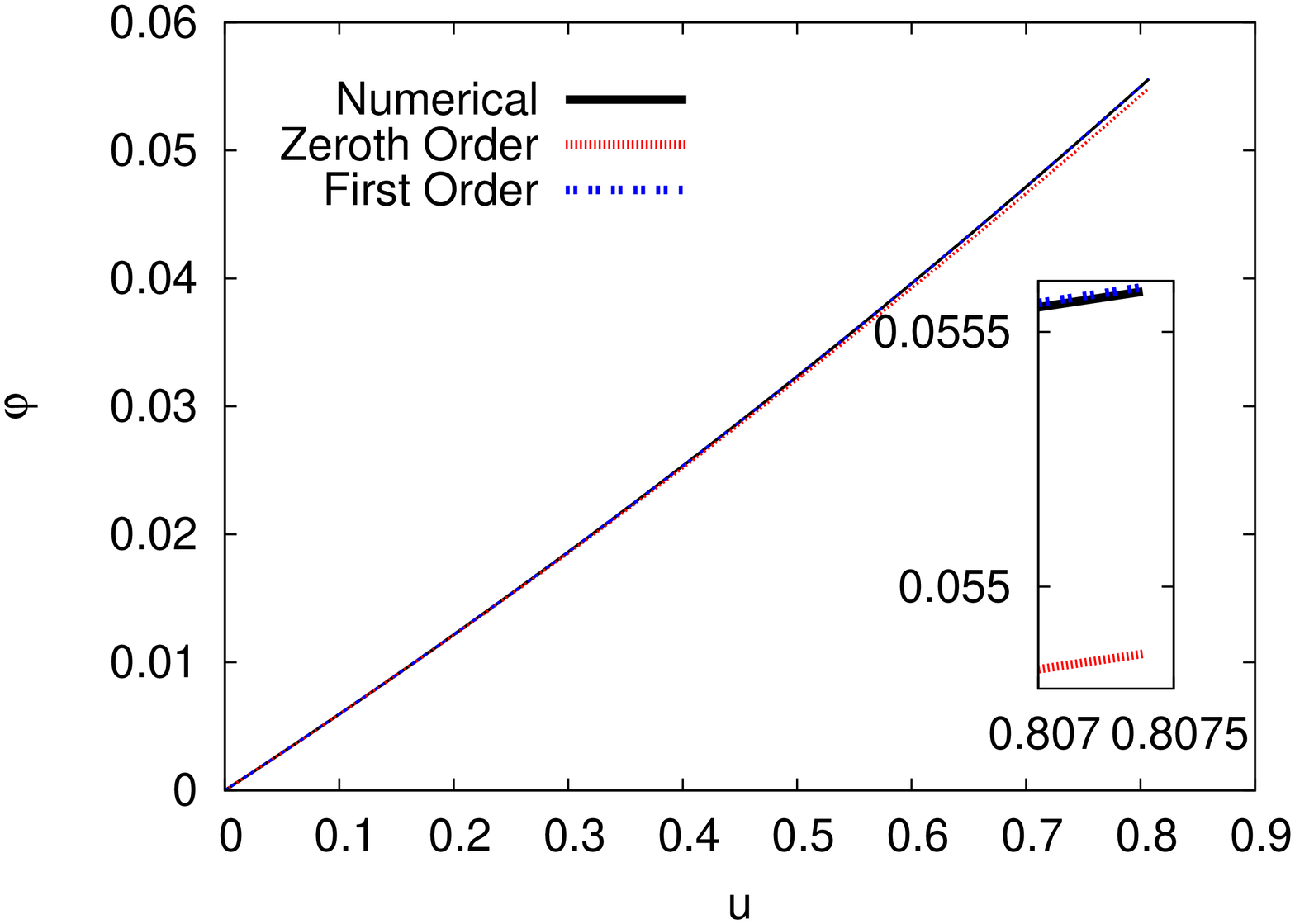,angle=270,width=0.4\hsize}
\epsfig{file=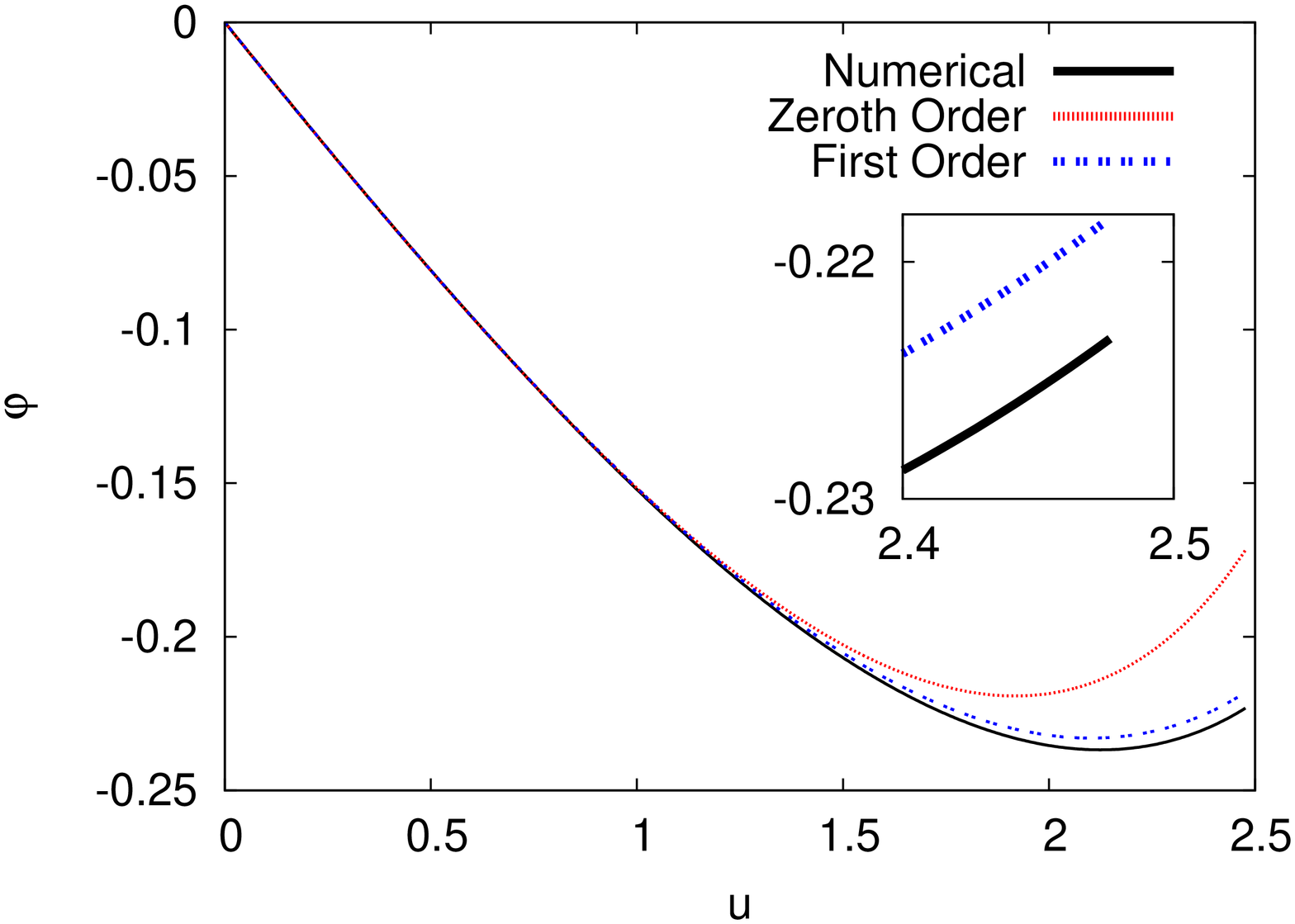,angle=270,width=0.4\hsize}
\caption{Comparison of $\varphi$ vs $u$ calculated perturbatively
and numerically for an incompressible star in quasi-Brans/Dicke
theory with $a_{0}^{2}=0.1$ and $b_{s}=4$, $\PressC=0.1$ (top
left); $b_s = 4$, $\PressC=1$ (top right); $b_{s}=-4$,
$\PressC=0.1$ (bottom left); and $b_s = -4$, $\PressC=1$ (bottom
right). All curves terminate at the stellar exterior, $u=U$.}
\label{Fig10} \label{Fig11}}

Of more practical interest is a similar comparison of the accuracy
of the perturbative expressions for plots that directly relate
observable quantities to one another, such as plots of $a_\ssA$ vs
$s$. Examples of these are given in figures \pref{Fig12}, which
give $s$, $\A = a_\ssA/a_0$, $\F = (\phi_\infty - \phi_0)/a_0$ and
$\M$ as functions of the central pressure, $\PressC = P_0/\rho_0$
for the special case of Brans-Dicke theory ($b_s = 0$) with $a_0^2
= a_s^2 = 0.1$. These again show good agreement between
perturbative and numerical calculations, with the biggest
deviations arising in the most relativistic settings (largest
$\PressC$).

\FIGURE[ht]{
\epsfig{file=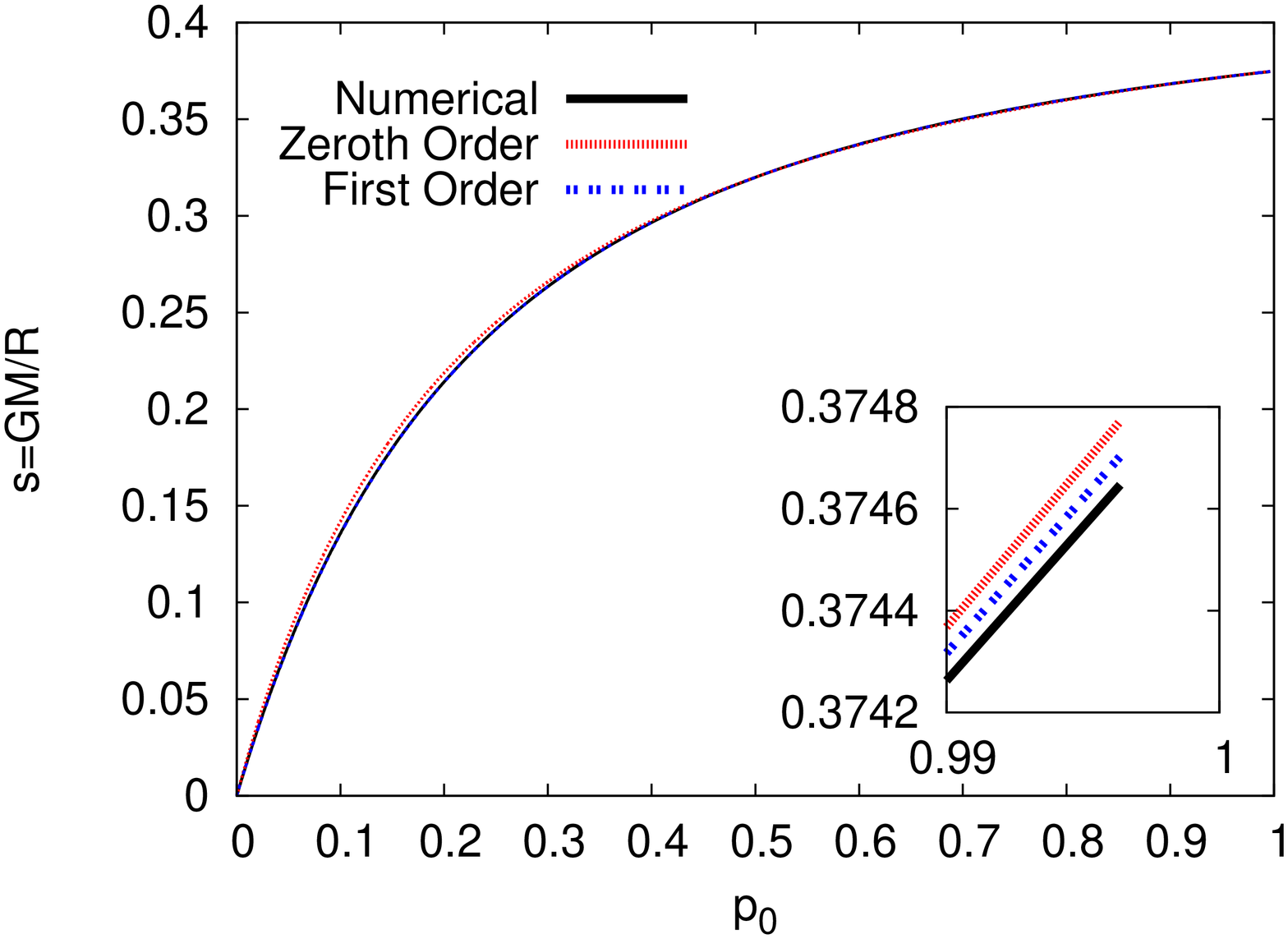,angle=270,width=0.4\hsize}
\epsfig{file=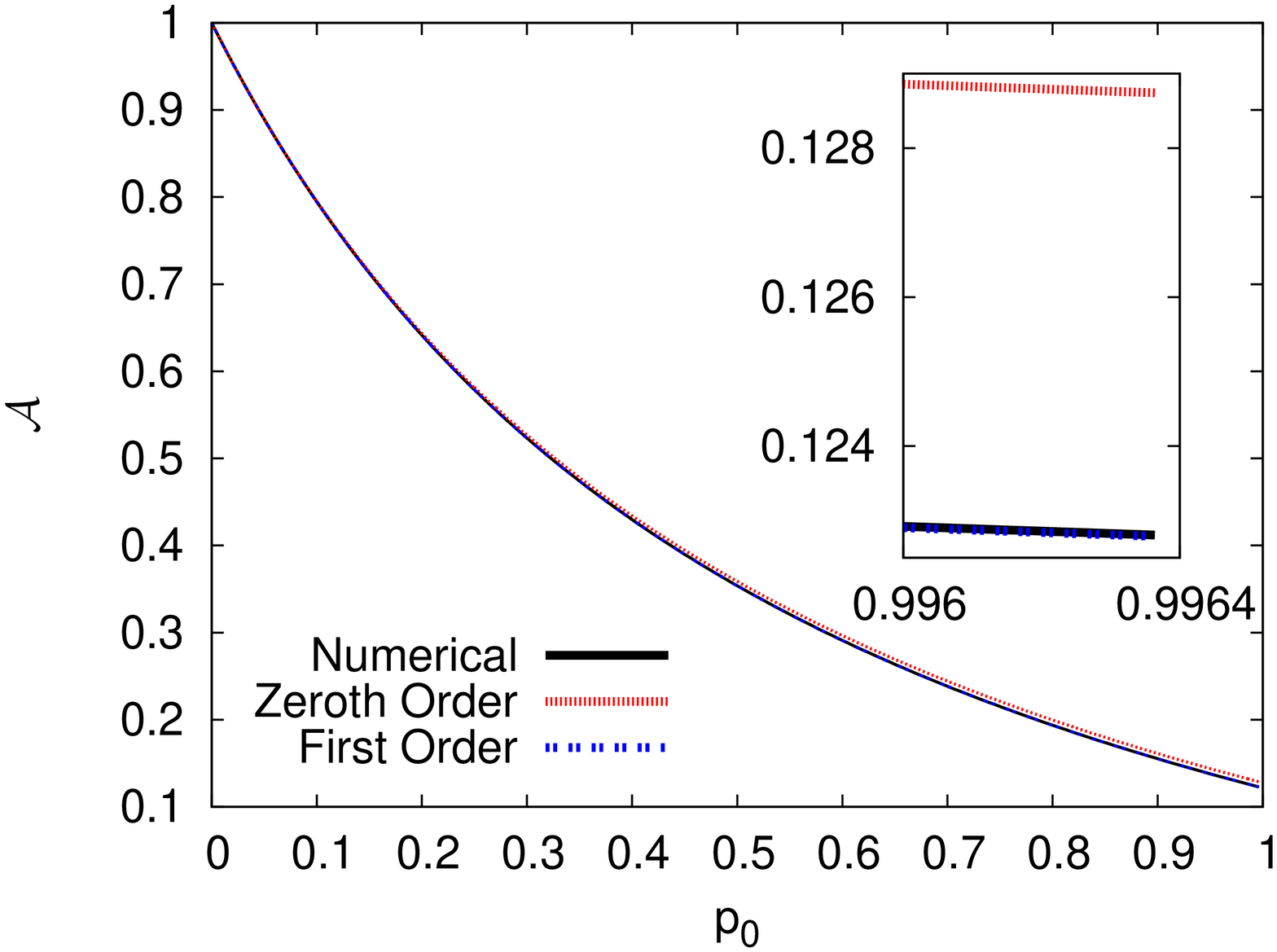,angle=270,width=0.4\hsize}
\epsfig{file=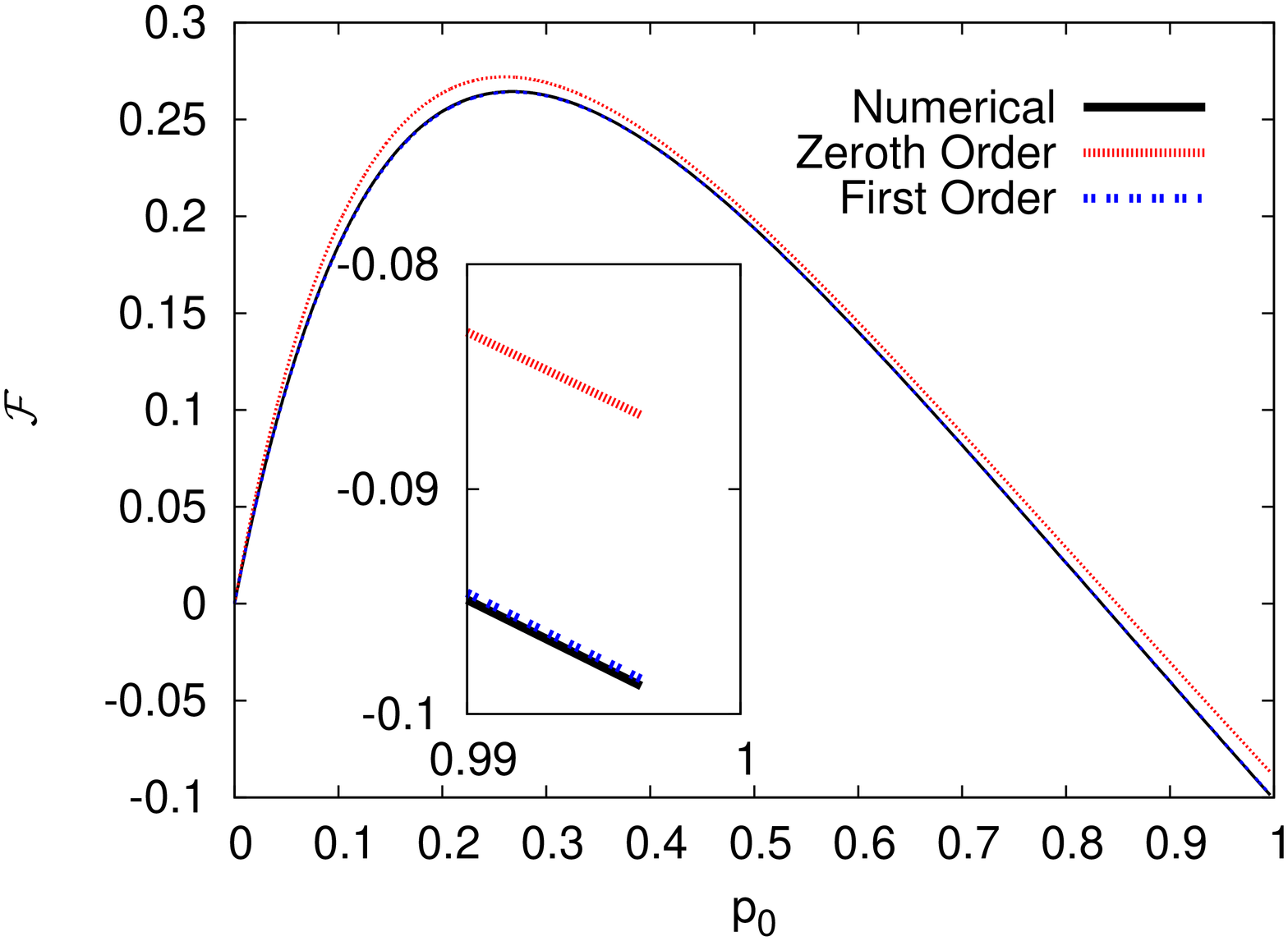,angle=270,width=0.4\hsize}
\epsfig{file=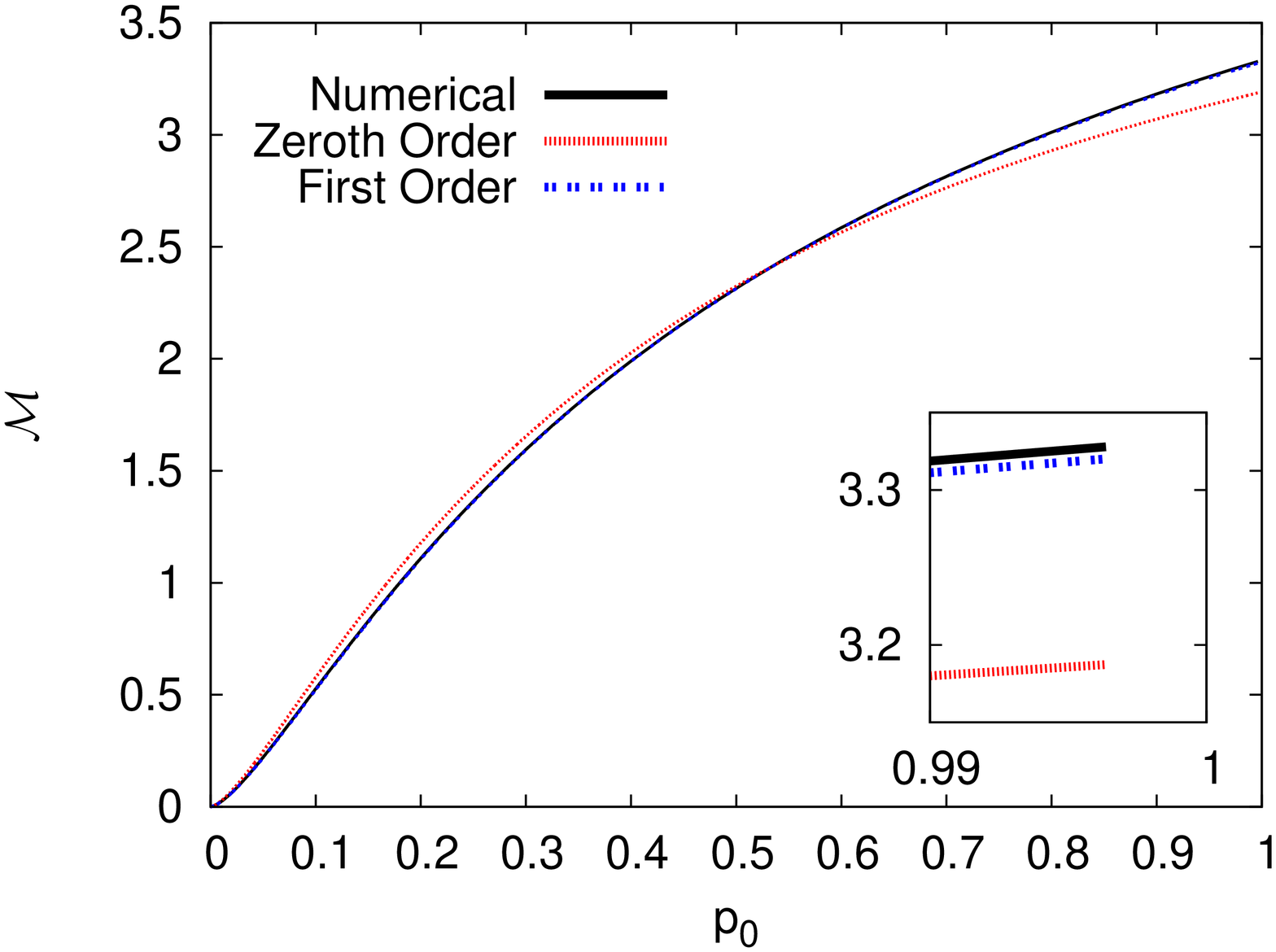,angle=270,width=0.4\hsize}
\caption{Comparison of physical quantities as functions of central
pressure, $\PressC = P_0/\rho_0$, calculated perturbatively and
numerically for an incompressible star in Brans-Dicke theory ($b_s
= 0$) with $a_{s}^{2}=0.1$. The plots show compactness, $s$ (top
left); external coupling, $\A = a_\ssA/a_s = Q/M a_s$, (top
right); $\F = (\phi_\infty - \phi_0)/a_s$ (bottom left); and $\M$
(bottom right). } \label{Fig12}}

\section{Conclusions}

In this paper we set up the equations of stellar
structure, with the stellar interior modeled as a
spherically symmetric, static fluid, and with gravity
described by a scalar-tensor theory with a
single light scalar coupling to matter only
through its coupling to a Jordan frame metric. For
practical reasons, and for the purposes of making
contact with earlier workers, we focus on the
special case where the scalar-matter coupling
function does not vary strongly with the
field, $a(\phi) \simeq a_s + b_s \phi$.

We seek solutions to these equations, for a variety
of equations of state, in the
special case where the scalar-matter
coupling at the stellar center is small,
$a^2_0 = a^2(\phi_0) \ll 1$. We obtain
solutions as perturbations to those
of General Relativity. By comparing these
solutions with explicit numerical
integration we verify that the perturbative
approximation works well throughout most
of the star.

These perturbative solutions
have the merit of being very simple to
integrate numerically, and of allowing
analytic solutions for some choices of
equation of state. This is very convenient
for efficiently exploring different
choices for the scalar properties, and
scalar-matter couplings.

We use these solutions to compute the
form of the observable relations that
are imposed among the external properties
of the stars by the condition that they
match continuously to the stellar interior.
There are two such relations among the
four external variables, $M$, $R$, $Q$ and
$\phi_\infty$, and our semi-analytic
approach allows a simple exposition of
how these relations depend on scalar
properties. These properties ultimately
underly any tests of scalar-tensor
theories using astrophysical systems,
such as binary pulsars.

Finally, these methods are applied to
the illustrative case of an incompressible
star, for which the density is constant.
In this case the solutions generated by
the $a_0$ expansion may be found analytically,
making the comparisons with numerical
results particularly simple. Again we
find that the perturbative expressions
agree well with the solutions obtained
by numerical integration.

\section*{Acknowledgements}
We thank Nemanja Kaloper and Maxim Pospelov for useful
discussions. This research was supported in part by funds from the
Natural Sciences and Engineering Research Council (NSERC) of
Canada. Research at the Perimeter Institute is supported in part
by the Government of Canada through Industry Canada, and by the
Province of Ontario through the Ministry of Research and
Information (MRI).

\appendix

\section{Expansions for large $b_s$}

This appendix evaluates the large-$b_s$ expansion for the
properties of incompressible stars.

The coefficients $c_{r}$ of the power series (\ref{heunseries})
can be written explicitly as
\begin{equation}
 c_{r} = \sum_{i=0}^{r} a_{i}^{(r)}b_{s}^{i}\,,
\end{equation}
where $a_{i}^{(r)}$ is a polynomial in $\PressC$ of degree $r$. It
follows from equation (\ref{recrel_new}) that $a_{r}^{(r)}$
satisfy a two-term recurrence relation, which can be solved
explicitly, giving
\begin{equation}\label{coef1}
 a_{r}^{(r)} = \frac{[6(1-3\PressC)]^{r}}{(2r+1)!}\,.
\end{equation}

The coefficients $a^{(r)}_{0}$ vanish for $r \geq 1$, and the
remaining coefficients are given by
\begin{equation}
\label{coef_higher} a^{(r)}_{r-k} = \frac{r! \cdot
[6(1-3\PressC)]^{r-2k} } {(2r+1)!(r-k-1)!}
\sum_{j=0}^{2k-1}P_{k,j}(\PressC)r^{j}\,, \qquad (k \geq 1,\ r-k
\geq 1)
\end{equation}
where $P_{k,j}$ is a polynomial of degree $2k$ with rational
coefficients. The first of these polynomials are given by
\begin{eqnarray}
P_{1,0}(\PressC) &=& 2(63\PressC^{2}-18\PressC+7)\,,
\\
P_{1,1}(\PressC) &=& 8(9\PressC^{2} + 9\PressC+4)\,,
\\
P_{2,0}(\PressC) &=&\textstyle{\frac{6}{5}} (-3321\PressC^{4} +
2916\PressC^{3}+6426\PressC^{2}+4644\PressC+679)\,,
\\
P_{2,1}(\PressC) &=& \textstyle{\frac{2}{5}} (5913\PressC^{4}
-21708\PressC^{3} + 29502\PressC^{2} + 17028\PressC + 2593)\,,
\\
P_{2,2}(\PressC) &=& \textstyle{\frac{48}{5}}
(729\PressC^{4}-729\PressC^{3}-1989\PressC^{2}-1251\PressC-176)\,,
\\
P_{2,3}(\PressC) &=& 32(9\PressC^{2}+9\PressC+4)^{2}\,,
\end{eqnarray}
and the higher ones can be computed from the relation
\begin{eqnarray}
&\displaystyle
\frac{r!}{12(r-l-1)!}\sum_{i=0}^{2l-1}P_{l,i}(\PressC)r^{i}
=\qquad\qquad
\nonumber\\
&\displaystyle \qquad\qquad =24 \sum_{j=l}^{r-1}
\frac{(2j+1)(j+2)(1+3\PressC)(1-3\PressC)^{2}j!}{(j-l)!}
\sum_{i=0}^{2l-5} P_{l-2,i}(\PressC)(j-1)^{i+1}
\nonumber\\
&\displaystyle \qquad\qquad +
\sum_{j=l}^{r-1}\frac{(1-9\PressC-6j\PressC)(1-3\PressC)jj!}{(j-l)!}
\sum_{i=0}^{2l-3} P_{l-1,i}(\PressC)j^{i}
\nonumber\\
\label{p_rec_rel} &\displaystyle \qquad\qquad
 + 4\sum_{j=l+1}^{r-1} \frac{(2j+1)(1+3\PressC)j!}{(j-l-1)!}
\sum_{i=0}^{2l-3}P_{l-1,i}(\PressC)(j-1)^{i}\,,
\end{eqnarray}
which holds for $l \geq 3$ and $r \geq l+1$.

From the identities \cite{gradshteyn}
\begin{eqnarray}
\sum_{k=0}^{n} (-1)^{k} \binom{n}{k} (k+a)^{n} &=& (-1)^{n}n! \,,
\qquad (n \geq 0,\ a \in \mathbb{R})
\\
\sum_{k=0}^{N}(-1)^{k}\binom{N}{k}(k+a)^{n-1} &=& 0 \,, \qquad (N
\geq n \geq 1,\ a \in \mathbb{R})
\end{eqnarray}
it follows that
\begin{equation}
P_{k,2k-1}(\PressC) = \frac{[8(9\PressC^{2}+9\PressC+4)]^{k}}{k!}
\,. \qquad (k \geq 1)
\end{equation}

Write
\begin{equation}
P_{k,j}(\PressC) = \sum_{l=0}^{2k} q_{k,j}^{(l)} \PressC^{l} \,.
\end{equation}

By means of telescopic summation, the identity (\ref{p_rec_rel})
can be used to derive recurrence relations for the coefficients
$q_{k,j}^{(l)}$. In general, these recurrence relations are very
complicated, but for $l=2k$, they are particularly simple:

\begin{eqnarray}
(k+1)q_{k,0}^{(2k)} + \sum_{i=1}^{2k-1}q_{k,i}^{(2k)} &=& 0  \,,
\\
(k+2)q_{k,1}^{(2k)} + \sum_{i=2}^{2k-1} (i+1) q_{k,i}^{(2k)} &=&
324 q_{k-1,0}^{(2k-2)} \,,
\\
(k+j+1)q_{k,j}^{(2k)} + \sum_{i=j+1}^{2k-1} \binom{i+1}{j}
q_{k,i}^{(2k)} &=& 216 q_{k-1,j-2}^{(2k-2)} + 324
q_{k-1,j-1}^{(2k-2)} \,,
\end{eqnarray}

where $k \geq 2$ and $2 \leq j \leq 2k-2$.

Now, $\varphi_{(0)}$ can be expanded in the form
\begin{eqnarray}
\varphi_{(0)}(z) &=& \frac{1}{b_{s}} \biggl[
 (a_{1}^{(1)}b_{s} + a_{0}^{(1)})z
 + (a_{2}^{(2)}b_{s}^{2} + a_{1}^{(2)}b_{s} + a_{0}^{(2)})z^{2}
\nonumber\\
&&\hphantom{\frac{1}{b_{s}} \biggl[} + (a_{3}^{(3)}b_{s}^{3} +
a_{2}^{(3)}b_{s}^{2} + a_{1}^{(3)}b_{s} + a_{0}^{(3)})z^{3} +
\ldots \ \biggr]
\nonumber\\
&=& \frac{1}{b_{s}}\biggl(
 a_{1}^{(1)} (b_{s}z) + a_{2}^{(2)} (b_{s}z)^{2} +
a_{3}^{(3)} (b_{s}z)^{3} + \ldots \biggr)
\nonumber\\
&&+\frac{1}{b_{s}^{2}} \biggl(  a_{0}^{(1)} (b_{s}z) + a_{1}^{(2)}
(b_{s}z)^{2} + a_{2}^{(3)} (b_{s}z)^{3} + \ldots \biggr) + \ldots
\nonumber\\
\label{beta_expansion} &\equiv&
\frac{\varphi^{(1)}_{(0)}(b_{s}z)}{b_{s}} +
\frac{\varphi^{(2)}_{(0)}(b_{s}z)}{b_{s}^{2}} + \ldots\,.
\end{eqnarray}
It follows from equations (\ref{coef1}) and (\ref{coef_higher})
that

\begin{eqnarray}
\label{t_intro} \varphi^{(1)}_{(0)}(b_{s}z) &=& \frac{\sinh
\sqrt{t}}{\sqrt{t}} -1 \,,
\\
\varphi_{(k+1)}^{(0)}(b_{s}z) &=&
\frac{1}{[6(1-3\PressC)]^{2k}}\sum_{j=0}^{2k-1}P_{k,j}(\PressC)f_{k,j}(t)\,,
\qquad (k \geq 1)
\end{eqnarray}
where
\begin{eqnarray}
f_{k,j}(t)&=&\sum_{r=k+1}^{\infty} \frac{r! \cdot
t^{r}r^{j}}{(2r+1)!(r-k-1)!} \nonumber
\\
&=& \left( t \frac{d}{dt} \right)^{j} \left(
-\frac{i\sqrt{t}}{2}\right)^{k+1} j_{k+1}(i \sqrt{t}) \,,
\end{eqnarray}
where
\begin{equation}
j_{n}(x) = \sqrt{\frac{\pi}{2x}} J_{n + \frac{1}{2}}(x) = (-x)^{n}
\left( \frac{1}{x} \frac{d}{dx}\right)^{n} \frac{\sin x}{x}
\end{equation}
are the spherical Bessel functions, and $t = 6b_{s}(1-3\PressC)z =
3b_{s}(1-3\PressC)(1-\sqrt{1-u/3})$.

The spherical Bessel functions satisfy the recursion relations
\begin{eqnarray}
x j_{l-1}(x) + x j_{l+1}(x) = (2l+1)j_{l}(x) \,, \nonumber
\\
lj_{l-1}(x) - (l+1)j_{l+1}(x) = (2l+1) j_{l}'(x) \,.
\end{eqnarray}

Therefore, the $f_{k,j}$ satisfy the recursion relations
\begin{eqnarray}
f_{k,1} = f_{k+1,0} + (k+1)f_{k,0} \,, \nonumber
\\
t f_{k-1,0} - 4 f_{k+1,0} = 2(2k+3)f_{k,0} \,.
\end{eqnarray}

The first of these relations can be used to write $f_{k,j}$
explicitly in terms of $f_{l,0}$:

\begin{eqnarray}
f_{k,j} = f_{k+j,0} + \sum_{m=1}^{j} \left[
\tilde{\sum}_{l_{1},\ldots,l_{m}=1}^{j-m+1} (k+l_{1}) \ldots
(k+l_{m})\right] f_{k+j-m,0} \,,
\end{eqnarray}
for $j \geq 1$. The tilde on the sum means that all numerical
factors which come from overcounting should be deleted. For
example,

\begin{equation}
\tilde{\sum}_{l_{1},l_{2}=1}^{2}(k+l_{1})(k+l_{2}) = (k+1)^{2} +
(k+1)(k+2) + (k+2)^{2} \,.
\end{equation}

Equations (\ref{match0_constrho})-(\ref{match1_constrho}) then
become
\begin{eqnarray}
\label{g0_app} \A_{(0)} &=& \frac{1+\PressC}{\PressC} \left\{
\frac{1}{2b_{s}} \left( \cosh \sqrt{T} - \frac{\sinh
\sqrt{T}}{\sqrt{T}}\right) + \sum_{k=1}^{\infty}\sum_{j=0}^{2k-1}
\frac{P_{k,j}(\PressC)f_{k,j+1}(T)}
{b_{s}^{k+1}[6(1-3\PressC)]^{2k}} \right\} \,,
\\
\label{f0_app} \mathcal{F}_{(0)} &=& \frac{1}{b_{s}} \left( \left[ 1 +
\frac{1+\PressC}{2\PressC}L \right] \frac{\sinh\sqrt{T}}{\sqrt{T}}
- \frac{1+\PressC}{2\PressC}L \cosh \sqrt{T} - 1\right)
\nonumber\\
&& + \sum_{k=1}^{\infty}\sum_{j=0}^{2k-1} \left( f_{k,j}(T) -
\frac{1+\PressC}{\PressC}Lf_{k,j+1}(T)\right)
\frac{P_{k,j}(\PressC)}{b_{s}^{k+1}[6(1-3\PressC)]^{2k}} \,,
\end{eqnarray}
where $T=6b_{s}\PressC(1-3\PressC)/(1+3\PressC)$ is the value of
the variable $t$ corresponding to $u=U$, and
$L=\log(1+\PressC)-\log(1+3\PressC)$. If $b_{s}>0$, then $T \leq
(6-4\sqrt{2})b_{s} \sim 0.34 b_{s}$. If $b_{s}<0$, then $T \geq
-(6-4\sqrt{2})|b_{s}| \sim -0.34 |b_{s}|$. The above expressions
(\ref{g0_app}) and (\ref{f0_app}) can be used to calculate the second
constraint (\ref{constraint2_fo}).

These perturbative solutions have a limited regime of
applicability. They break down as $b_{s} \to 0$, because the
$1/b_{s}$ expansion fails. But they also break down at the onset
of spontaneous scalarization.

%\pagebreak
%\section{Table of Notation}

\TABULAR[b]{|l|l|l|l|}{\hline Symbol & Definition & Equation &
Meaning
\\
\hline $G$ & & (\ref{action}) & Einstein-frame gravitational
constant
\\
\hline $g_{\mu \nu}$ & & (\ref{action}) & Einstein-frame metric
\\
\hline $\phi$ & & (\ref{action}) & Einstein-frame scalar field
\\
\hline $\tilde{g}_{\mu \nu}$ & $A^{2}(\phi) g_{\mu \nu}$ &
(\ref{action})& Jordan-frame metric
\\
\hline $a(\phi)$ & $d (\log A(\phi))/d\phi$ & (\ref{fieldeq2}) &
Scalar-matter coupling function
\\
\hline $b(\phi)$ & $d a(\phi) / d\phi$ & (\ref{ppn}) & Derivative
of scalar-matter coupling function
\\
\hline $a_{s},b_{s}$ & $a(\phi)=a_{s}+b_{s}\phi$ & (\ref{quadmod})
& Parameters of the quasi-Brans/Dicke model
\\
\hline $a_0$ & $a(\phi_{0})$ & (\ref{eqssnew1}) & Scalar-matter
coupling in centre of star
\\
\hline $\varphi$ & $(\phi - \phi_{0})/a(\phi_{0})$ &
(\ref{chvar1}) & Shifted and scaled Einstein-frame scalar field
\\
\hline $r$ & & (\ref{sscoords}) & Schwarzschild coordinate radial
variable
\\
\hline $u$ & $8 \pi G \rho_{0} A^{4}(\phi_{0})r^{2}$ &
(\ref{chvar1}) & Radial variable
\\
\hline $z$ & $(1-\sqrt{1-u/3})/2$ & (\ref{chvar_bn0}) &
Dimensionless radial variable for constant-density stars
\\
\hline $t$ & $6b_{s}(1-3\PressC)z$ & (\ref{t_intro}) &
Dimensionless radial variable for constant-density stars
\\
\hline $\PressC$ & $P_{0}/\rho_{0}$ & (\ref{ics_1}) &
Pressure-to-density ratio in centre of star
\\
\hline $\mu(r)$ & $(1-g^{rr})/2$ & (\ref{sscoords}) & Related to
$rr$ component of Einstein-frame metric
\\
\hline $\nu(r)$ & $\log(-g_{tt})$ & (\ref{sscoords}) & Related to
$tt$ component of Einstein-frame metric
\\
\hline $n(r)$ & & (\ref{barmass}) & Baryon number density
\\
\hline $R$ & $P(R)=0$ & (\ref{match1}) & Schwarzschild coordinate
radius of star
\\
\hline $U$ & $8\pi G\rho_{0}A^{4}(\phi_{0})R^{2}$ & & Value of $u$
corresponding to $r=R$
\\
\hline $T$ & $6b_{s}\PressC(1-3\PressC)/(1+3\PressC)$ & & Value of
$t$ corresponding to $r=R$
\\
\hline $M$ & $g_{tt} = -1 + 2GM/r + \ldots $ & (\ref{match1}) &
ADM Mass of Einstein-frame metric
\\
\hline $s$ & $GM/R$ & & Self-gravity, or compactness, of star
\\
\hline $M_\ssB$ & & (\ref{barmass}) & Baryonic mass of star
\\
\hline $Q$ & $\phi = \phi_{\infty} - GQ/r + \ldots$ &
(\ref{match2}) & Scalar charge of star
\\
\hline $a_{\ssA}$ & $Q/M$ & & Effective scalar-matter coupling of
star
\\
\hline $\A$ & $a_{\ssA}/a(\phi_{0})$ &  & Rescaled effective
scalar-matter coupling of star
\\
\hline $\phi_{\infty}$ & $\phi(r=\infty)$ & (\ref{match3}) &
Asymptotic value of Einstein-frame scalar field
\\
\hline $\phi_{0}$ & $\phi(r=0)$ & (\ref{ics_1}) & Value of
Einstein-frame scalar field in centre of star
\\
\hline $\mathcal{F}$ & $(\phi_{\infty}-\phi_{0})/a(\phi_{0})$ & &
Change in $\phi$
\\
\hline $P$ & & (\ref{perf_fluid}) & Jordan-frame pressure
\\
\hline $\Press $ & $P/\rho_{0}$ & (\ref{eqss1}) & Rescaled
Jordan-frame pressure
\\
\hline $\Dens(\Press ;\PressC)$ & $\rho/\rho_{0}$ & (\ref{eqss1})
& Equation of state
\\
\hline $\rho$ & & (\ref{perf_fluid}) & Jordan-frame mass-energy
density
\\
\hline}{Table of Notation}

\clearpage

\end{document}